\documentclass[
aps,%
12pt,%
final,%
notitlepage,%
oneside,%
onecolumn,%
nobibnotes,%
nofootinbib,%
superscriptaddress,%
noshowpacs,%
centertags]
{revtex4}
\usepackage[russian]{babel}
\usepackage{graphicx}
\usepackage{amsmath}
\usepackage{xcolor}
\graphicspath{ {figs/} }
\begin{document}
\selectlanguage{english}
\title{Sources of confusion noise in the infrared wavelength range}
\author{\firstname{A.A.}~\surname{Ermash}}
\email[]{aermash@asc.rssi.ru}
\affiliation{
Lebedev Physical Institute, Russian Academy of Sciences, Leninskii pr. 53, Moscow, 119991 Russia
}
\author{\firstname{S.V.}~\surname{Pilipenko}}
\email[]{spilipenko@asc.rssi.ru}
\affiliation{
Lebedev Physical Institute, Russian Academy of Sciences, Leninskii pr. 53, Moscow, 119991 Russia
}
\author{\firstname{E.V.}~\surname{Miheeva}}
\email[]{helen@asc.rssi.ru}
\affiliation{
Lebedev Physical Institute, Russian Academy of Sciences, Leninskii pr. 53, Moscow, 119991 Russia
}
\author{\firstname{V.N.}~\surname{Lukash}}
\email[]{lukash@asc.rssi.ru }
\affiliation{
Lebedev Physical Institute, Russian Academy of Sciences, Leninskii pr. 53, Moscow, 119991 Russia
}

\begin{abstract}

In this paper we use the model of extragalactic background light to investigate the factors that have influence on the confusion noise.
It was shown that
(1) Large-Scale Structure of the Universe is an important factor;
(2) gravitational lensing does not have a significant effect on the confusion noise;
(3) lower redshift limit of objects that contribute to the confusion noise does not depend on the wavelength and is about $z_{min}\sim 0.5-0.6$, while upper redshift limit gradually changes from  $\sim4$ to $\sim3$ with the increase of wavelength from 70$\mu m$ up to 2000$\mu m$;
(4) at rather short wavelengths ($\simeq70\mu m$) galaxies with luminosities in the range $10^7L_\odot$ -- $10^9L_\odot$ give the most contribution to the confusion noise, while at larger wavelengths (650-2000$\mu m$) their luminosities are greater than $L\geq10^{10}L_\odot$; 
(5) contribution from objects with different color characteristics is considered;
(6) the variability of the extragalactic background on the timescale from 1 day to 1 year is noticeable at short wavelengths (70--350$\mu m$) and manifests at fluxes ${}^<_\sim$ 1~mJy.

\end{abstract}

\maketitle
\section{Introduction}
In recent years many papers were dedicated to the Cosmic Infrared Background that is created by distant unresolved galaxies.
This background has significant variations on angular scale from arcseconds to arcminutes which leads to the so-called ``confusion'' that makes extraction of individual sources a challenging task.

Estimation of parameters of this noise is very important for observational strategies of future instruments.
If multiwavelength observations are available it is possible to use several different approaches to extract sources from the maps 
(see., e.g., \cite{2012A&A...542A..81M,2013A&A...560A..63M,2017A&A...607A..64M}).
The method of using differential maps worth special mention.
It allows one to extract objects with chosen peculiar color characteristics (see, e.g. \cite{2016MNRAS.462.1989A} ¨ \cite{2014ApJ...780...75D}).

Earlier we have created a model of Extragalactic Background Light  (see. \cite{2020AstL...46..298E}) and compared it's results with predictions of other models and observational data in the following aspects:
1) differential source number counts
2) change of source number counts with redshift
3) EBL spectrum
4) prediction of confusion noise.

The current paper is a result of further research in this area.
In order to develop methods of extracting sources below the confusion limit it is necessary to get information about the redshift, luminositites and colors of objects that give the most significant contribution to the confusion  noise.
An interesting example of an attempt to extract sources below confusion using the information from several wavelengths can be found in~\cite{2018ApJ...853..172L}.
For further optimization of the model we need to know whether the Large Scale Structure of the Universe has significant influence on the estimates of the confusion noise.
Another important question is the variability of the EBL.
This is crucial for estimation of perspectiveness of observation of astronomical transients and minor bodies in solar system.

Currently there are two main approaches of modeling the EBL.
The first one that is often called ``phenomenological'' is based on the estimations of luminosity functions and spectra of galaxies 
(see., e.g., \cite{2013MNRAS.428.2529H}).
It is also sometimes called ``backward evolution model'' because it is based on the local luminosity function that is evolved ``backwards'' in time with increase of redshift.
The evolution parameters are estimated by fitting the observational data (number counts, luminosity functions of different populations, source redshift distributions etc.).
The main advantage of such an approach is quite accurate reproduction of observational data.
But at the same time they lack prediction power at wavelengths and redshifts for which we do not have observational data.
Models of this kind can be found in \cite{2011MNRAS.418..176R,2003ApJ...585..617D,2001ApJ...556..562C}.

The second approach that is often called ``semianalytical'' is based on the numerical simulations of dark matter distribution in the universe.
To every halo of dark matter a galaxy with certain parameters is assigned.
A certain spectral energy distribution is set for each galaxy.
Such a model allows to estimate source number counts, EBL spectrum and the confusion noise.
The following papers present models obtained with such approach:
\cite{2010MNRAS.405....2L,2015MNRAS.446.1784C,2008MNRAS.391..420S,2010MNRAS.405..705F,2015A&A...575A..32C,2015A&A...575A..33C}.
It should be noted that such a classification is significantly simplified and there are a lot of different models that utilize both approaches to some extent.
Our model presented in \cite{2020AstL...46..298E} can be classified as semianalytical because it is based on the numerical evolution of the dark matter halos.

Due to the fact that the confusion noise affects the processing of observations, its adequate modeling is vital for success of future space telescopes, such as Millimetron that is currently is in stage of active development.
It will be launched in 2029.
Platform with 10m cooled mirror will move on the halo-orbit in the vicinity of the Lagrange point L2 of the Sun - Earth system.
The more detailed description of the mission can be found in the following papers:
~\cite{2017ARep...61..310K,2014PhyU...57.1199K,2012SPIE.8442E..4CS}
and on the official website of the project\footnote{millimetron.ru}.

Millimetron will contain two matrix detectors~--- SACS~(Shortwave Array Camera Spectrometer) and LACS~(Longwave Array Camera Spectrometer) for short and long wavebands, respectively.
SACS and LACS will have four bands each: 70, 110, 250, 350$\mu m$ and 650, 850, 1100, 2000$\mu m$, respectively.

Short wave matrix spectrometer (SACS) will consist of two main parts~-- the matrix photometer operating in the whole frequency range,
 which is divided into several sub-bands by dichroic beam splitter, and a matrix spectrometer, the spectral resolution of which will be determined by the input optical filter.
A similar approach was used in the PACS receiver\footnote{http://www.cosmos.esa.int/web/herschel/science-instruments} that successfully operated as a part of the Herschel Space Observatory.
LACS is similar to the SPIRE receiver.

\section*{Model of the Extragalactic Background}
The model of the Extragalactic  Background was based on galaxy formation and evolution model eGalICS (\cite{2015A&A...575A..32C} ¨ \cite{2015A&A...575A..33C})
and is a further development of ideas discussed in \cite{2020AstL...46..298E}.
For model galaxies we created a SED library with help of the GRASIL code~\cite{1998ApJ...509..103S} and each disk or bulge component was 
assigned a certain spectrum.
For active galactic nuclei we used the AGN type 1 SED that was proposed by~\cite{2017ApJ...841...76L}. 
The effects of gravitational lensing were taken into account with the approximation of point lens model.
As a final step we created a model cone from simulation cubes.

One of the main results are the model maps of the EBL for the wavelengths of the detectors of Millimetron.
Figure~1 shows an example of such a map for 850$\mu m$ waveband with angular size $10\times10$ arcmin.
As can be clearly seen, there are many closely placed objects and their complete photometrical separation poses a challenging task.
The problem is so severe that recently authors of~\cite{2013MNRAS.434.2572H} proposed a new term~-- ``Submillimeter source'' instead of ``Submillimeter galaxy''.

\begin{figure}[h]
\centering
\includegraphics[width=0.5\textwidth]{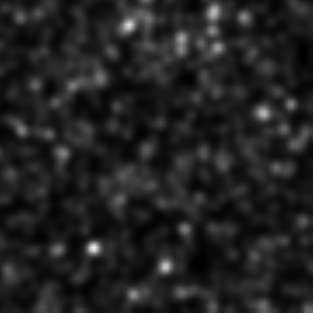}
\caption{
Model map of the extragalactic background of the 850$\mu m$ waveband created for a telescope with 10m main mirror.
Size of the map is $10\times10$ arcmin.
}
\label{fig:model_map}
\end{figure}

\begin{figure}[h]
\centering
\includegraphics[width=0.5\textwidth]{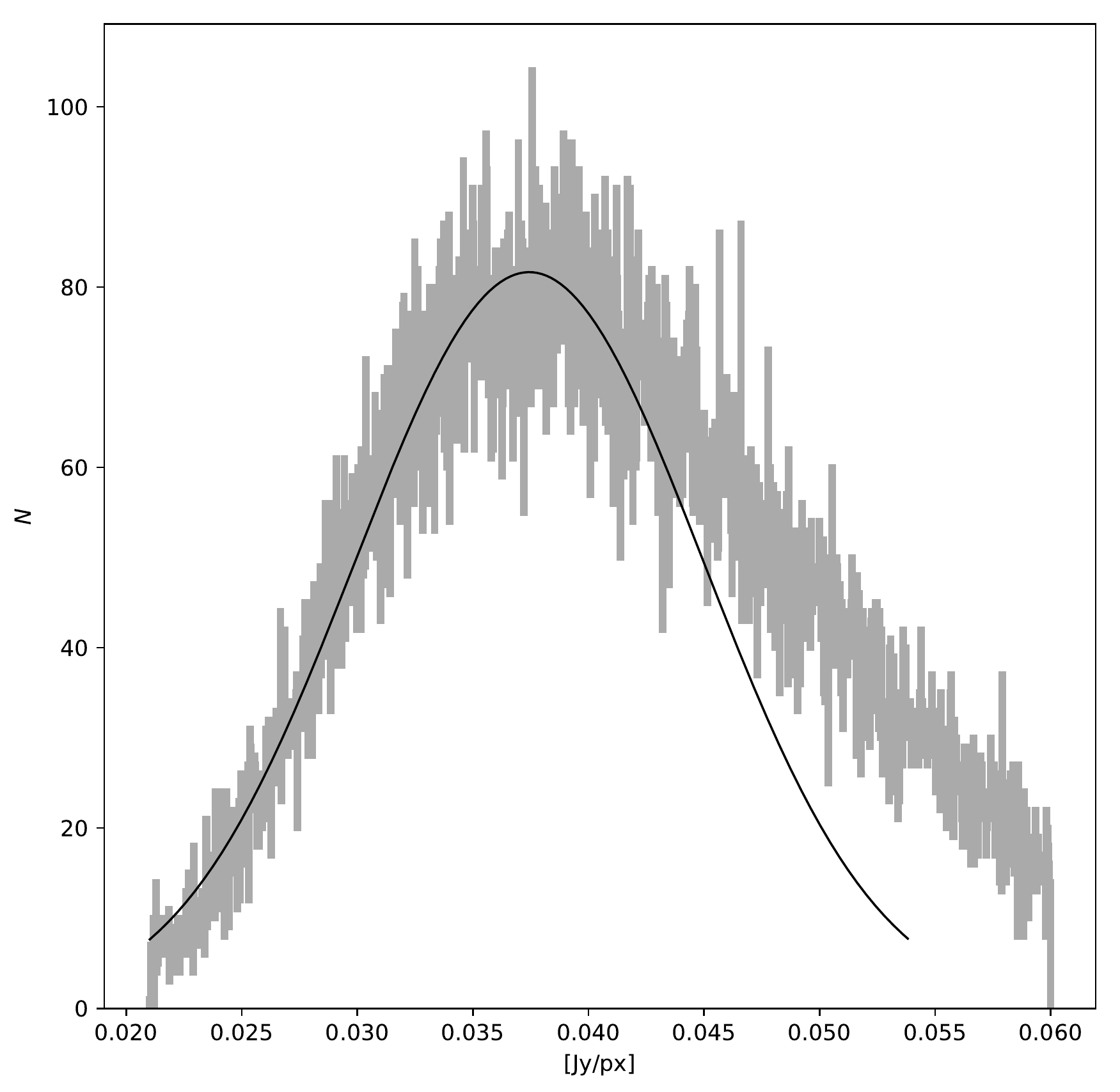}
\caption{
Pixel histogram of the model map of the extragalactic background on the 2000$\mu m$ wavelength.
The X axis shows the flux in pixel, the Y axis is the number of pixels with certain flux.
Distribution derived from the model map is shown in gray, black line is the Gauss curve fitted to the left side of the pixel histogram.
}
\label{fig:pixel_hist}
\end{figure}

Currently there are several approaches of estimating the confusion noise.
Some utilize only the information about the number counts curve, others are based on the analysis of maps.
In the simplest case the confusion noise is estimated as $\sigma$  of the flux in pixels, see, e.g. \cite{2008A&A...481..885F}.
On the other hand, as was shown in \cite{2010A&A...518L...5N}, \cite{2009ApJ...707.1729M} and \cite{2015A&A...579A..93L}, the flux distribution in pixels does not always follow the Gaussian curve.
An example of such deviation is shown on fig.~\ref{fig:pixel_hist}.
That is why in some papers (see, e.g. \cite{2009ApJ...707.1729M}) it was proposed to define the confusion noise as a dispersion of the Gaussian obtained 
by fitting only the left side of the pixel histogram.
Such an approach allows to exclude the influence of resolved bright objects that contribute to the flux distribution, but not to the confusion noise.

But there is a certain obstacle in using such an approach for our task.
If we create a map that contains objects with certain interval of some parameter, e.g. redshift $z$, the histogram might deviate from the Gaussian shape even in
its left side.
In this case the $\sigma$ value significantly depends on the part of the histogram selected for fitting.
In this case we used the Full Width at Half Maximum (FHWM) of the left side of the histogram.
In case where there is no prominent maximum on the histogram we did not perform the estimation.
The values of the confusion noise for the detectors of the Millimetron mission are given in Table~1.

\begin{table} 
\caption{
Confusion noise estimations for the wavebands of the Millimetron telescope.
}
\label{tab:conf}  
\begin{center} 
 \begin{tabular}{|c|c|c|} 
 \hline
  Wavelength,     & \multicolumn{2}{|c|}{Confusion noise}   \\
  {$\mu m$}       & $\sigma$, mJy  & $FWHM/2.355$, mJy       \\
 \hline
70   &$(2.51\pm0.07)\times10^{-5}$  & $(2.79\pm0.08)\times10^{-5}$ \\
110  &$(2.40\pm0.07)\times10^{-4}$  & $(2.68\pm0.08)\times10^{-4}$ \\
250  &$(9.69\pm0.48)\times10^{-3}$  & $(1.07\pm0.05)\times10^{-2}$ \\
350  &$(4.44\pm0.46)\times10^{-2}$  & $(4.88\pm0.53)\times10^{-2}$ \\
650  &$(3.26\pm0.17)\times10^{-1}$  & $(3.59\pm0.20)\times10^{-1}$ \\
850  &$(2.99\pm0.15)\times10^{-1}$  & $(3.21\pm0.18)\times10^{-1}$ \\
1100 &$(2.19\pm0.07)\times10^{-1}$  & $(2.34\pm0.08)\times10^{-1}$ \\
2000 &$(7.95\pm0.44)\times10^{-2}$  & $(8.19\pm0.53)\times10^{-2}$ \\
 \hline
 \end{tabular}
 \end{center} 
\end{table} 

As can be seen, for regular model maps the estimations of the confusion noise as $\sigma$ of the Gaussian curve and as $FWHM/2$ are equal within
the errorbars because the shape of the left side of the histogram follows the Gaussian.
Errors were estimated as a standard deviation of 10 test realizations of the model  cone. 

\section*{Influence of the Large Scale Structure and the gravitational lensing} 

\label{sec:lss_influence} 
During the creation of the model maps from simulation cubes a problem of the ``effect of perspective'' arises.
It is caused by the fact that the Large Scale Structure evolves slowly and in the cone the same part of the structure hits the line of sight several times.

Currently there are several ways to deal with this effect.
If for the scientific task model maps of moderate size suffice then it is possible to set such an angle of line of sight that it will cross a single area of a
model cube only once.
Such an approach was successfully implemented in, e.g.~\cite{2017AstL...43..644P}.
Another way to deal with this effect is to rotate each cube randomly 10 degrees around two coordinate axes (see \cite{2017A&A...607A..89B}).

Another effective way to deal with this problem was proposed in \cite{2005MNRAS.360..159B}.
Each cube underwent the following transformations: random shift along each axis, random rotation to 0, 90, 180 or 270 degrees, mirroring.
We chose this approach because it proved its effetiveness in eliminating the repeated structures on model maps.

\begin{figure}[h] 
\centering
\includegraphics[width=\textwidth]{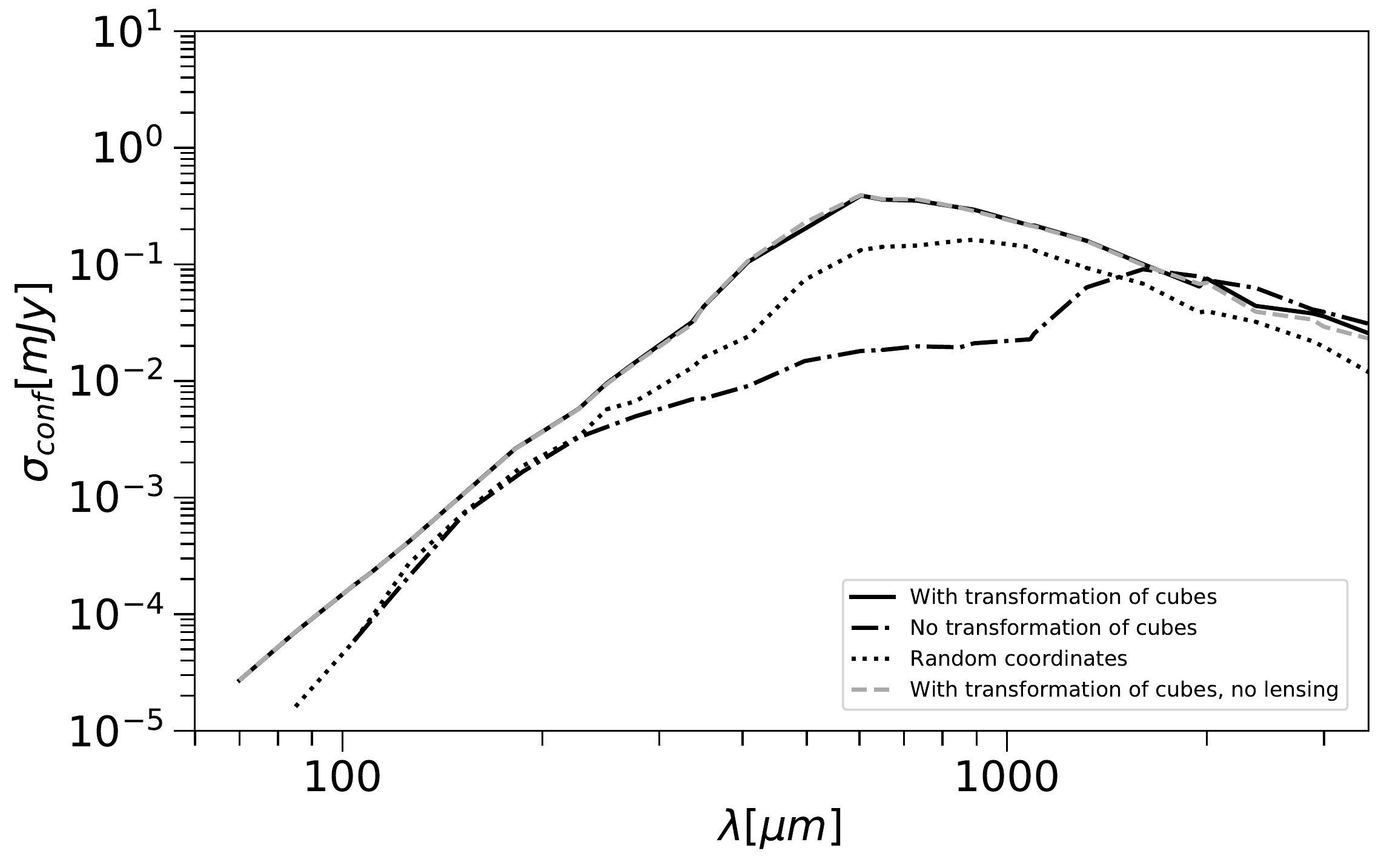} 
\caption{
Influence of the Large Scale Structure of the Universe on the confusion noise in different wavebands.
The X axis~-- wavelength in micrometers, the Y axis~-- value of the confusion noise in mJy.
Black solid line shows the estimation for maps with correction of the perspective effect applied.
Black dot-dashed line~-- without any correction.
Dotted line~--- after elimination of the Large Scale Structure.
Gray dotted line shows the confusion in case when effects of perspective were corrected, but lensing effects were not included in calculations.
} 
\label{fig:lss} 
\end{figure}

In order to understand the confusion noise we must estimate the significance of influence of the Large Scale Structure on the confusion noise.
We estimated the confusion noise on the maps created without any correction.
Also we created model maps from the cone, where coordinates of objects were set randomly within each cube thus eliminating the Large Scale Structure.
Fig.~\ref{fig:lss} illustrates the results.

The characteristic size of the cells of the Large Scale Structure (distance between major walls) is about 100 Mpc.
This value corresponds to the double correlation length of initial perturbations of the velocity field of matter, see e.g.\cite{2004A&A...422..423D} (equation (4)).

Let us consider the influence of the Large Scale Structure of the Universe on the confusion noise more closely.
Thick solid black line on fig.~\ref{fig:lss} (partially covered by gray dashed line) shows the values of confusion noise derived from cone that was assembled 
with transformation of cubes applied.
Black dot-dashed line shows the case when no transformations were applied to cubes of simulation.
The deviation between different models reaches significant values.

We also created a model cone in which coordinates of all objects were set random within each cube.
The confusion noise for this case is shown as thin dotted line.
The shape of this curve in general corresponds to the case when transformations were applied but lies about a half an order of magnitude lower.
It should be stressed that number counts of sources are equal in these three cases.
So in order to correctly estimate the confusion noise the Large Scale Structure must be correctly taken into account.

Figure~3 shows the results for one single realization of the $10\times10$ arcmin map.
In order to estimate the dependence of aforementioned effects on the area of survey we did the following.
Large map one square degree in size was split in 4, 9, 16 and so on square parts and on each of these parts the confusion noise was estimated.
The error was estimated as deviation of the confusion values from square to square for each division number.
The obvious expected result is that the confusion value is the same within errorbars, while the errors themselves will increase.
This is depicted on Figure~4.
The data for the case when no lensing was included in model is not shown because of the equality of the trends.
Left panel contains data for models with Large Scale Structure present in the cone while on the right panel we show trend for cone with random coordinates within model cubes.
As can be clearly seen, at small area maps the dispersions of estimations of the confusion noise are significantly larger.
It can be explained by the presence of areas of various density on small angular scales.
This is especially prominent for two shortest wavebands (70 and 110$\mu m$).
This again accentuates the necessity of including the Large Scale Structure into the model.
Also areas of different density should be included in the analysis.

\begin{figure}
\centering
\includegraphics[width=0.49\textwidth]{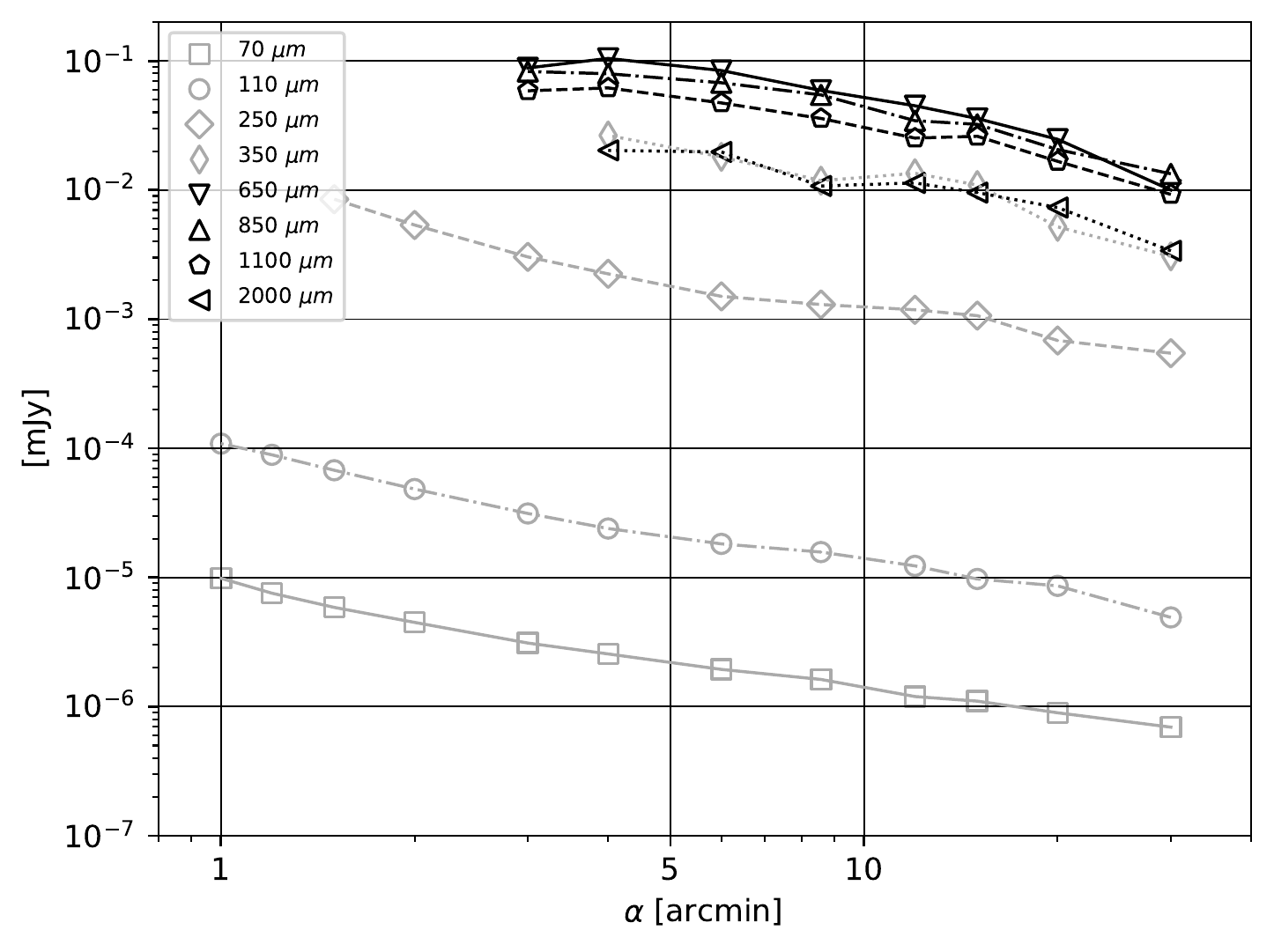}
\includegraphics[width=0.49\textwidth]{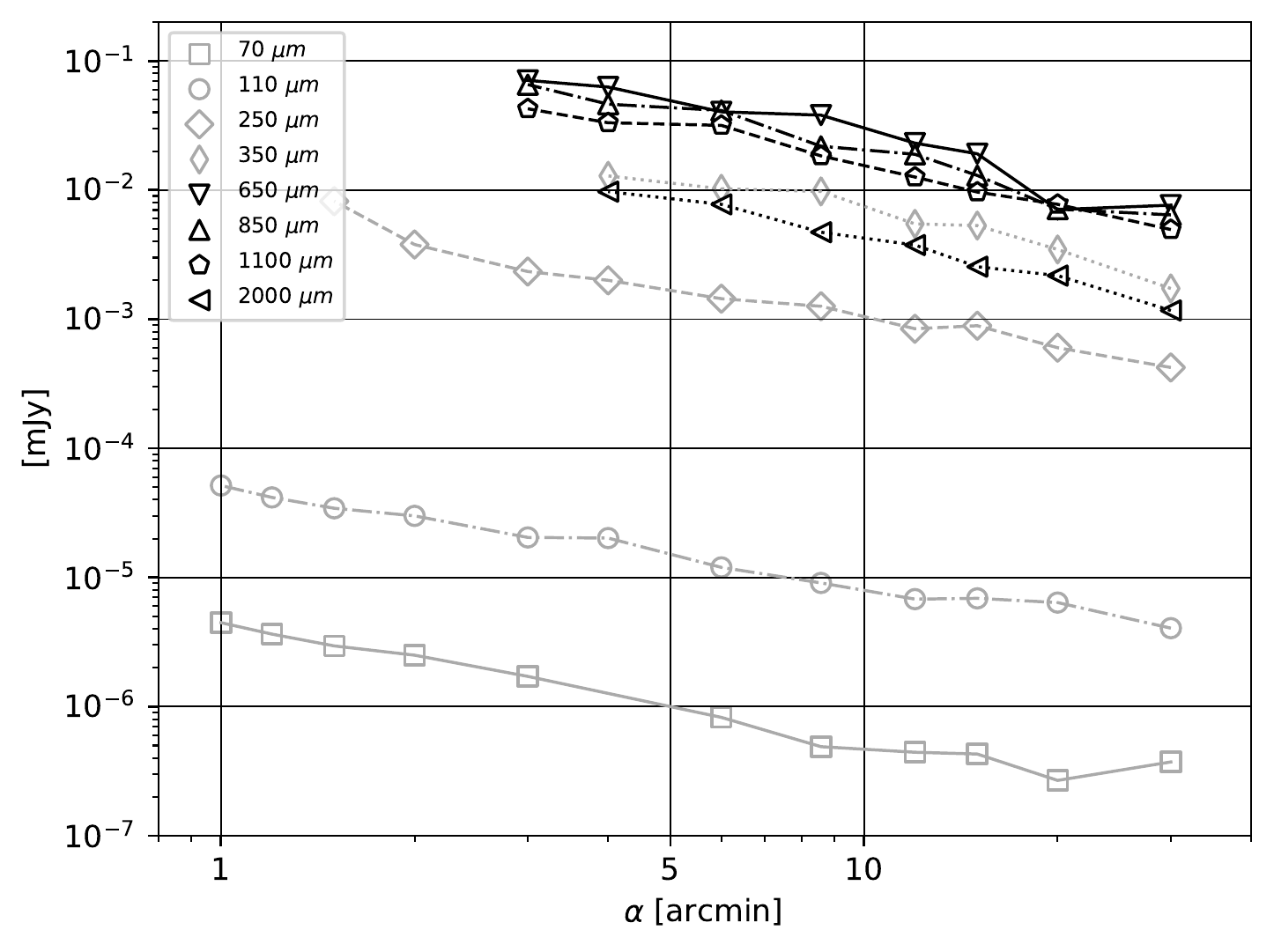}
\caption{
The dependence of the estimations of confusion noise on the area of model maps.
The X axis~-- side of the model map square.
The Y axis~-- the error in confusion noise.
Model with Large Scale Structure is shown on the left panel, model with random coordinates of objects is on the right.
}
\label{fig:small_squares}
\end{figure}

Gravitational lensing increases the brightness of sources that in turn builds up the number of sources brighter than certain flux value.
This effect must be taken into account especially in case when the number counts curve has steep slope in certain range of fluxes.

Let us consider for example a simple case when the number counts curve is a horizontal flat line.
In this case we have $N$ objects with flux 1mJy per $dex$ per unit of sky area and the same amount of objects with flux 2mJy.
The probability of 2x magnification is 0.1\%.
Then the amount of objects with 2mJy flux will be:
\begin{equation}
N\times(1-1/1000)+N\times1/1000=N.
\end{equation}
Situation changes drastically if the number counts curve has a steep slope.
If the magnification probability and number of objects with 2mJy flux are the same, but the amount of objects with 1mJy flux is much greater~-- $10000N$,
the resulting amount of objects with 2mJy flux will be:
\begin{equation}
N-N\times1/1000+10000N\times1/1000=10.999N
\end{equation}
the amount of sources increased one order of magnitude.
Figure~\ref{fig:lss} illustrates the effect of gravitational lensing. 
Black solid line and gray dashed line show the estimations of confusion noise with and without gravitational lensing, respectively.
As can be seen, the difference between these curves does not exceed the thickness of lines on the plot in all the wavelength range.
The gravitational lensing does not play crucial role for confusion noise analysis.

\section*{Contribution into the confusion noise from various sources} 

\subsection*{Contribution into the confusion noise from objects at different redshifts}

\begin{figure}
\includegraphics[width=0.32\textwidth]{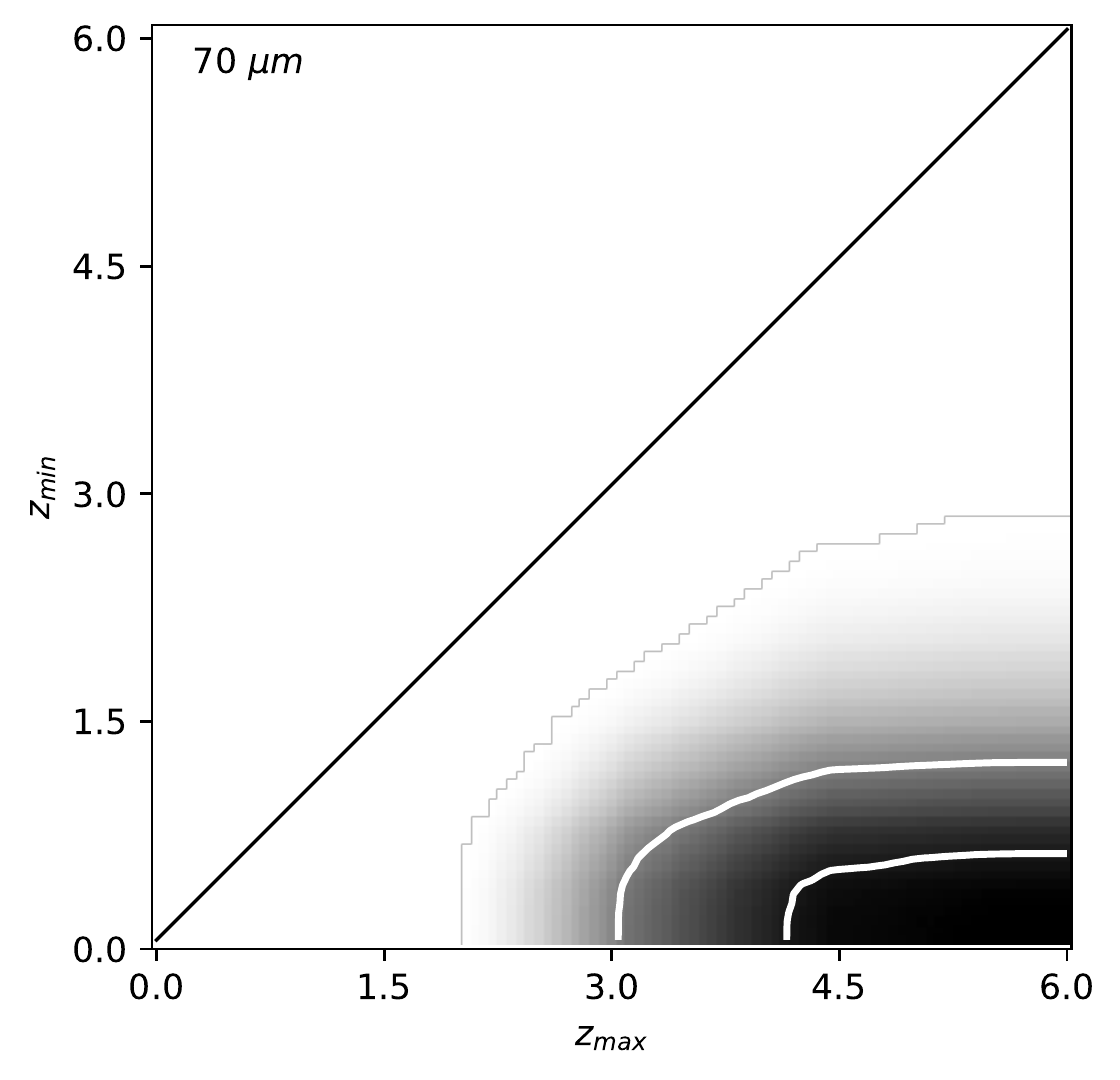}
\includegraphics[width=0.32\textwidth]{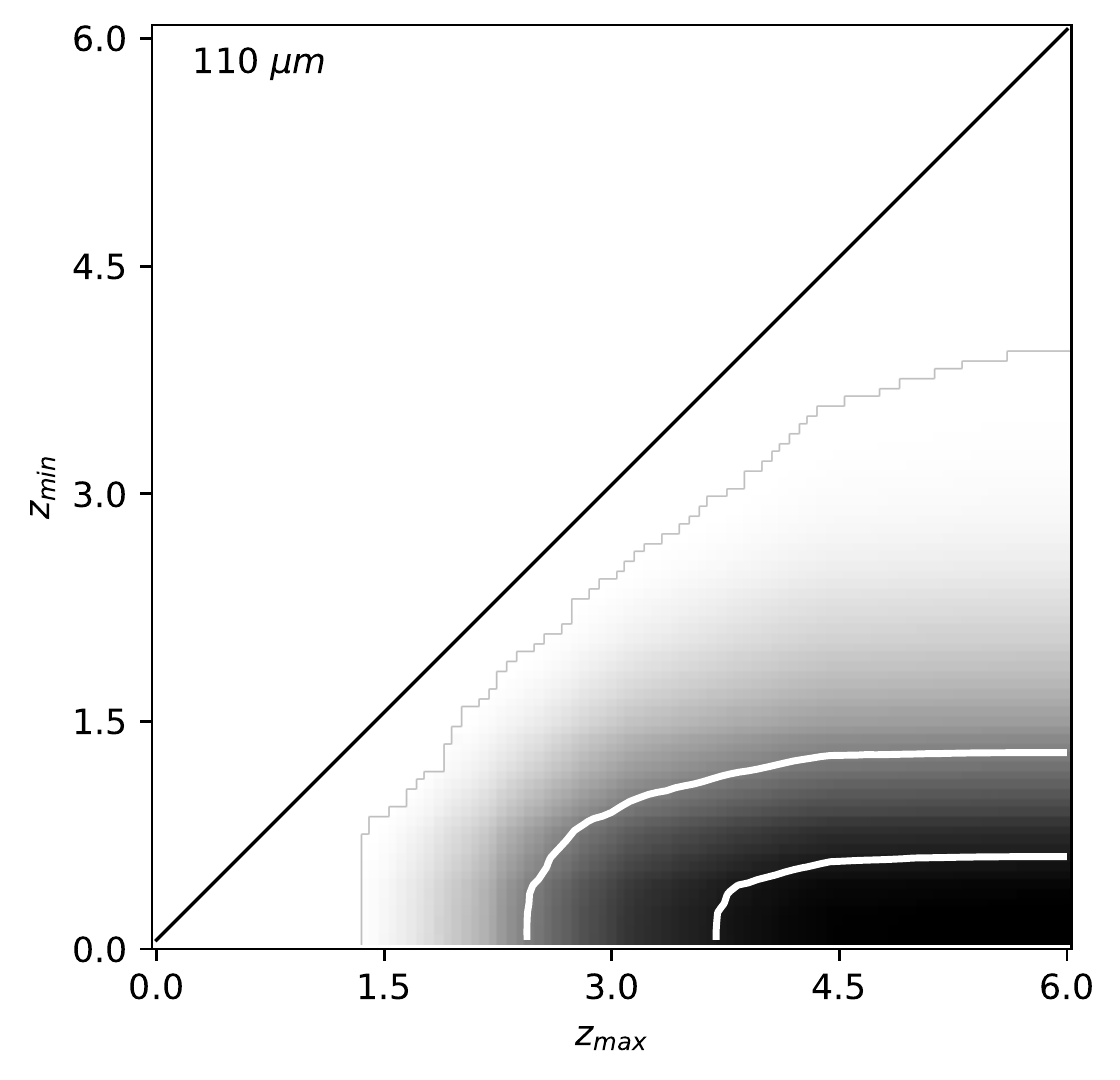}
\includegraphics[width=0.32\textwidth]{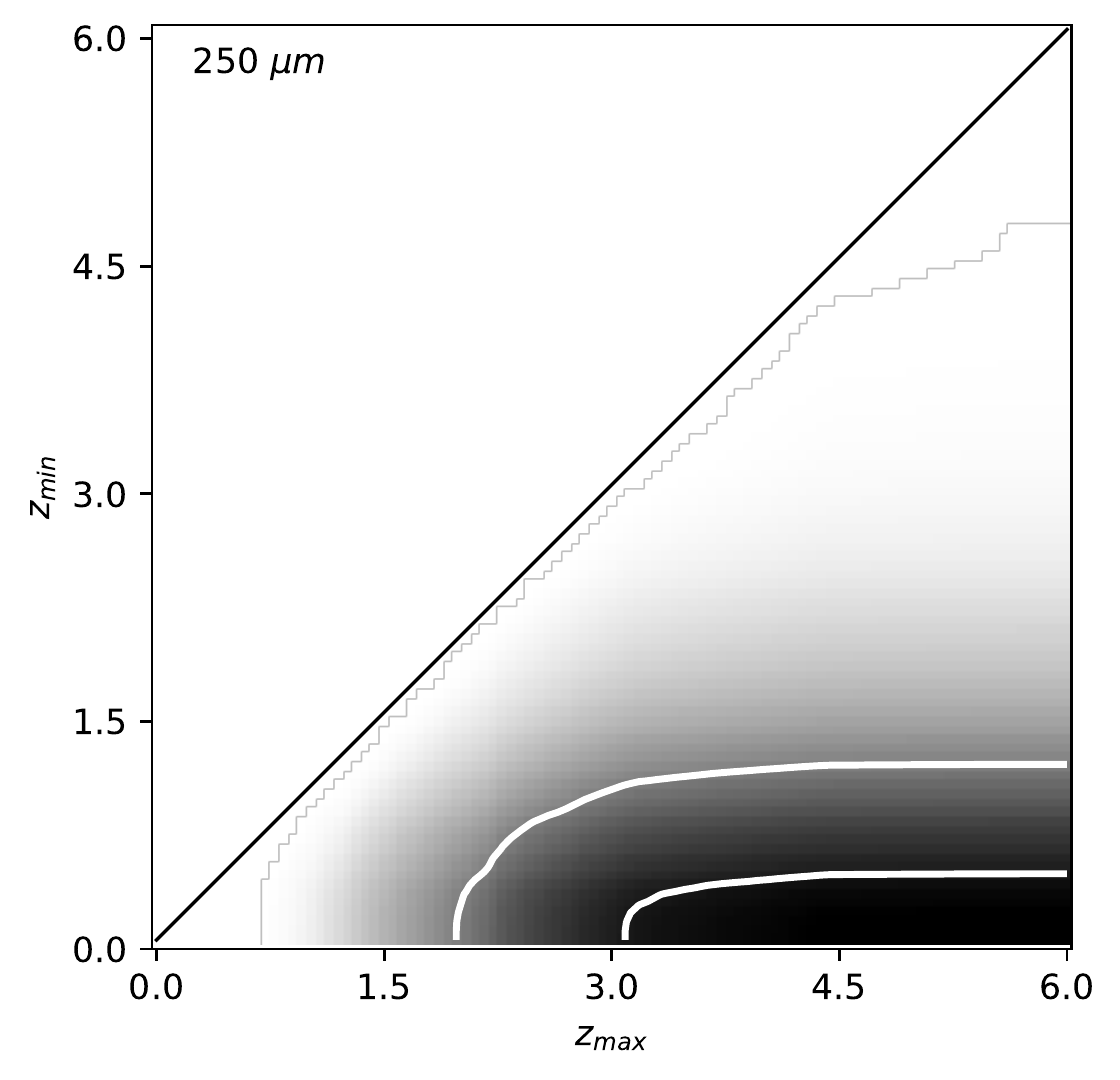}\\
\includegraphics[width=0.32\textwidth]{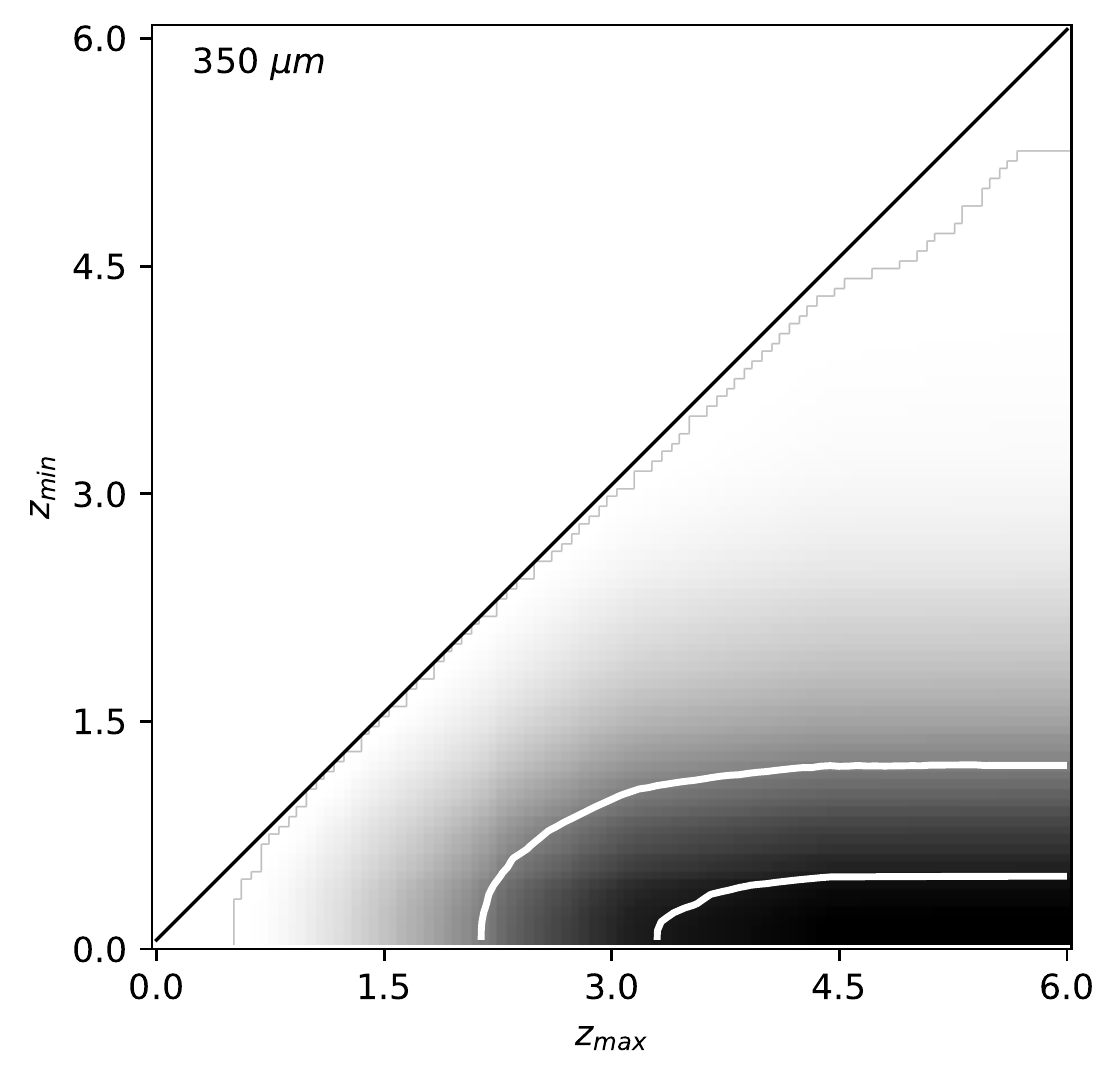}
\includegraphics[width=0.32\textwidth]{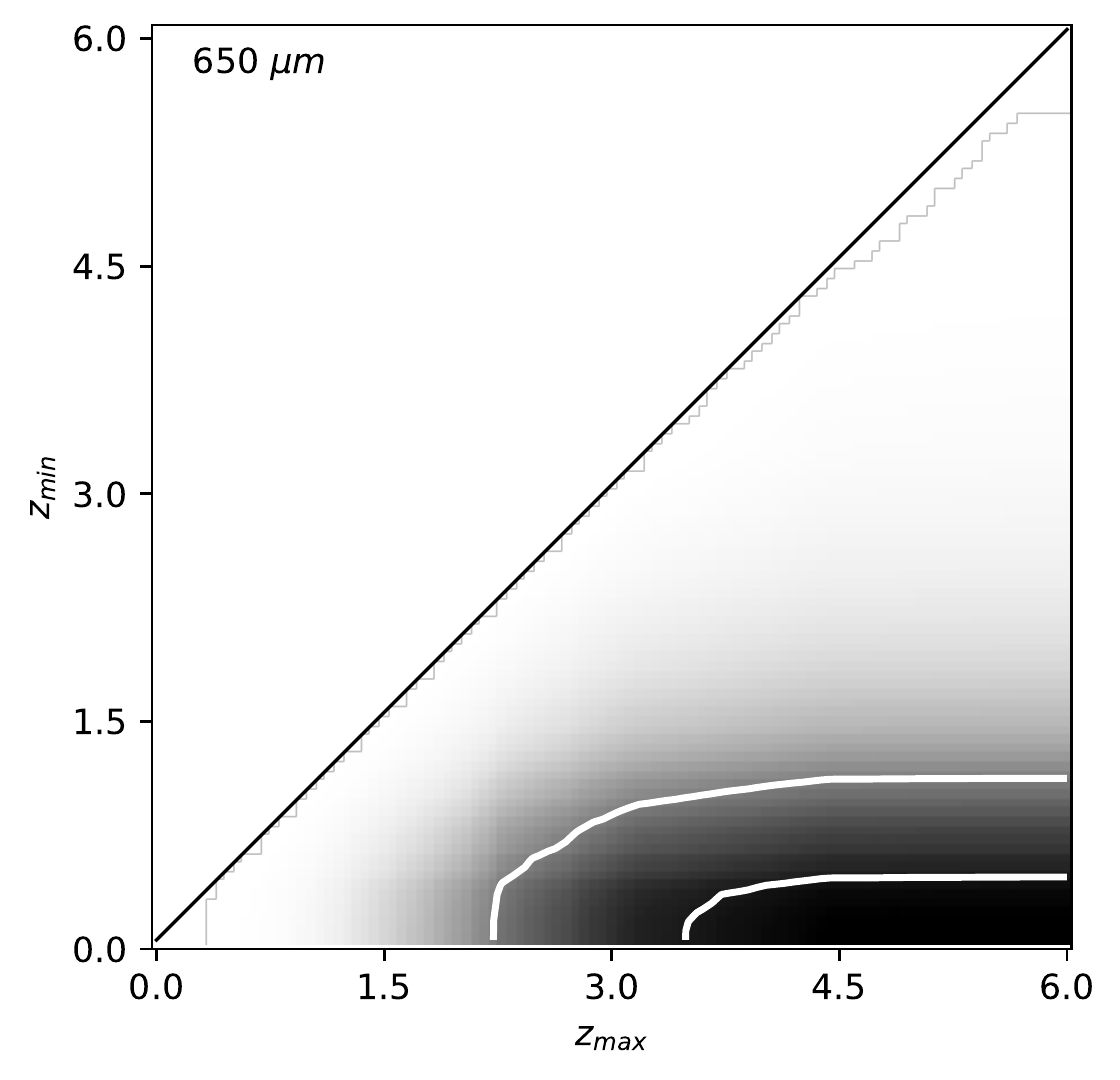}
\includegraphics[width=0.32\textwidth]{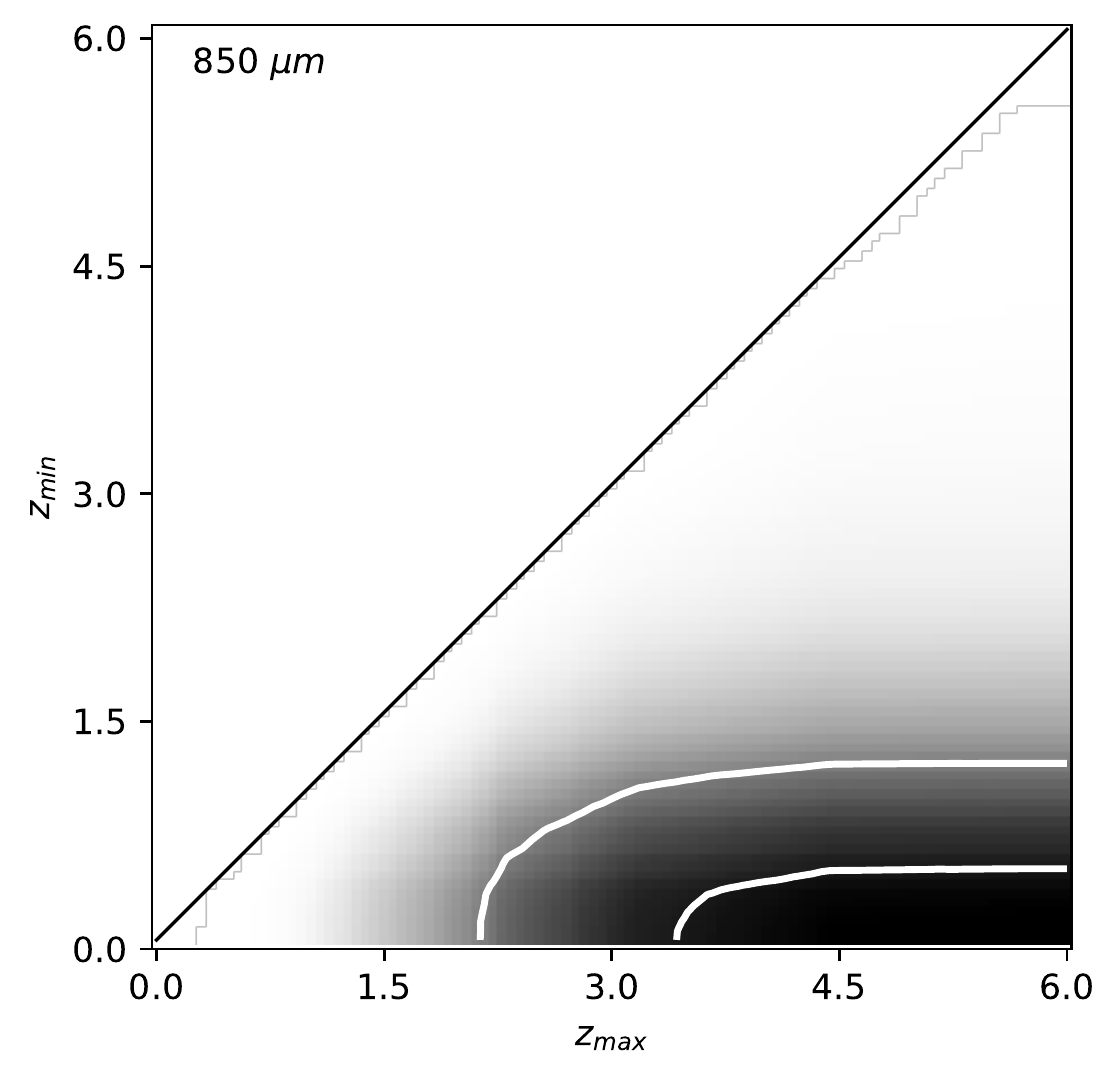}\\
\includegraphics[width=0.32\textwidth]{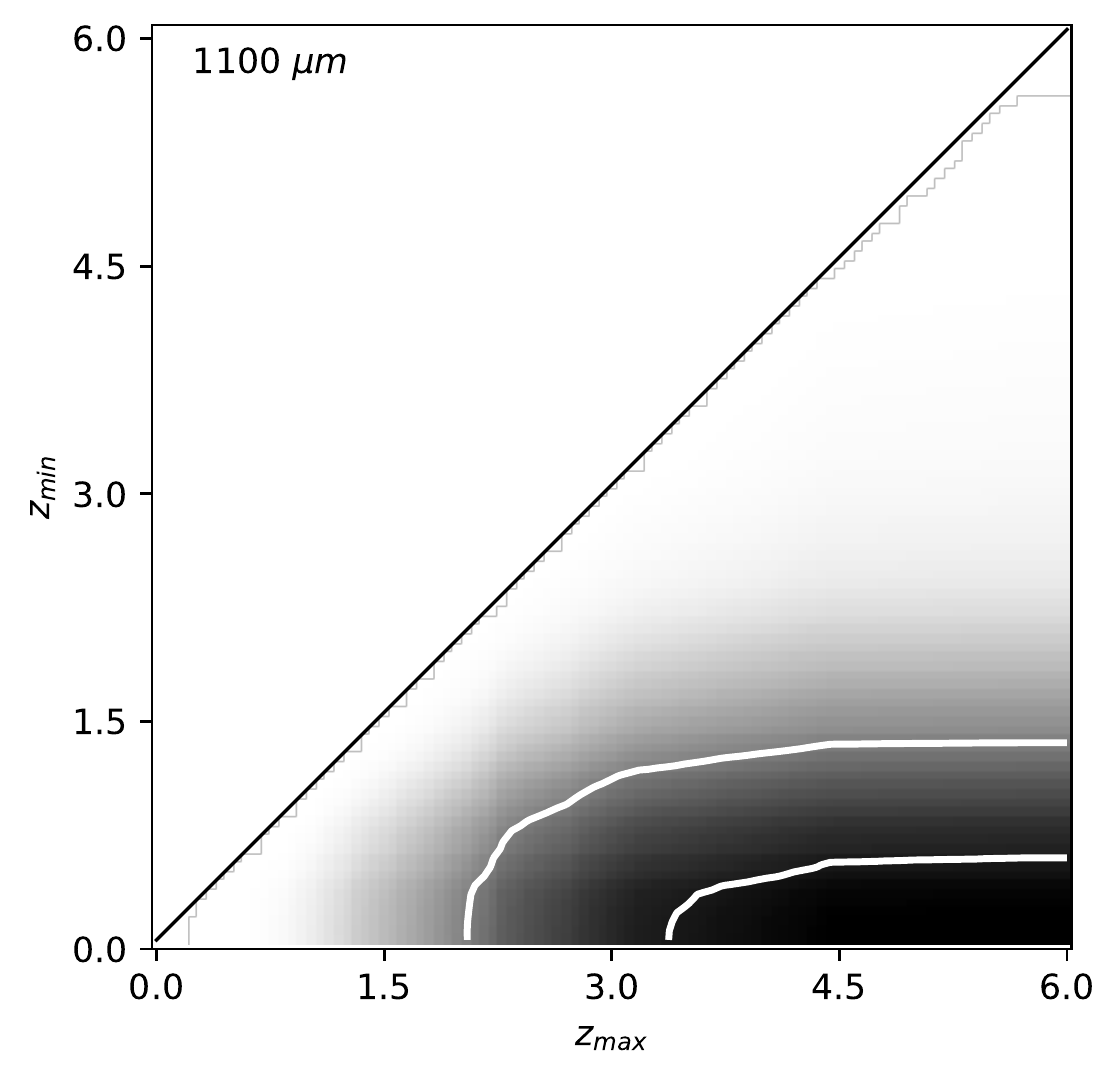}
\includegraphics[width=0.32\textwidth]{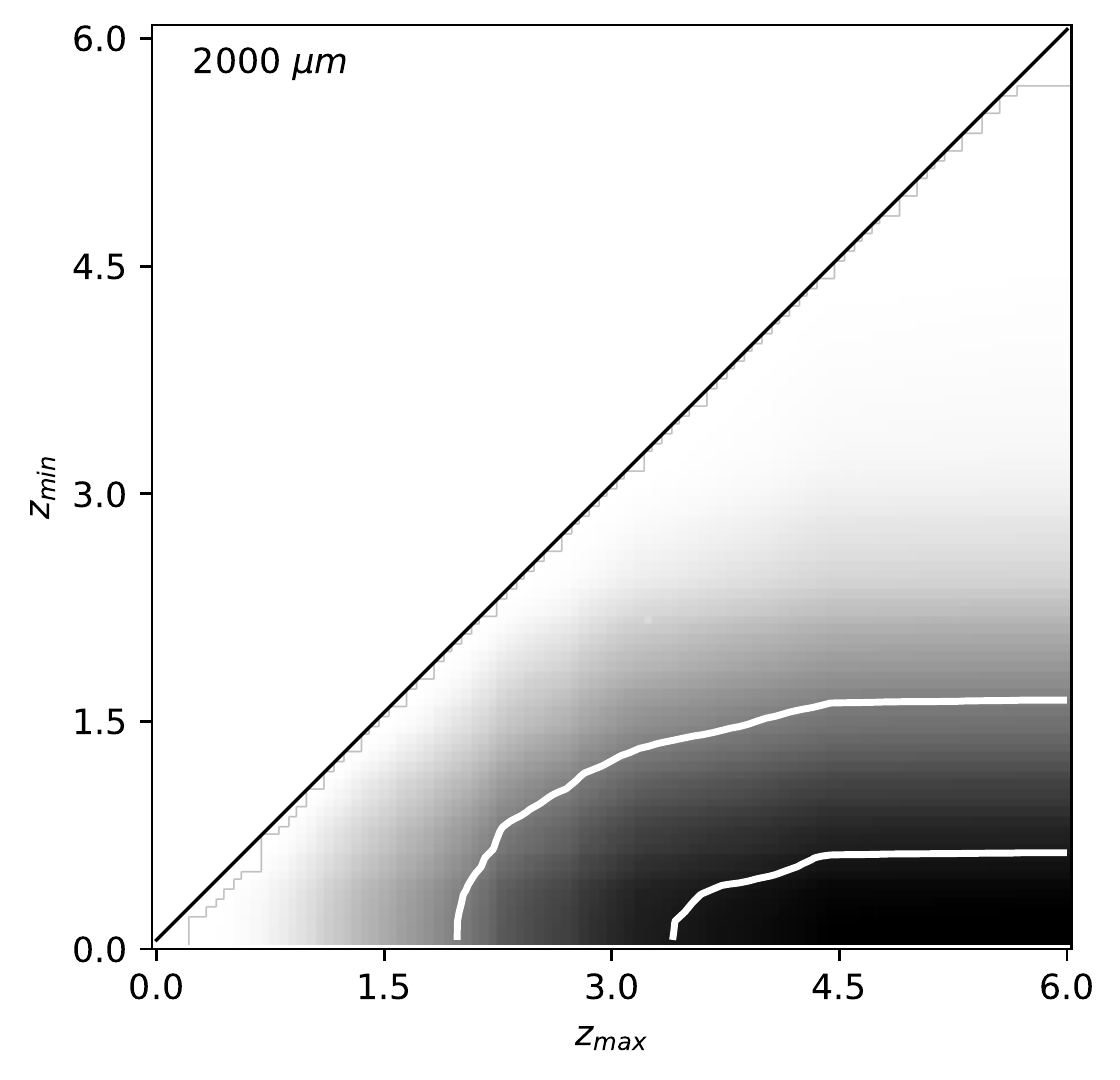}
\caption{
Contribution into the confusion noise from objects in redshift interval from $z_{min}$ to $z_{max}$. 
Eight panels from left to right, from top to bottom correspond to eight bands of SACS and LACS detectors:
70, 110, 250, 350, 650, 850, 1100 ¨  2000~$\mu m$.
The X axis is maximum redshift $z_{max}$  of model objects, Y axis~-- minimum redshift $z_{min}$. 
Color on this 2d plane gives the confusion noise value of the map that contains objects with redshifts $z_{min}<z<z_{max}$. 
Color intensity illustrates the fraction of the confusion noise relatively to the map containing all objects.
Black is unity.
White areas show redshift intervals where there is no confusion.
Upper triangle area in not filled because $z_{max}<z_{min}$.
White contours encompass areas of 90\% and 50\% of maximum value.
} 
\label{fig:redshifts} 
\end{figure} 

Analysis of the extragalactic background often includes creation of redshift distribution of certain numerical characteristic.
Sometimes number counts on certain redshift cuts are plotted (see \cite{2011A&A...532A..49B}),
or the dependence of number of objects with flux greater than certain value on redshift.
In some papers the contribution into the extragalactic background SED from objects on different redshifts is plotted (see \cite{2012A&A...542A..58B}).
In general, contribution of distant galaxies to the background increases with wavelength due to the K-correction.
At 870$\mu m$ the observable flux does not depend on redshift if $z=1-10$ and is defined only by physical properties of the galaxy~\citep{2014ApJ...780...75D,2013MNRAS.432L..85H}.

We created series of maps for given pairs of $z_{max}$ and $z_{min}$ and estimated the confusion noise in them.
The results for eight wavebands of Millimetron detectors are shown on fig.~\ref{fig:redshifts}.
Value of confusion noise created by objects within $z_{min}-z_{max}$ is shown as shade of gray.
White solid lines encompass 50\% and 90\% of the total confusion created by all objects.

The following conclusions can be made.
For a large wavelength range shape and location of curves that encompass percentiles of confusion noise are similar.
In the areas where these lines are horizontal the confusion noise does not depend on the upper limit of the redshift interval.
Likewise in the areas where the lines are vertical the noise value does not depend on the lower limit of redshift interval.
The redshift interval $z_{min}<z<z_{max}$ of objects that contribute the most to the confusion noise can be derived from these diagrams.
We define these values as interceptions of the 90\% percentile with coordinate axes.
The numerical values are given in Table~2.
First column contains the wavelength in micrometers, second and third contain the limits of the redshift interval.
The result obtained is quite nontrivial.
The $z_{min}$ is almost constant for all the wavelength considered, while $z_{max}$ gradually decreases from $\sim 4$ to $\sim 3$ with increase of wavelength from 70 to 2000$\mu m$.

\begin{table}
 \caption{
  Interval of redshift of objects that give the most contribution to the confusion noise.
 }
 \begin{center}
 \begin{tabular}{|c|c|c|}
  \hline
  $\lambda$[$\mu m$]   & $z_{min}$ & $z_{max}$  \\
  \hline
   70       & 0.62      & 4.11       \\
   110      & 0.60      & 3.65       \\
   250      & 0.49      & 3.06       \\
   350      & 0.47      & 3.27       \\
   650      & 0.47      & 3.45       \\
   850      & 0.52      & 3.39       \\
   1100     & 0.59      & 3.34       \\
   2000     & 0.63      & 3.37       \\
  \hline
 \end{tabular}

 \end{center}
 \label{tab:zconf}
\end{table}

\subsection*{Contribution to the confusion noise from objects with different luminosities}

\begin{figure}[ht!]
 \includegraphics[width=0.32\textwidth]{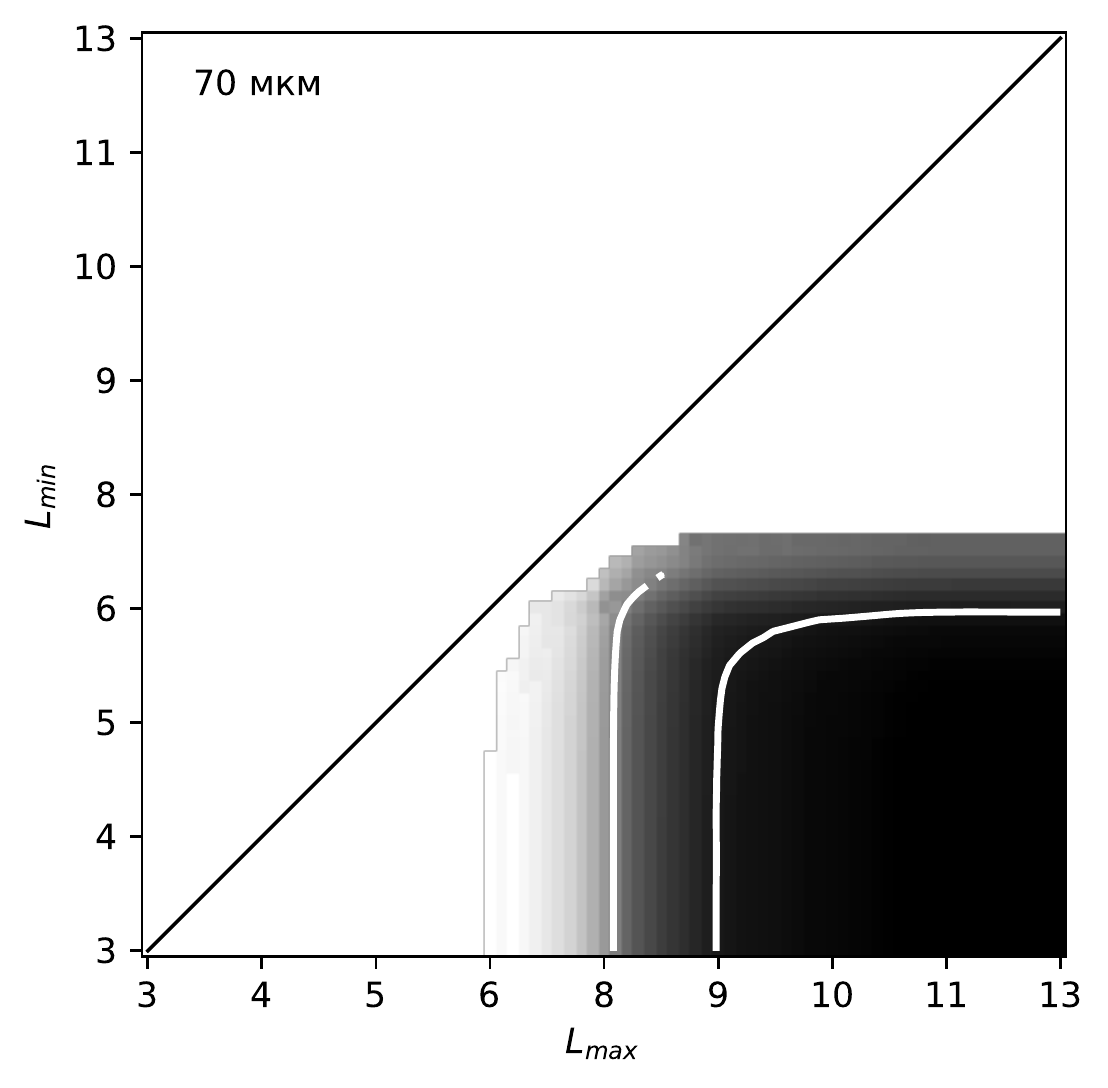}
 \includegraphics[width=0.32\textwidth]{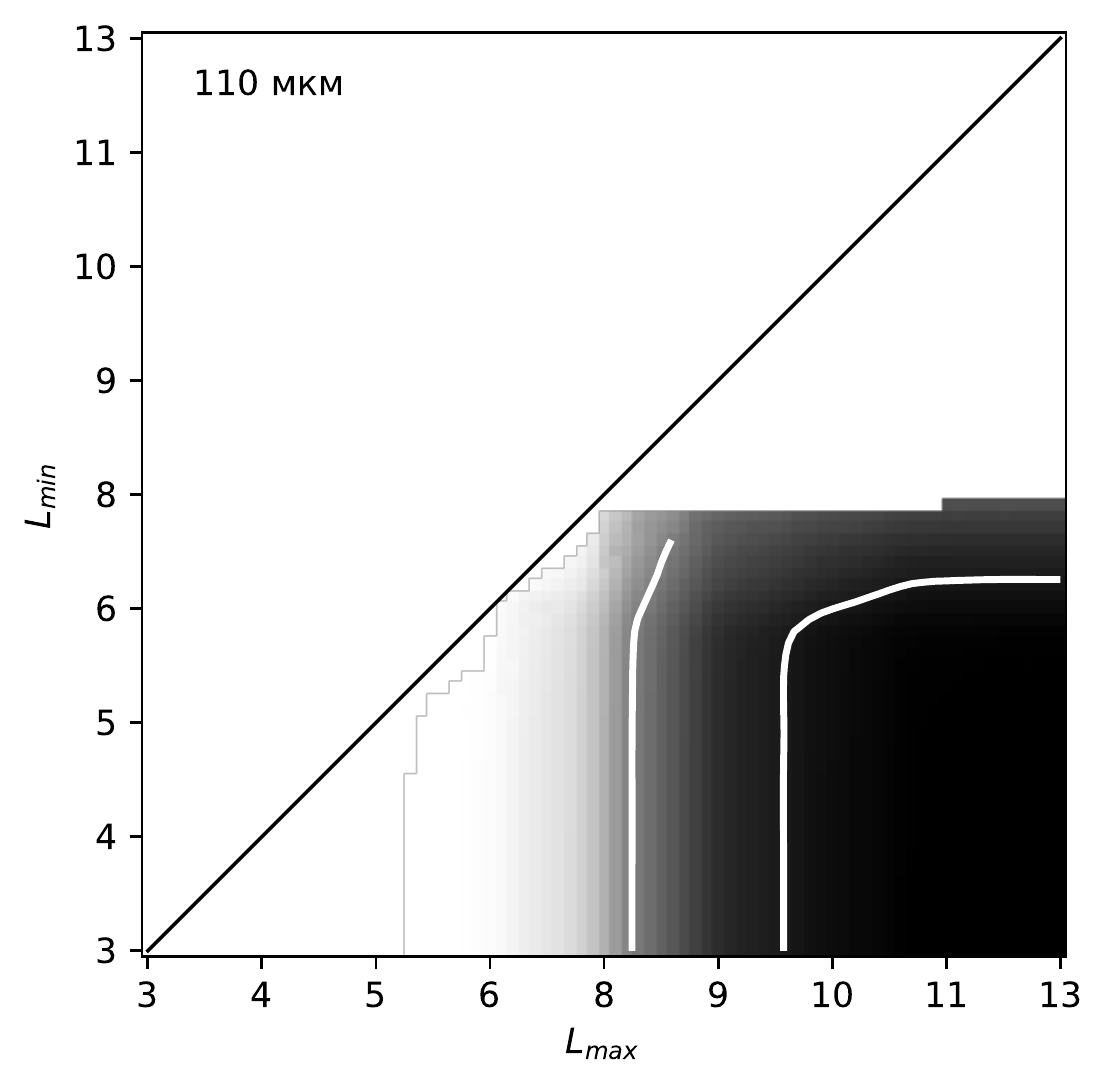}
 \includegraphics[width=0.32\textwidth]{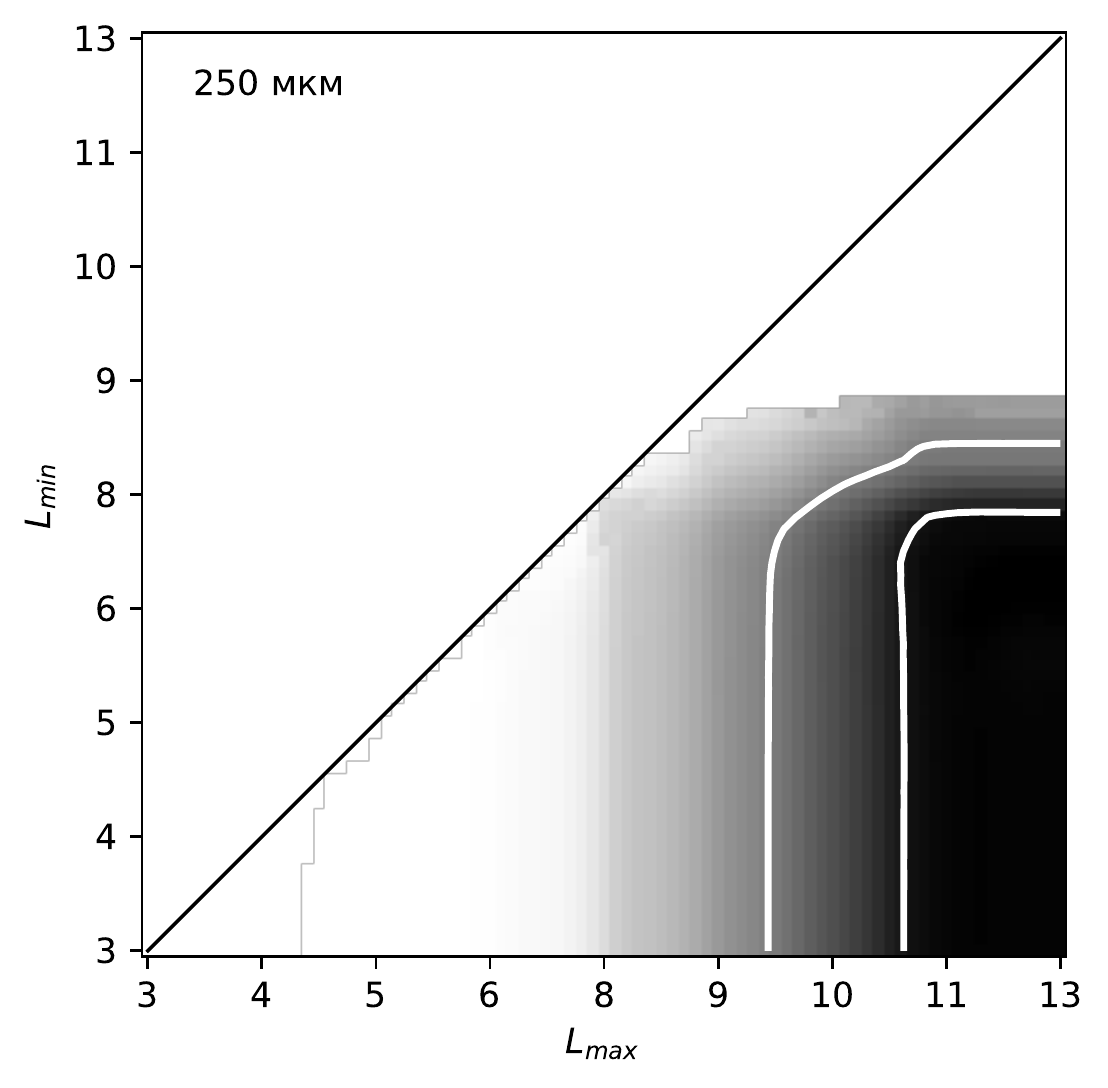}\\
 \includegraphics[width=0.32\textwidth]{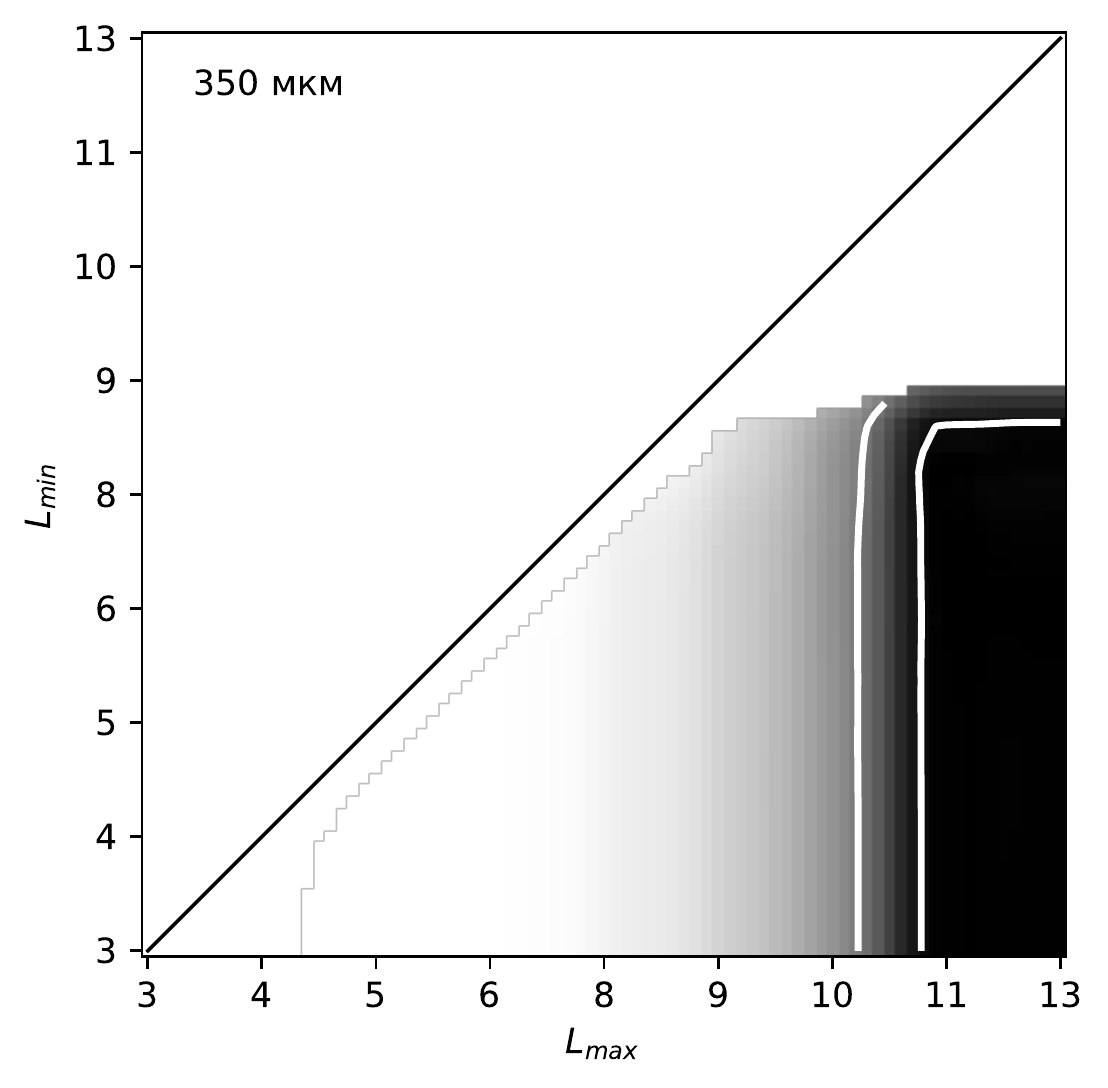}
 \includegraphics[width=0.32\textwidth]{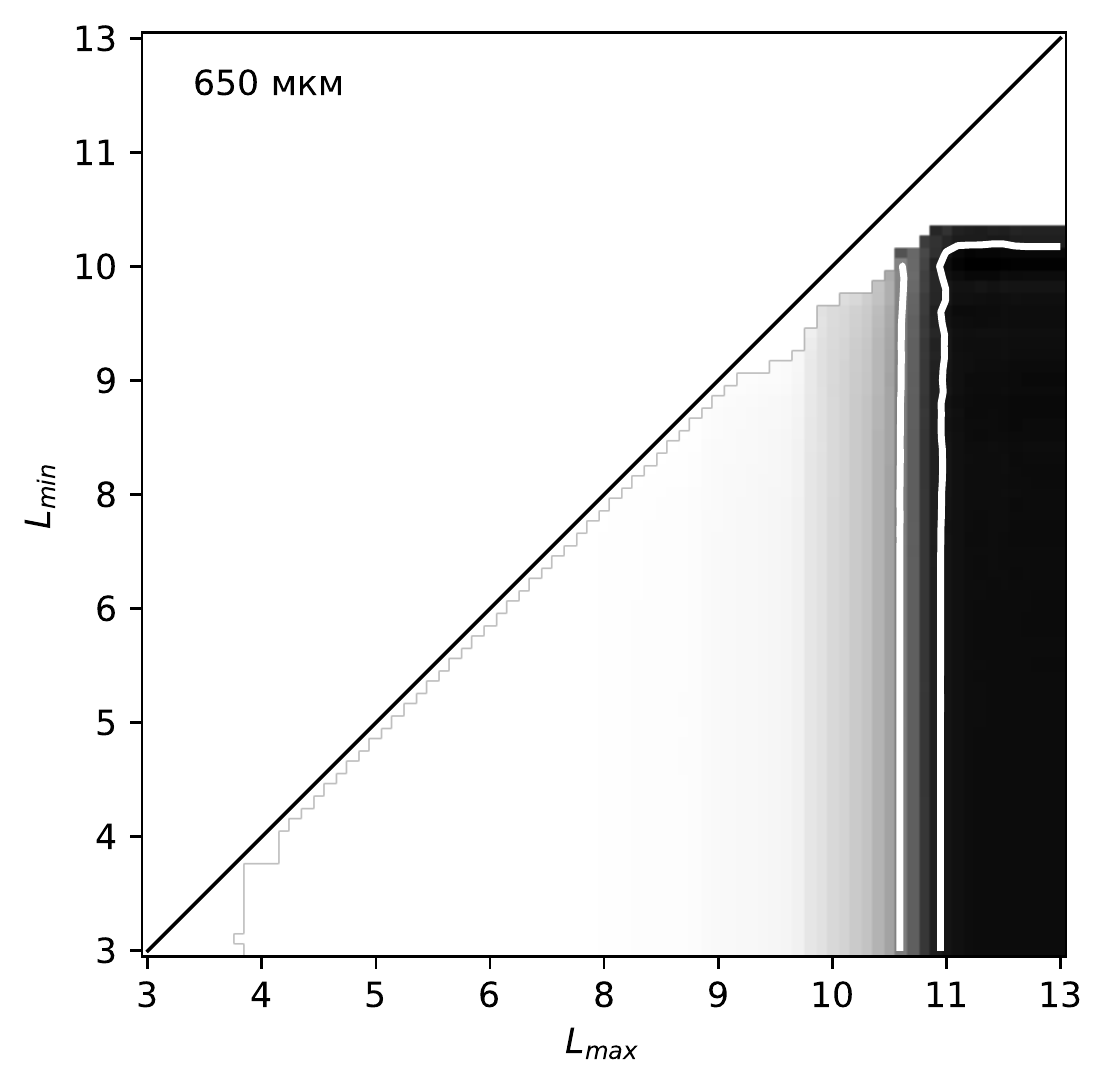}
 \includegraphics[width=0.32\textwidth]{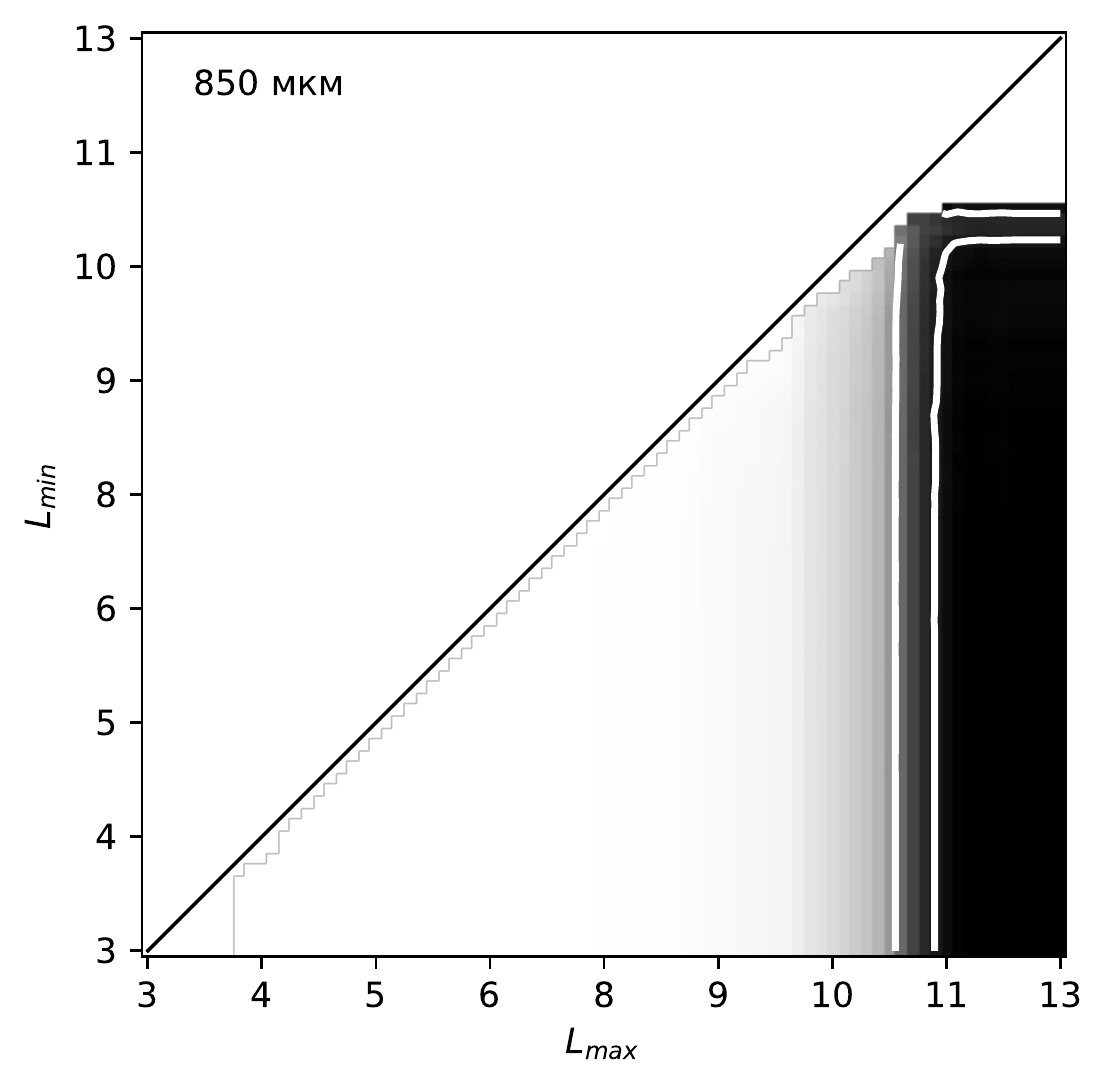}\\
 \includegraphics[width=0.32\textwidth]{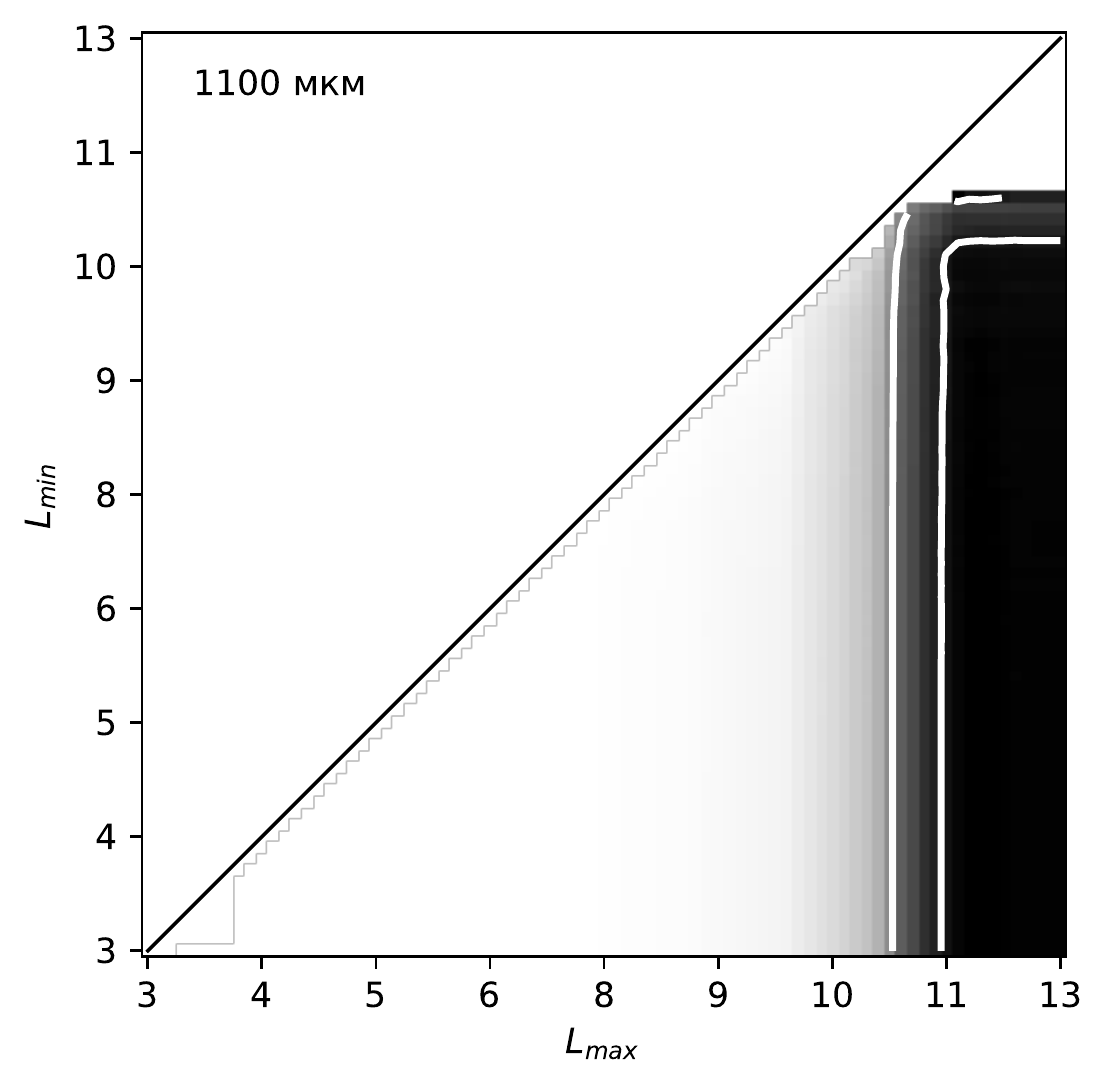}
 \includegraphics[width=0.32\textwidth]{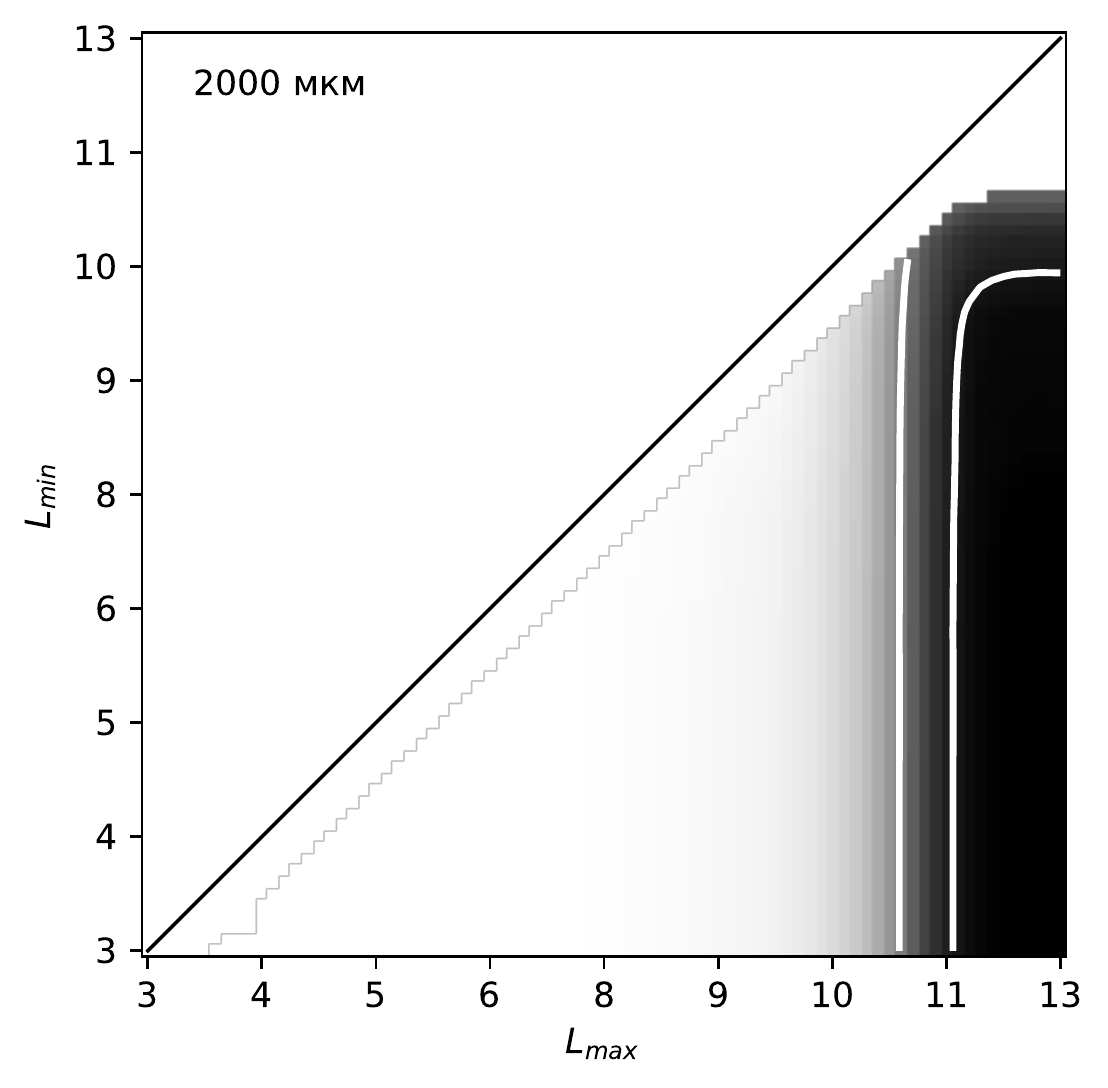}
 \caption{
 The confusion noise for different intervals of the infrared luminosity from $L_{min}$ up to $L_{max}$.
 Estimations were made for the parameters of the Millimetron space observatory.
 Eight panels correspond to eight bands of the SACS and LACS detectors.
 Contours encompass 90\% and 50\% levels of the total confusion noise.
 Black color corresponds to 100\%, white is zero.
 } 
 \label{fig:lums} 
\end{figure}

In this section we analyze the contribution into the confusion from objects with different luminosities.
In the model cone we used to create model maps galaxy luminosities span from $L_{min}=10^{3}L_\odot$ to $L_{max}=10^{13}L_\odot$.
The infrared luminosity is defined as luminosity in wavelength interval from 8$\mu m$ to 1000$\mu m$.
Diagrams showing the contribution from galaxies within given luminosity ranges are shown on Figure~\ref{fig:lums}.
Luminosities of objects that give the most contribution to the confusion are given in Table~3.
The following conclusions can be made.
Values $L_{min}$ and $L_{max}$ significantly depend on the wavelength.
On short wavelengths objects of moderate luminosities $L\simeq10^7L_\odot$--$10^9L_\odot$ create the confusion, while on large wavelengths the confusion is created by objects 
that have $L^>_\sim 10^{10.5}L_\odot$.
It should be noted that numerous faint galaxies with $L<10^7L_\odot$ do not significantly contribute to the confusion noise.
All the above said does not contradict with the statement that the confusion is created by distant unresolved objects.
The angular resolution decreases with wavelength and brighter objects start to contribute into the confusion noise.

\begin{table}
 \label{tab:l_conf}
 \caption{
 The luminosity range of galaxies that make the major contribution (90\%) to the confusion noise at different wavelengths.
 } 
 \begin{center}
 \begin{tabular}{|c|c|c|}
  \hline
  $\lambda$, $\mu m$ & $\log_{10}{(L_{min}/L_\odot)}$ & $\log_{10}{(L_{max}/L_\odot)}$ \\
  \hline
   70   & 6.9  & 9.1  \\
   110  & 7.1  & 9.9  \\
   250  & 8.1  & 11.2 \\
   350  & 9.3  & 12.0 \\
   650  & 10.7 & 11.6 \\
   850  & 11.1 & 11.6 \\
   1100 & 11.2 & 12.2 \\
   2000 & 10.4 & 11.7 \\
  \hline
 \end{tabular}
 \end{center}
\end{table}

\subsection*{Contribution to the confusion noise from objects with different spectral characteristics}

Eight photometric bands of SACS and LACS detectors provide 7 color indexes as difference in flux in adjacent bands.
We also added two additional wavebands~-- 50 and 3000$\mu m$.
Let us consider for example the 250$\mu m$ waveband.
It is possible to define two colors for it, namely:
$C(350\mu m-250\mu m)\equiv \log_{10}(S_{350}/S_{250})$ and
$C(250\mu m-110\mu m)\equiv \log_{10}(S_{250}/S_{110})$. 
For each of them it is possible to set minimum and maximum value, e.g.
$C_{max}(250\mu m-110\mu m)$ and $C_{min}(250\mu m-110\mu m)$
and estimate the confusion noise created by objects that have spectral colors within this range.

\begin{figure} 
\includegraphics[width=0.32\textwidth]{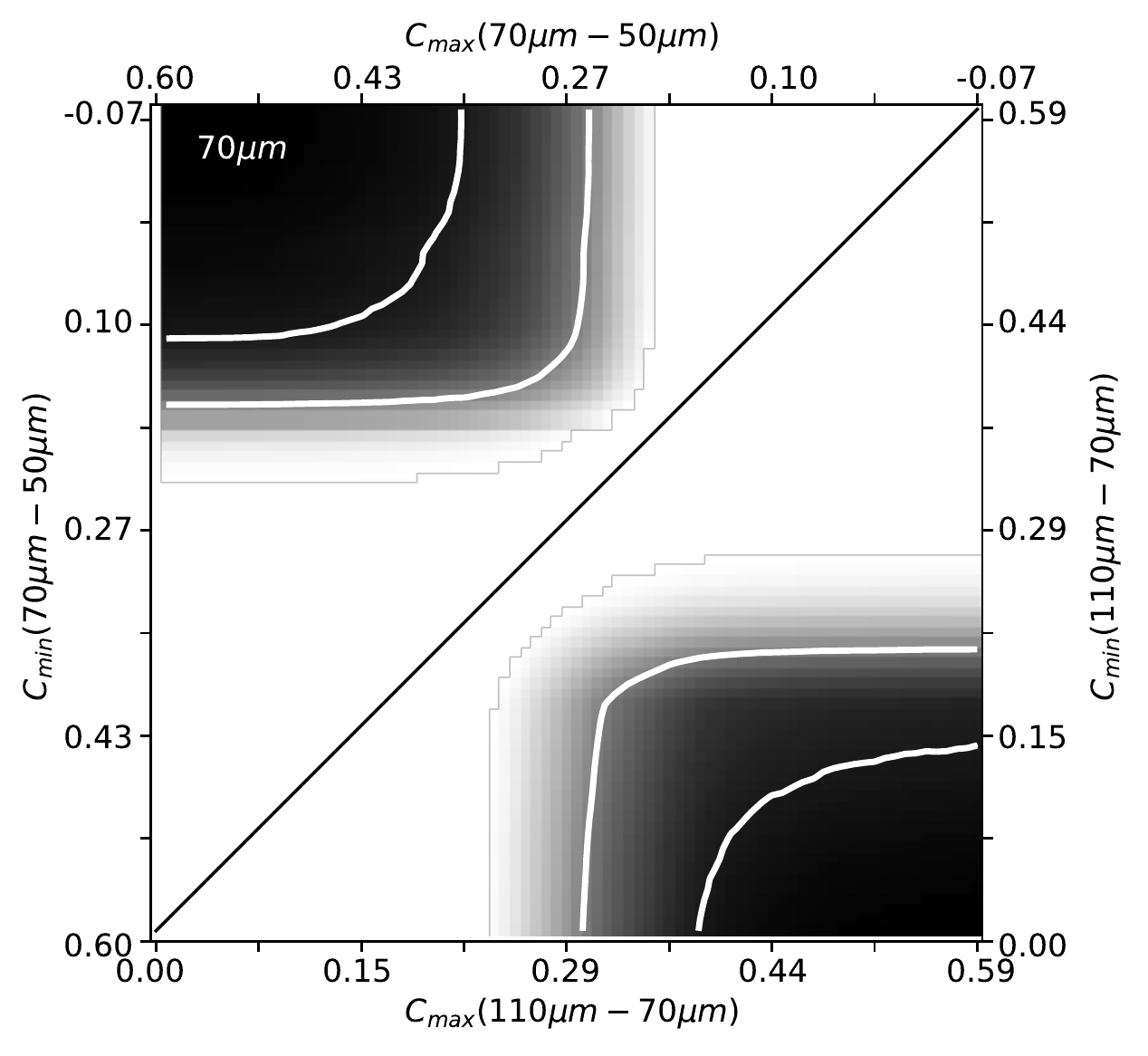}
\includegraphics[width=0.32\textwidth]{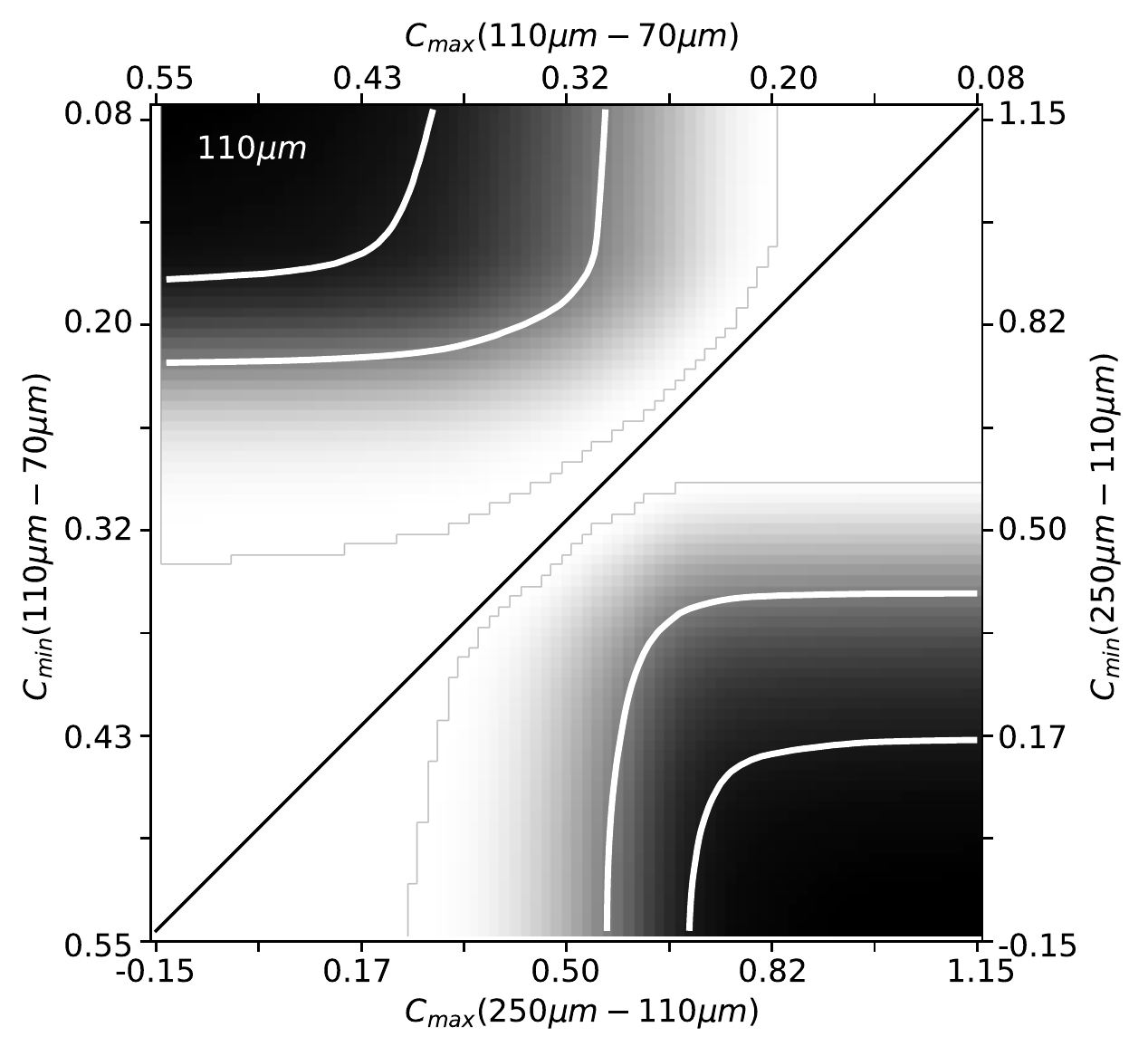}
\includegraphics[width=0.32\textwidth]{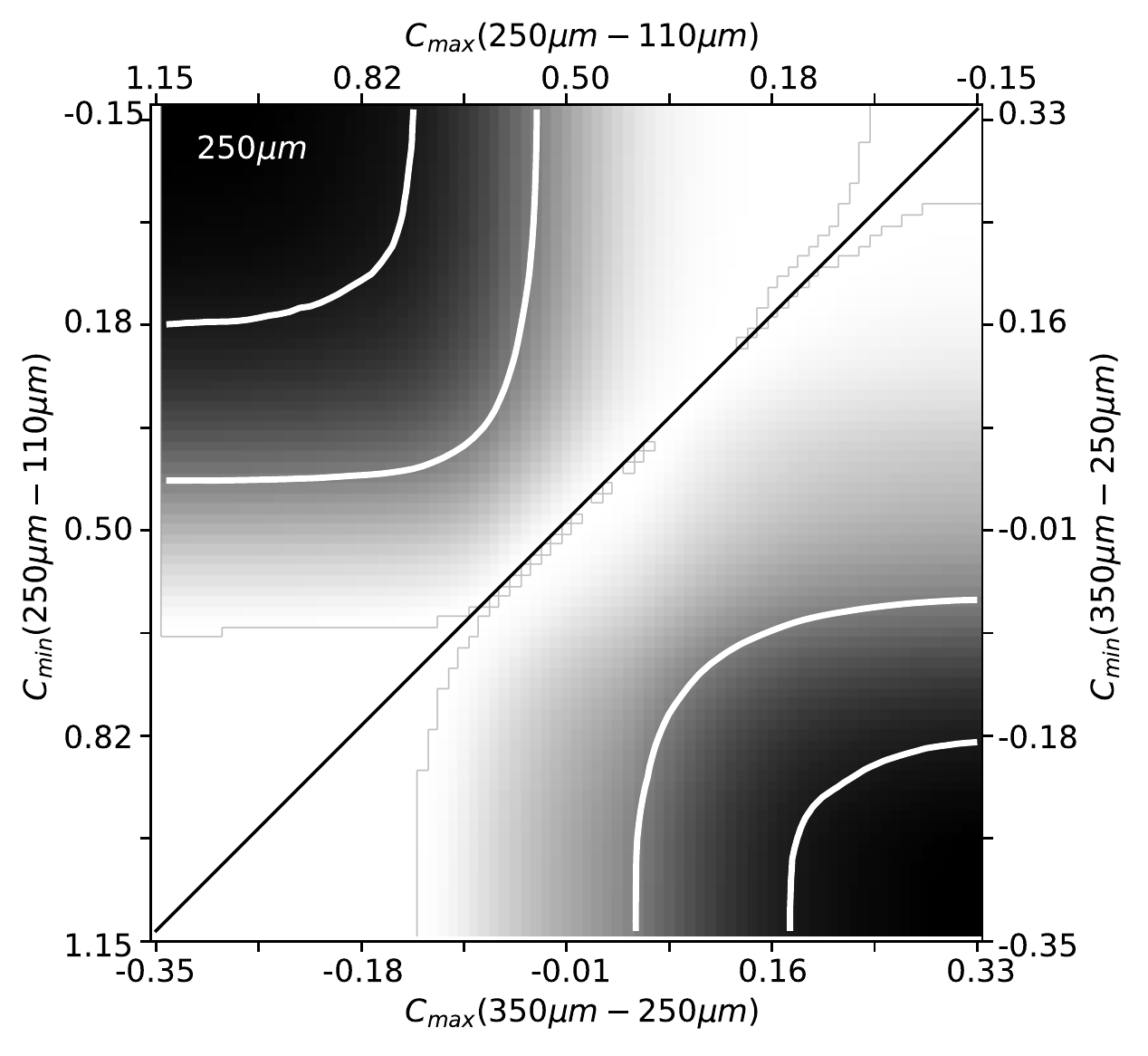}\\
\includegraphics[width=0.32\textwidth]{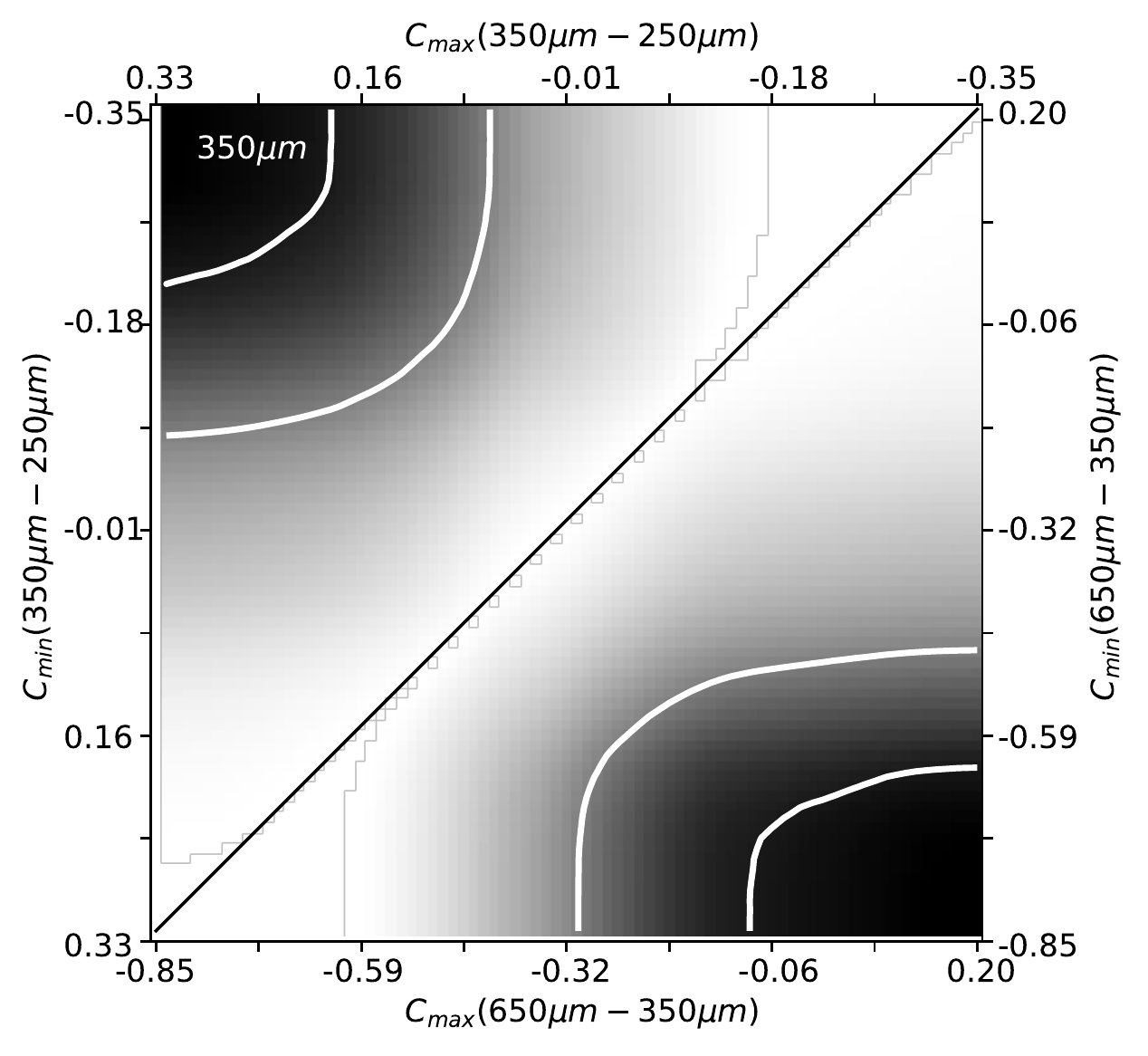}
\includegraphics[width=0.32\textwidth]{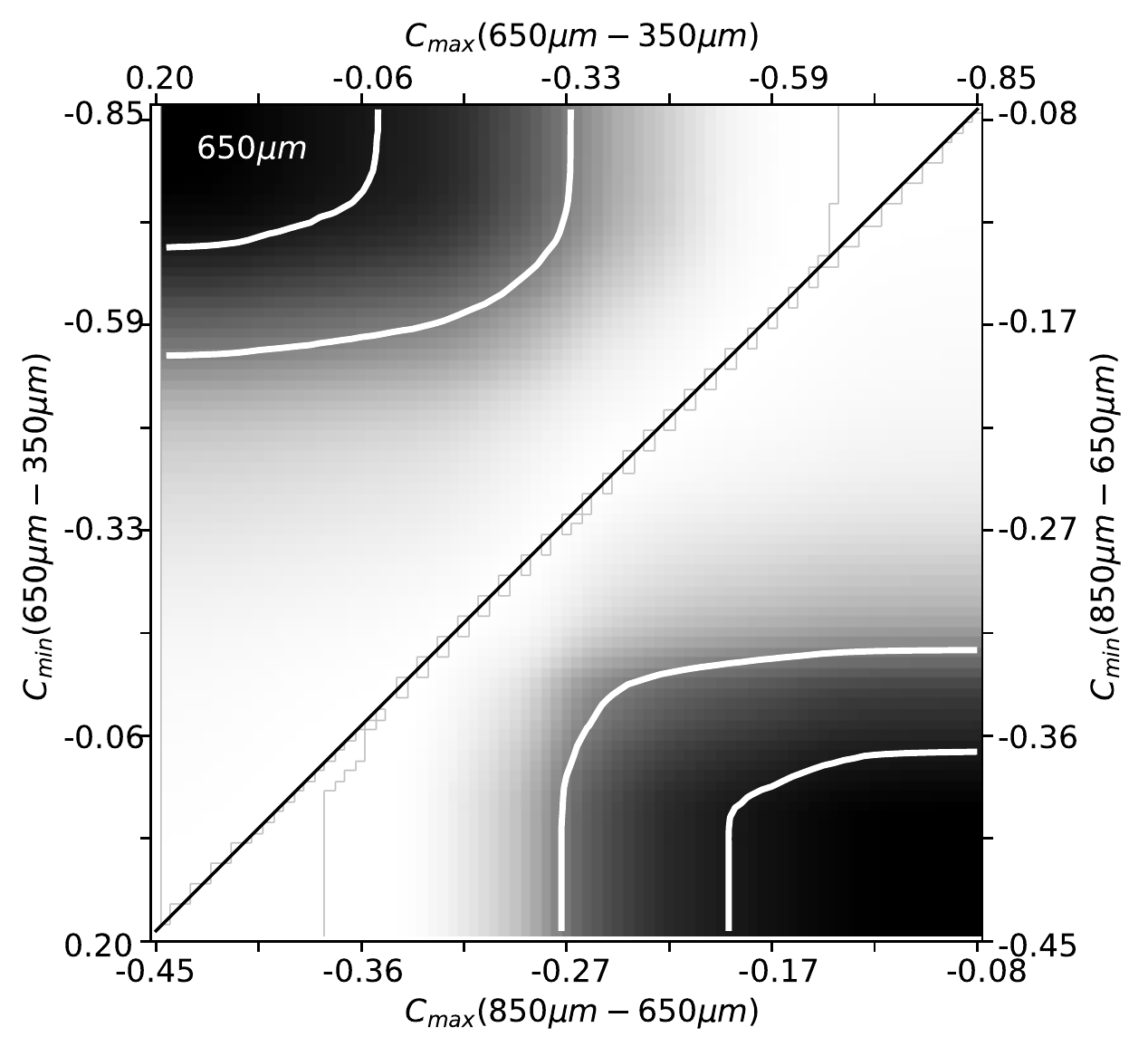}
\includegraphics[width=0.32\textwidth]{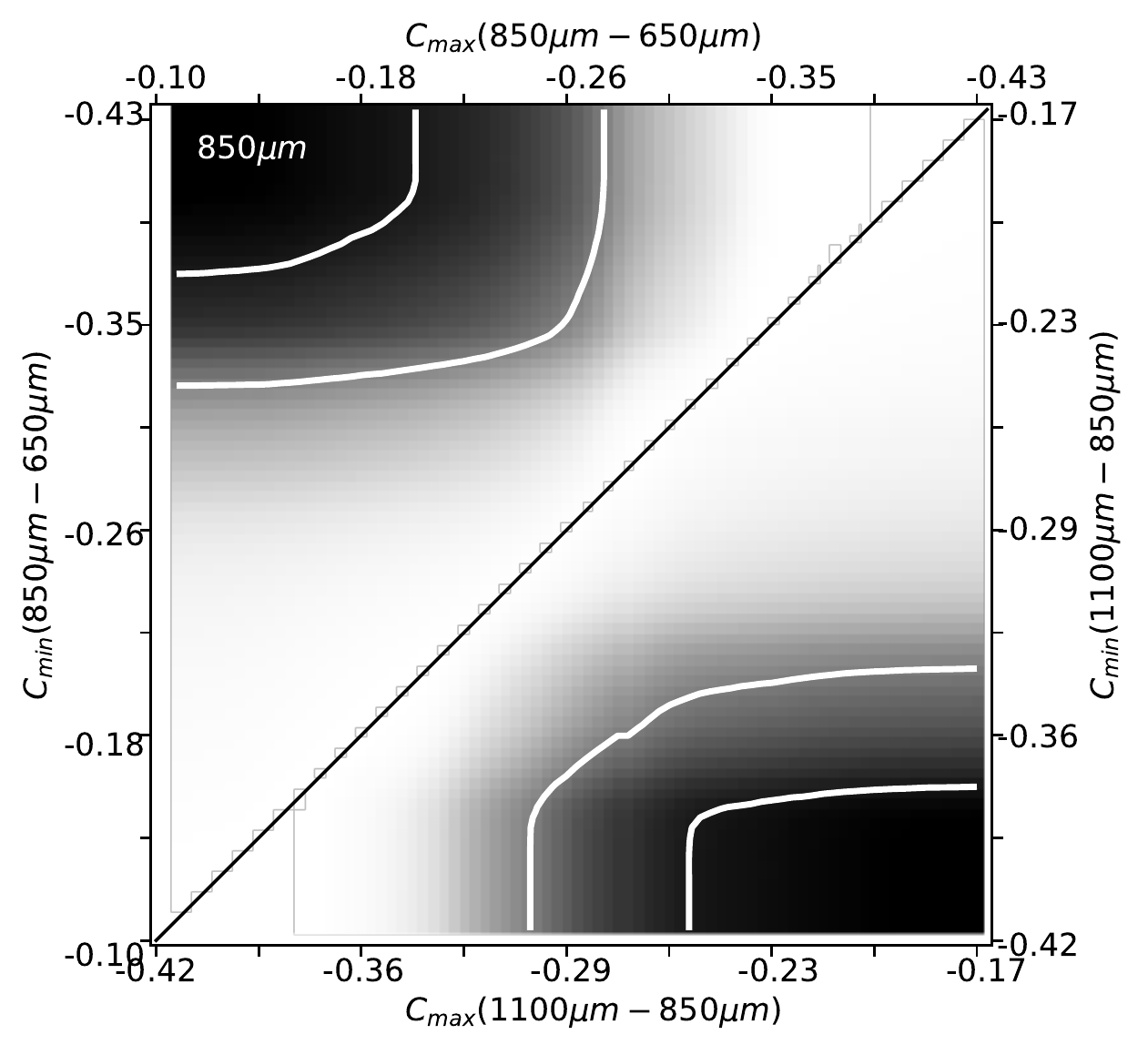}\\
\includegraphics[width=0.32\textwidth]{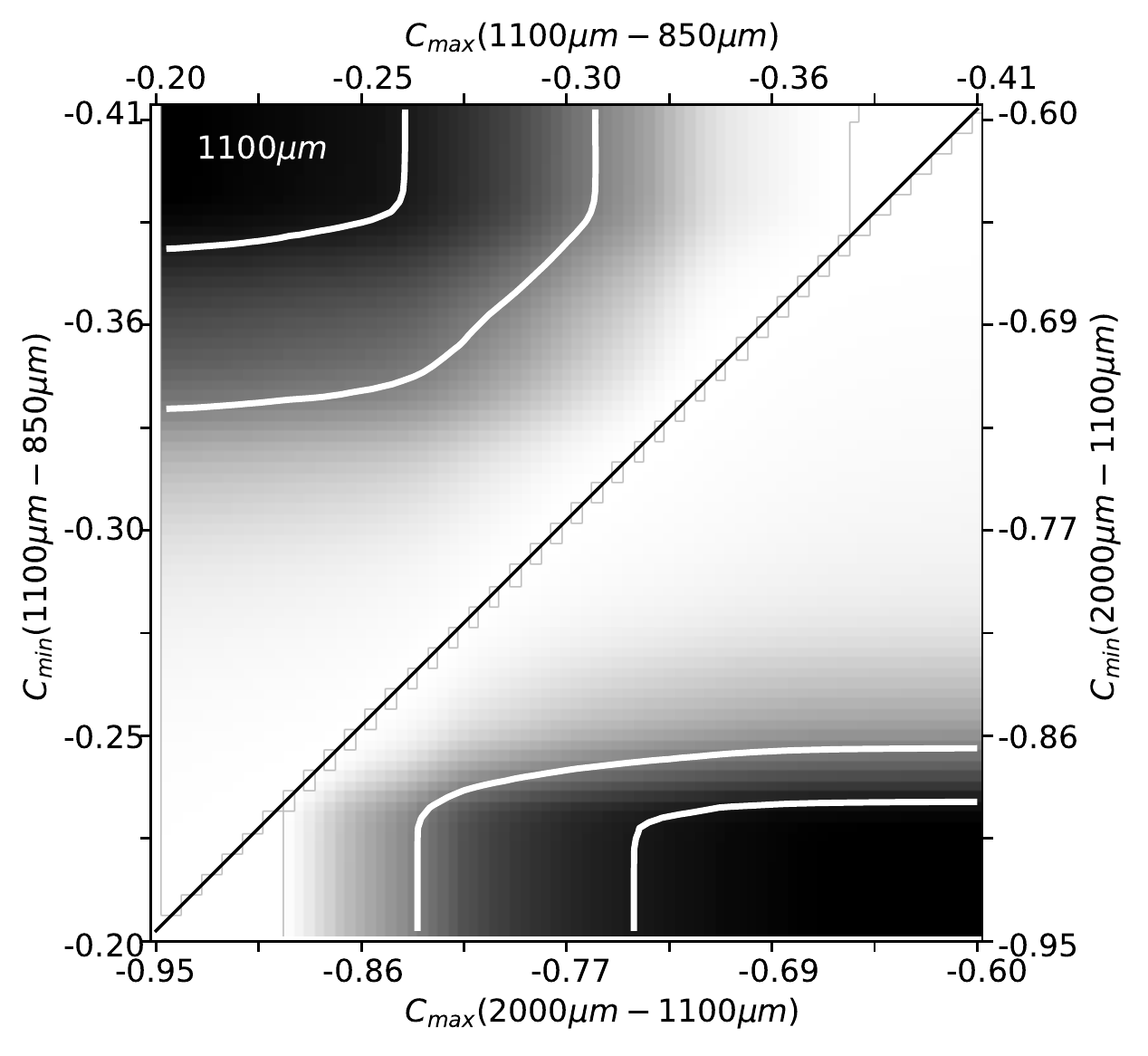}
\includegraphics[width=0.32\textwidth]{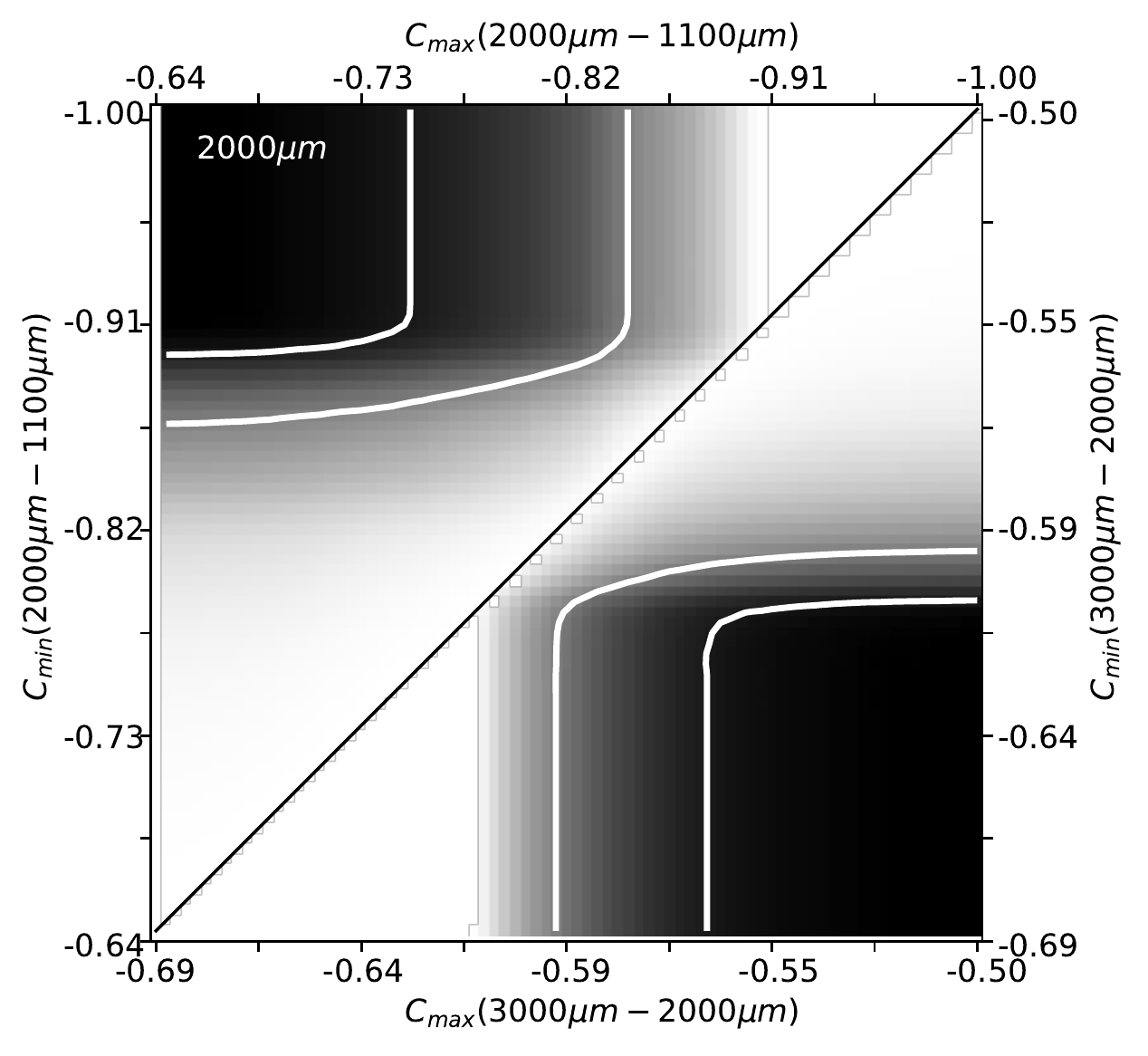}
\caption{
 Contribution to the confusion noise from objects with different color characteristics.
 Eight panels correspond to eight bands of the SACS and LACS detectors.
 Upper left panel shows the diagram for the 70$\mu m$ waveband.
 Upper part of the plot shows data for
 $C_{min}(70\mu m-50\mu m)\leq C(70\mu m-50\mu m)\leq C_{max}(70\mu m-50\mu m)$. 
 Lower parts show the confusion estimates for objects with
 $C_{min}(110\mu m-70\mu m)\leq C(110\mu m-70\mu m)\leq C_{max}(110\mu m-70\mu m)$. 
 Panel number $N$ shows estimates for the $N$-th band of Millimetron's detectors.
 Lower part of the plot shows confusion noise estimates for different ranges of flux differences between
 the $N+1$ (adjacent longwave) and $N$ bands.
 Upper panel~-- between $N$ and $N-1$ (adjacent shortwave).
 White solid lines correspond to 50\% and 90\% of total value of confusion noise.
} 
\label{fig:colors}
\end{figure}

The diagrams plotted on Figure~\ref{fig:colors} help to clarify whether the same objects contribute co confusion noise in adjacent wavebands.
If, for example, the main contributors to the confusion noise in the 250$\mu m$ waveband are objects with significant positive color coefficient
$C(250\mu m-110\mu m)$ it means that different objects contribute to the confusion in the 110$\mu m$ waveband.

Let us consider in detail the results plotted on the panels of Figure~\ref{fig:colors}.
Eight panels correspond to eight wavebands of the detectors of the Millimetron mission.
Each panel is divided into two triangular parts.
On the upper part the values of the confusion noise created by objects within given color interval defined as flux difference with the adjacent shortwave band,
e.g. 250$\mu m$ for 350$\mu m$, 650$\mu m$ for 850$\mu m$ and so forth are shown.
On the upper part of the $70\mu m$ panel the data for the 50-70$\mu m$ color is shown.

Lower part of each panel shows data for the adjacent longwave band, e.g. 350$\mu m$ for 250$\mu m$ etc.
For the 2000$\mu m$ the confusion noise created by objects within the certain color range of 3000 -- 2000$\mu m$ is shown.

Black color corresponds to 100\% of the confusion noise created by all objects in the model cone.
White lines show 90\%  and 50\% contours.

The range of color indexes of objects that give major contribution to the confusion noise can be defined as interval defined by intersection of 90\% percentile 
with coordinate axes.
These values are given in Table~\ref{tab:color_conf}.
The first column is the wavelength $\lambda_r$ for which the map was created and the confusion noise estimations were performed.
The color indexes were calculated as difference of fluxes of objects between
a)  $\lambda_1$ and $\lambda_r$
b)  $\lambda_r$ and $\lambda_0$.
$\lambda_1$  and $\lambda_0$  are given in columns 3 and 2, respectively.
Values of $C_{min}$ and $C_{max}$ in the 4th and 5th column encompass objects that create 90\% of the total confusion noise.

\begin{table}
\caption{
 Contribution to confusion noise from objects with various color characteristics.
 Details see in text.
 }  
 \label{tab:color_conf} 
 \begin{center} 
 \begin{tabular}{|c|c|c|c|c|}
  \hline
   $\lambda_r$[$\mu m$] & $\lambda_{0}$[$\mu m$] & $\lambda_{1}$[$\mu m$] & $C_{min}$ & $C_{max}$\\
   \hline
    70   & 70   & 110  & 0.14  & 0.38  \\ 
    70   & 50   & 70   & 0.11  & 0.35  \\ 
    110  & 110  & 250  & 0.16  & 0.68  \\ 
    110  & 70   & 110  & 0.18  & 0.39  \\ 
    250  & 250  & 350  & -0.19 & 0.17  \\ 
    250  & 110  & 250  & 0.19  & 0.75  \\ 
    350  & 350  & 650  & -0.63 & -0.10 \\ 
    350  & 250  & 350  & -0.21 & 0.19  \\ 
    650  & 650  & 850  & -0.37 & -0.20 \\ 
    650  & 350  & 650  & -0.68 & -0.08 \\ 
    1100 & 1100 & 2000 & -0.89 & -0.75 \\ 
    1100 & 850  & 1100 & -0.37 & -0.26 \\ 
    2000 & 2000 & 3000 & -0.61 & -0.56 \\ 
    2000 & 1100 & 2000 & -0.89 & -0.75 \\ 
   \hline
  \hline
 \end{tabular} 
 \end{center} 
\end{table} 

\begin{figure}
\begin{center}
 \includegraphics[width=\columnwidth]{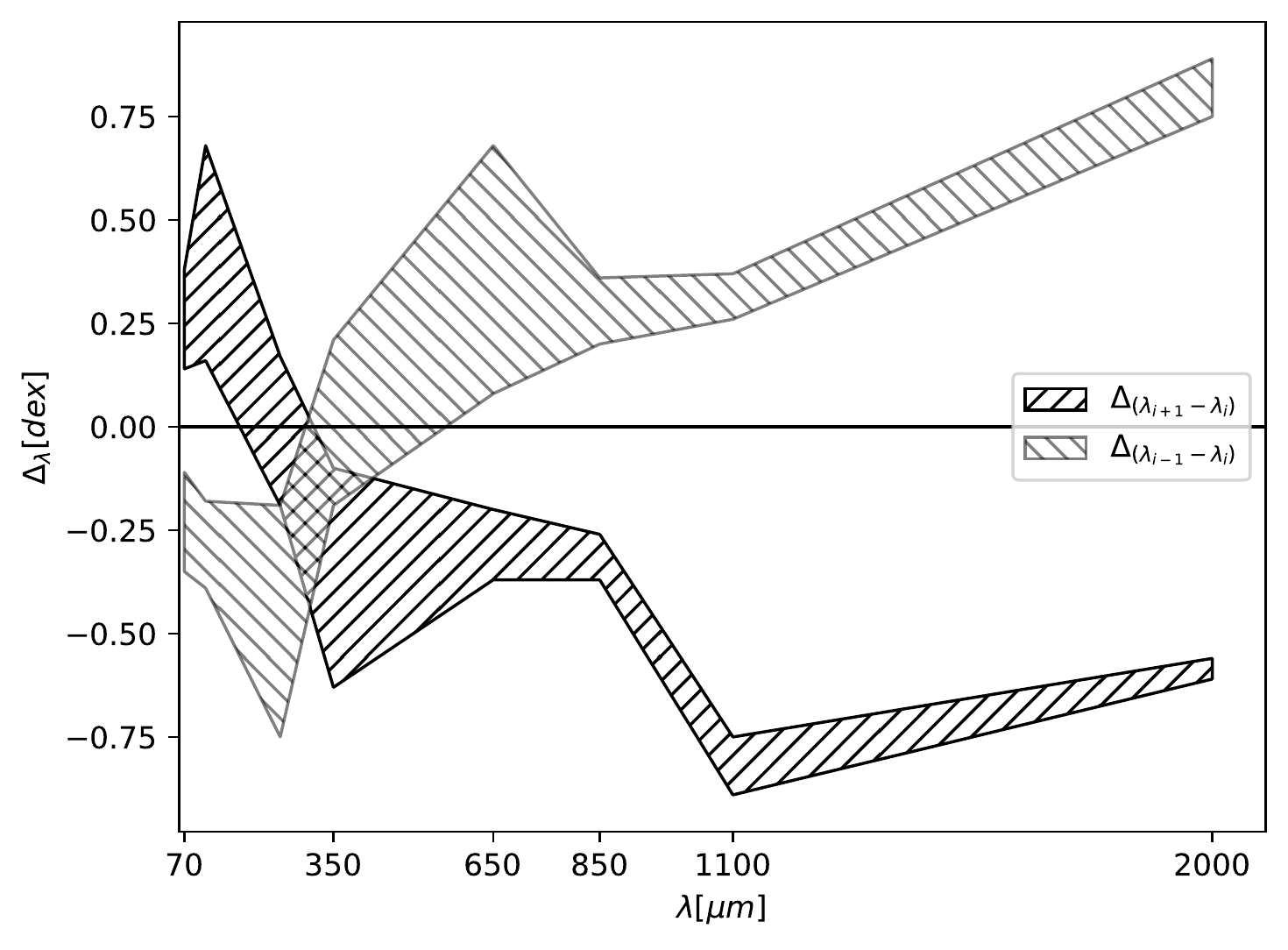}
\end{center}
\caption{
The dependence of the range of color indexes of objects that give 90\% contribution to the confusion noise on wavelength.
The black dashed area is the range of values of the color index defined as the difference in flux with the adjacent longwave band (e.g. $250\mu m-110\mu m$ for the 110$\mu m$ map).
Gray dashed area shows the color index range for the closest shortwave band (e.g. $70\mu m-110\mu m$ for the 110$\mu m$).
 }
\label{fig:colors_2}
\end{figure}

In order to better illustrate the conclusions that can be derived from Figure~\ref{fig:colors} and Table~\ref{tab:color_conf} we additionally created the 
Figure~\ref{fig:colors_2} that illustrates the change of the range of color indexes of objects that give the most contribution to the confusion noise with wavelength.

As can be seen from the plot, objects that create the confusion noise in 70$\mu m$ band have significant positive 110--70$\mu m$ color index.
In another words, objects that create confusion noise in the 70$\mu m$ waveband give significant contribution in the 110$\mu m$ background.

At the same time these objects on average have significant negative $C(50\mu m-70\mu m)$ color index, i.e. they do not significantly contribute to the EBL in the adjacent shortwave band.
If we go from 70$\mu m$ to the longest 2000$\mu m$, the situation gradually changes to the opposite.

The practical sense of this result is the following.
It is well known that in order to at least partially beat the confusion the photometry on the adjacent short waveband is crucial.
But for the band from 70 up to 350$\mu m$ the information from the longer wavebands is also very important.

\section*{The dependence of the confusion noise on the diameter of the main mirror of the telescope at different wavelengths}

The confusion noise first and foremost depends on the angular resolution of the telescope  which is defined by the well known formula
\begin{equation}
\theta\propto\dfrac{\lambda}{d},
\end{equation}
where $\theta$ is the angular resolution, $\lambda$~-- wavelength, $d$~-- diameter of the main mirror.
So it is natural to expect the confusion noise to increase with wavelength and the decrease of the diameter of the main mirror.
But these two trends are influenced by the shape of the number counts of sources and the actual spacial distribution of objects.

\begin{figure} 
\includegraphics[width=\columnwidth]{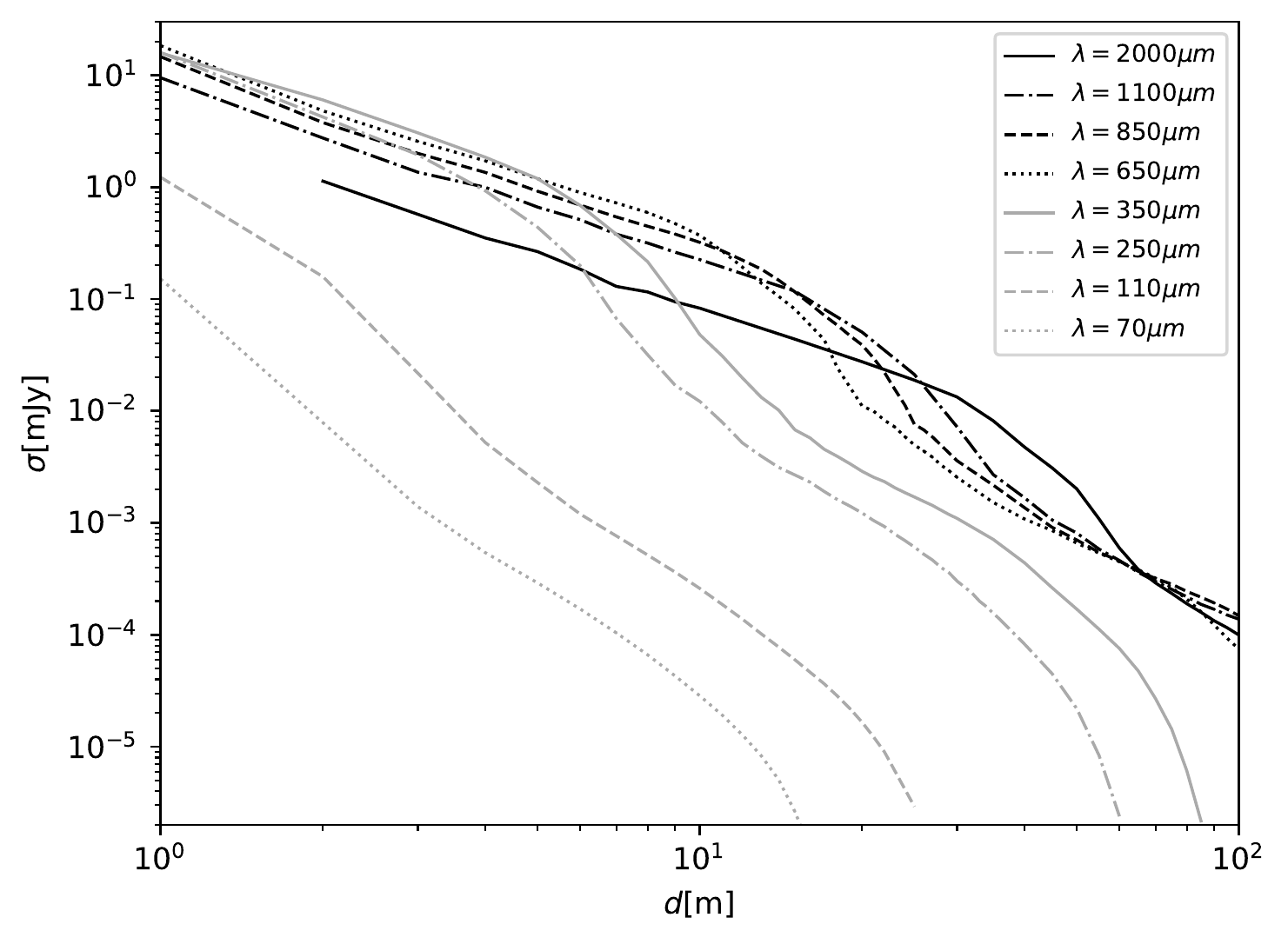} 
 \caption{
 The dependence of the confusion noise from the diameter of the main mirror of the telescope for eight wavebands.
 The X axis~-- diameter of the main mirror in meters, Y axis~-- the confusion noise in mJy.
 Black line correspond to the SACS bands, gray line corresponds to the LACS.
} 
 \label{fig:sigma_vs_d}
\end{figure}

The estimations of confusion noise for different wavelengths as a function of the diameter of the main mirror of a telescope are given on Figure~\ref{fig:sigma_vs_d}. 

As was previously shown in \cite{2020AstL...46..298E}, for a telescope with 10m main mirror the confusion noise has a peak at  $\lambda\simeq300\mu m$.
But if we consider larger diameters of the main mirror the situation changes and the confusion gradually increases with wavelength.
It must be also noted that for the four shortest wavebands considered here (70, 110, 250, 350$\mu m$) the confusion rapidly decreases with the increase of angular resolution of the telescope.
For longer wavebands the decrease is much more gradual and the value of confusion itself does not depend on wavelength.

Here one significant note must be made.
The confusion noise does not only depend on the number counts and the angular resolution, but also on the effective angle of the beam shape.
Even if the side lobes are relatively weak, they can have significant area and bright objects in them can affect the confusion noise estimates.
This problem is left for the future research when the beam parameters of the Millimetron telescope will be known with precision.

\section*{Variability of the extragalactic background light}
Various transient events are of significant scientific interest in the far infrared.
For example, minor bodies of the solar system further than 30 a.u. from the Sun have temperatures lower than 50K and the maximum of their thermal emission
is situated in the far infrared.
Observations of the thermal emission are used for determination of important characteristics of the transneptunian objects, such as size and albedo \cite{2012A&A...541A..94V,2018A&A...618A.136V}.
As can be easily estimated, an object with surface temperature of 50~K situated at 30~a.u. and 1~km in diameter will have flux density in the 70$\mu m$ band about $7\times 10^{-5}$~mJy.
Such a flux density is detectable by the Millimetron.
In order to separate an object from the background (see Table~1) one can use the fact that the TNO moves at a speed of about 20 arcsec a day.
If the background does not change with time the task of subtracting two sequential images from one another to detect the TNO is possible. 
In the following section we will try to estimate the variability of the extragalactic background.

The emission of stars, gas and dust in galaxies can be considered constant on the timescale of astronomical observations.
But the same can not be said about active galactic nuclei.
As was shown in some papers before, see, e.g. \cite{2020AstL...46..298E} and references therein, AGN give significant contribution in the EBL SED at 70--110$\mu m$.
In order to estimate the variability of the background we created series of model maps for a year with time step of one  day.
To simulate the changing contribution from AGN into the background we utilized the following approach.

The methods of simulating the AGN variability were developed in many different papers, see, e.g. \cite{2016MNRAS.459.2787K, 2016ApJ...826..118K}.
We have used the approach from that paper. 
The variability curve of AGN is defined as stochastic process by a covariation signal matrix as:
\begin{equation} 
 cov(\Delta t) = \sigma_s^2 e^{-\left(\dfrac{|\Delta t|}{\tau}\right)^\beta} 
\end{equation} 
where $\tau>0$ is the decorrelation timescale,
$\sigma_s^2$ the signal dispersion, 
$\Delta t=t_i-t_j$ -- difference in time between $i$-th and $j$-th point on the curve, the parameter $0<\beta<2$, where $\beta=1$ corresponds to the  DRW process (Damped Random Walk).

In order to simulate a variability curve with $N$ points an $N\times N$ covariation matrix of signal is created:
 \begin{equation}
  C_{ij} = \left(
   \begin{array}{cccc}
    \sigma_s^2                                                & \sigma_s^2 e^{-\left(\frac{|t_1-t_2|}{\tau}\right)^\beta}    & \ldots     & \sigma_s^2 e^{-\left(\frac{|t_1-t_N|}{\tau}\right)^\beta}\\
    \sigma_s^2 e^{-\left(\frac{|t_2-t_1|}{\tau}\right)^\beta} & \sigma_s^2                                                   & \ldots     & \sigma_s^2 e^{-\left(\frac{|t_2-t_N|}{\tau}\right)^\beta}\\
    \vdots & \vdots & \ddots & \vdots\\
    \sigma_s^2 e^{-\left(\frac{|t_N-t_1|}{\tau}\right)^\beta} & \sigma_s^2 e^{-\left(\frac{|t_N-t_2|}{\tau}\right)^\beta}    & \ldots     & \sigma_s^2
   \end{array}
  \right)
 \end{equation}
 
This matrix must be affected by Cholesky decomposition $\mathbf{C}=\mathbf{L}^T \mathbf{L}$, where $\mathbf{L}$~-- upper triangular matrix.
Is is calculated the following way:
 \begin{equation}
  \begin{array}{l}
   l_{11} = \sqrt{a_{11}}, \\
   l_{j1} = \dfrac{a_{j1}}{l_{11}}, \quad j \in [2,n], \\
   l_{ii} = \sqrt{a_{ii}-\sum\limits_{p=1}^{i-1}{l_{ip}^2}}, \quad i \in [2,n], \\
   l_{ji} = \dfrac{1}{l_{ii}}\left(a_{ji}-\sum\limits_{p=1}^{i=1}{l_{ip}l_{jp}}\right), \quad i \in[2,n-1],\quad j \in[i+1,n], \\
  \end{array}
 \end{equation}
where $l_{ij}$ - element of the Cholesky matrix, $a_{ij}$ - element of the covariation matrix.
The calculations are performed from top to bottom and from left to right.
Structural function is defined as:
\begin{equation} 
 SF_{obs}(\Delta t)=\sqrt{\dfrac{1}{N_{\Delta t pairs} }\sum_{i=1}^{N_{\Delta t pairs}}(y(t)-y(t+\Delta t))^2}, 
\end{equation} 
where $y(t)$ - variability curve. 
The summation is done for all pairs of observations separated by time interval $\Delta t$.
Asymptotic variability $SF_\infty$ in this case is relied to the variability amplitude as $SF_\infty=\sqrt{2}\sigma$.
So the variability curve can be defined as $\mathbf{y}=\mathbf{Lr}$, where $\mathbf{r}$ is the vector of Gaussian deviation with dispersion of unity.

Such a random process is created the following way.
The chain of data points starts with $s_1=G(\sigma^2)$, where $G(x^2)$~-- Gaussian deviation with $x^2$ dispersion.
Subsequent  points are defined as:
 \begin{equation}
 s_{i+1}=s_i e^{-\Delta t/\tau}+G[\sigma^2(1-e^{-2\Delta t/\tau})],
 \end{equation}
where $\Delta_t=t_{i+1}-t_i$~-- time interval.

In order to create simualated lightcurves of AGN we need to connect the mass of central black hole $M_{BH}$ and the infrared luminosity $L$ with parameters  $\sigma$ and $\tau$.
Such a relation was obtained by~\cite{2010ApJ...721.1014M}: 
 \begin{equation}
 \begin{array}{r}
  \log{f} = A + B\log{\left(\dfrac{\lambda_{RF}}{4000\mbox{\AA}}\right) + C(M_i+23)}+ \\
  +D\log{\left(\dfrac{M_{BH}}{10^9 M_\odot}\right)} + E\log{(1+z)}
 \end{array}
 \label{eq:macleod}
\end{equation}
where $f$~-- parameter $SF_\infty$ or $\tau$,
$\lambda_{RF}$~--wavelength in the source reference frame, $M_i$~-- absolute magnitude in the SDSS $i$ band ($7625\mbox{\AA}$), $z$~-- redshift.
Numerical values of these coefficients were taken from the same paper.

In order to derive absolute magnitude from the bolometric luminosity the information about the SED shape is required.
We have used the SED of the type 1 AGN from~\cite{2017ApJ...841...76L}.
Photometric magnitude is calculated by the well known formula:
\begin{equation}
 m_{AB} = -2.5\log_{10}{f_\nu}-48.60,
\end{equation}
where flux is expressed in units erg~s${}^{-1}$~cm${}^{-2}$~Hz${}^{-1}$.

\begin{figure}
 \includegraphics[width=0.49\columnwidth]{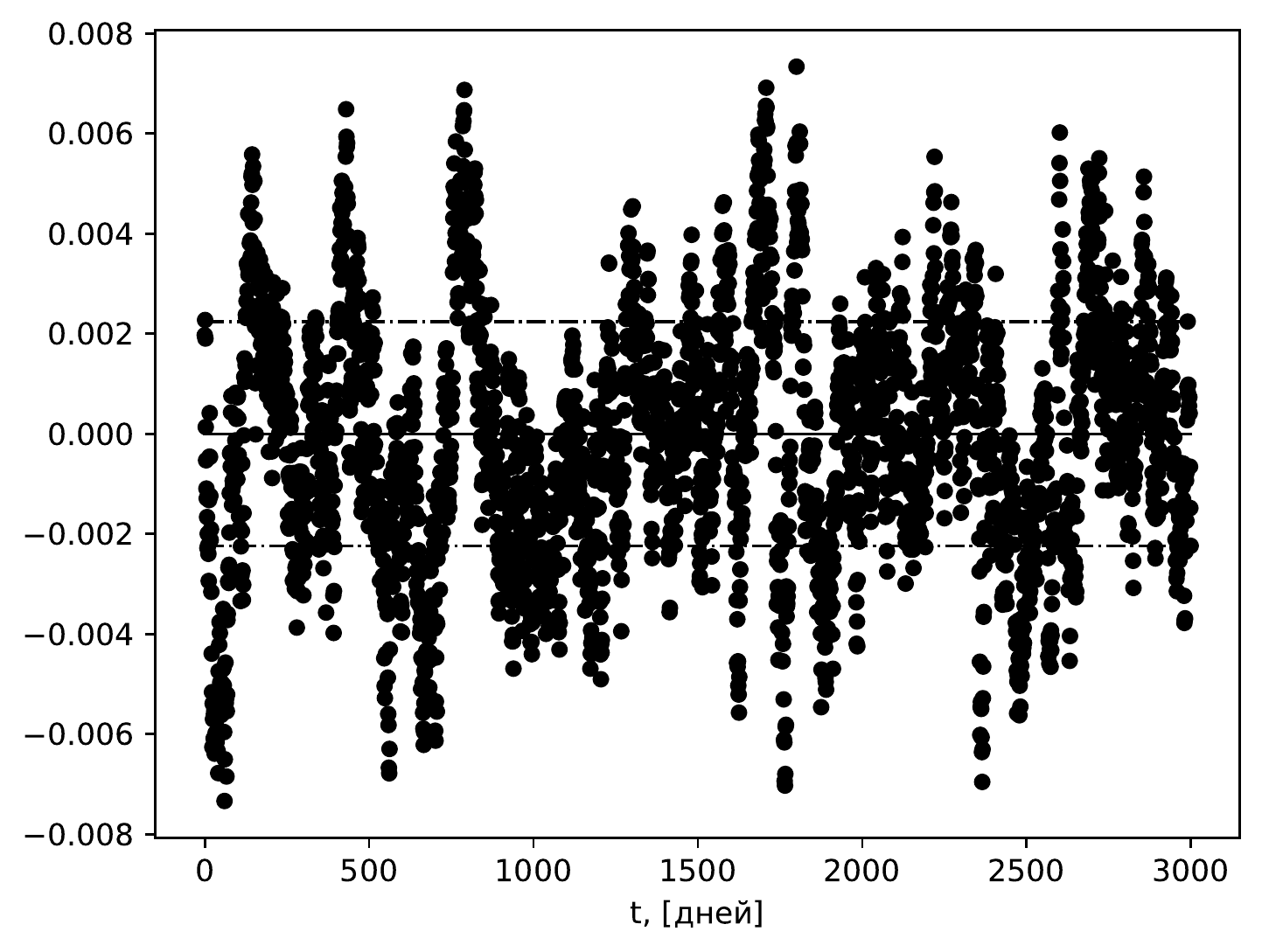}
 \includegraphics[width=0.49\columnwidth]{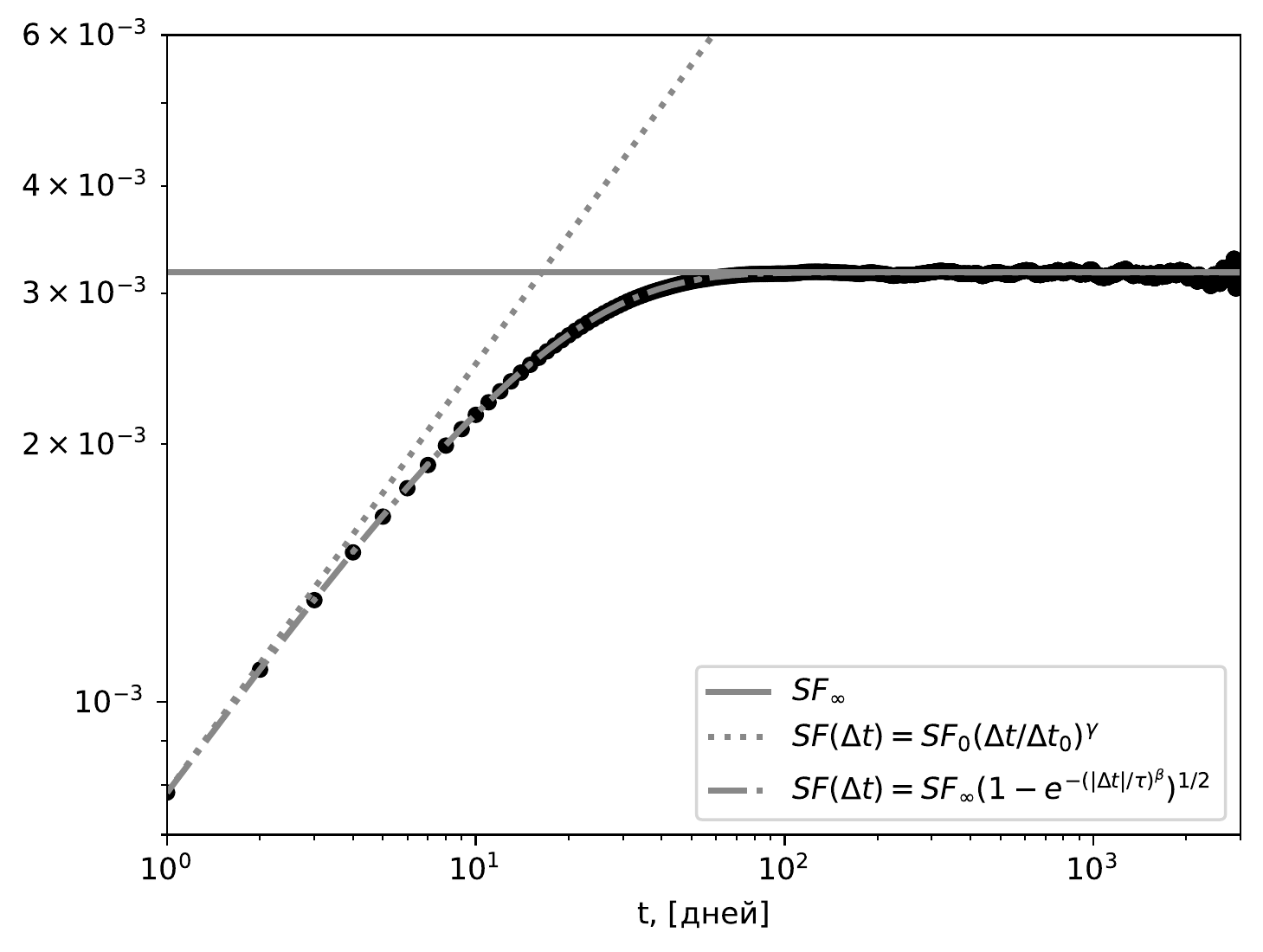}
 \caption{
 \textbf{Left panel}.
  An example of a model lightcurve of AGN with parameters: $\log{SF_\infty}=-2.8$ and $\log{\tau}=1.2$.
  The X axis~-- time in days, Y axis~-- change of absolute magnitude.
  Solid horizontal line shows $\Delta S=0$.
  Two dashed lines show $\Delta S=+\sigma_s$ and $\Delta S=-\sigma_s$.
  \textbf{Right panel}.
  Structural function derived from model data.
  Black circles~-- estimations from the model curve shown on the left.
  Horizontal solid line shows the $SF_\infty$ value.
  Gray dot-dashed line shown the theoretical prediction by equation~\ref{eq:SF_fit}.
  Dashed line is the asymptote for the $\gamma=0.5$ case.
 }
 \label{fig:var_model}
\end{figure}

As a next step we created a library of lightcurves for the following parameter grid: $-2.8\leq SF_\infty\leq1.0$, $0.6\leq\tau\leq3.7$, with 0.1 step.
For each pair of values a 100 random curves were created and for each object with certain parameters a random lightcurve was selected from these.
An example of such a simulated lightcurve is shown on the left panel of Figure~\ref{fig:var_model}, the corresponding structural function is shown on the right panel.
The structural function for the DRW process is defined as:
\begin{equation}
\label{eq:SF_fit}
SF(\Delta t)=SF_\infty(1-e^{-(|\Delta t|/\tau)^\beta})^{1/2}.
\end{equation}
As can be seen, theoretical structural function and the one obtained from a model curve agree with each other reasonably well.

After each objects was assigned a model lightcurve the series of maps were created.

In order to numerically evaluate the variability in each pixel of the map we utilized the following two criteria.
Firstly it is possible to simply measure the dispersion of flux in each pixel.
We also utilized the widely used variability criterium  defined as a width of distribution of flux difference between points with certain time separation (see, e.g.  \cite{Smith_2018}).
Figure~11 shows estimations for the intervals of 1, 2, 10 and 50 days.
An obvious trend can be seen on these plots. 
With the increase of the time interval the shape of the structural function approaches the dispersion value.

The following conclusion can be made.
For the Millimetron telescope the significant amount of pixels will show noticeable variability only at four short wavebands: 70, 110, 250 and 350$\mu m$.
It must be noted that this effect cannot be attributed only to the decrease of angular resolution with wavelength.
The lower right panel of Figure~11 demonstrates the variability estimates for the 70$\mu m$ waveband but with the resolution of 2000$\mu m$ band.
This corresponds to the diameter of the main mirror of 35cm.
As can be clearly seen, even at such low resolution the variability is clearly present.

For illustration purposes we also created plots of dispersion of logarythmic value of flux in pixels vs flux (see fig.~\ref{fig:sigma_vs_flux}).
This plot illustrates the decrease of variability with wavelength even more clearly, see the scale of vertical axes.
Another important conclusion is that not just several bright objects are variable, but many faint parts of the model map.

Summing up the above said we can conclude that for wavebands 70--350$\mu m$ the variability of extragalactic background on timescales from 1 day to 1 year manifests at fluxes about 1$\mu$Jy and lower.
It poses some limitations on the possibility of extraction of faint transient events, TNO or rogue planets from the background.
For observations of objects with expected fluxes lower than 1~$\mu$Jy it is necessary to devise a method to separate their signal from fluctuations of the extragalactic background.

 \begin{figure}
 \centering
 \includegraphics[width=0.32\textwidth]{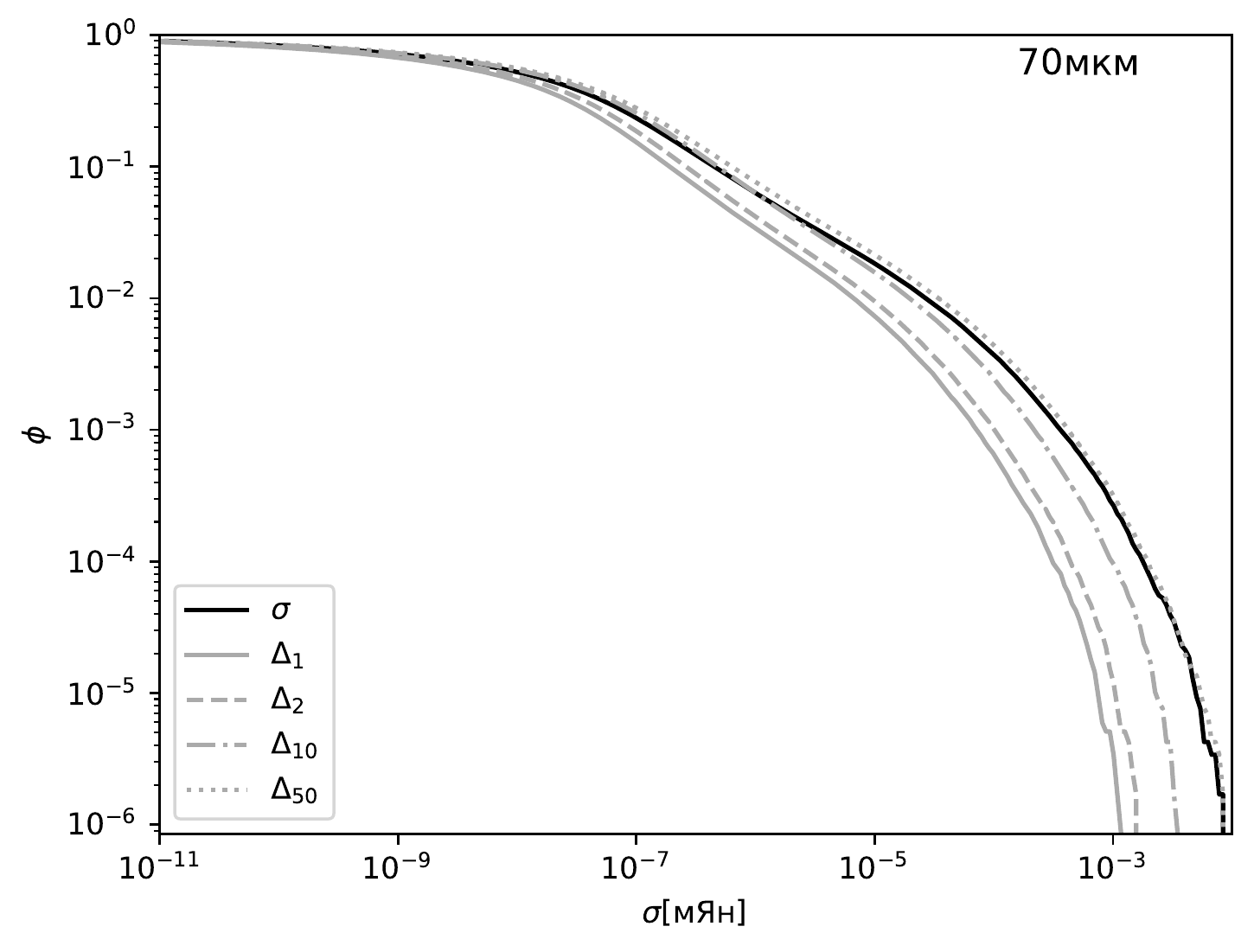}
 \includegraphics[width=0.32\textwidth]{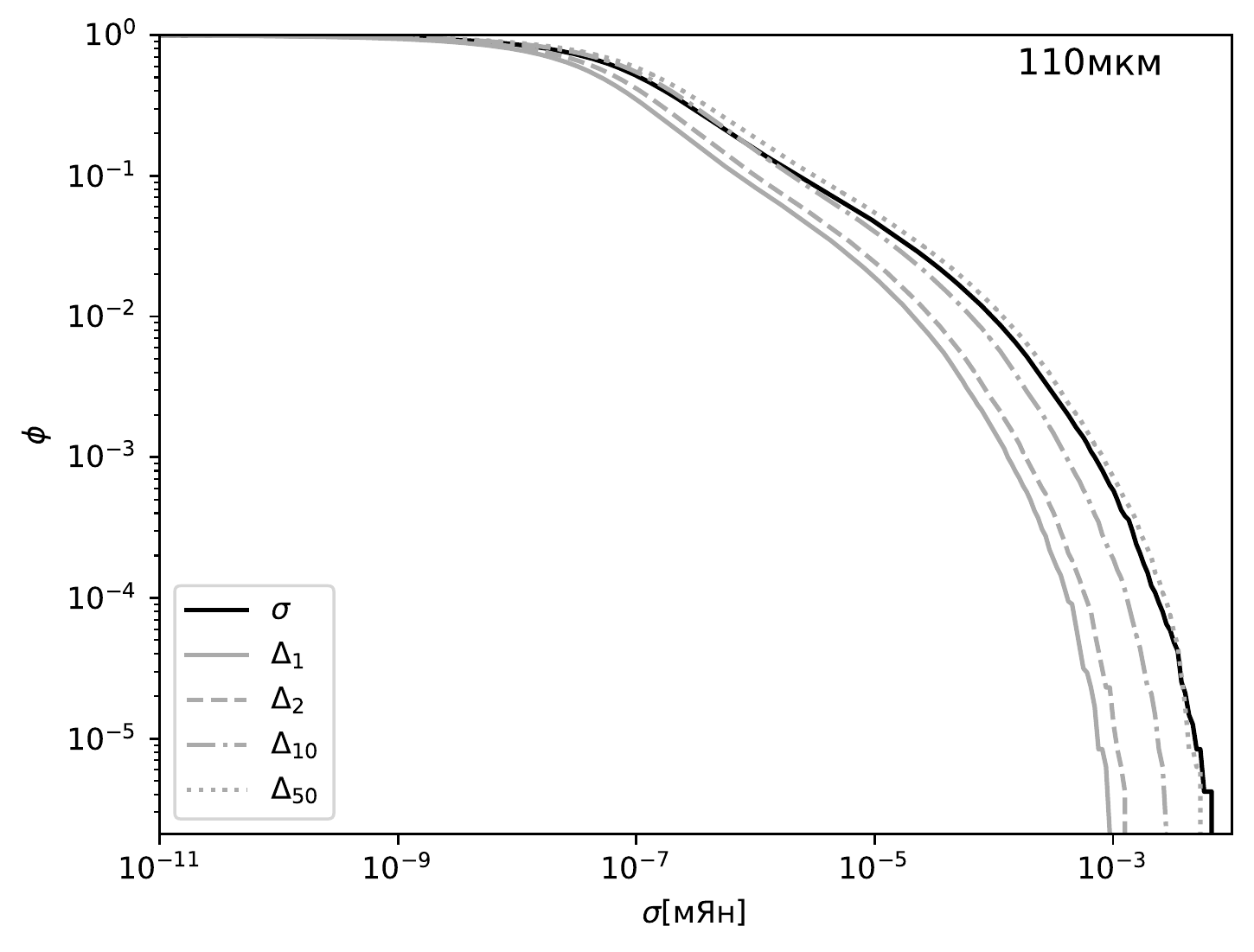}
 \includegraphics[width=0.32\textwidth]{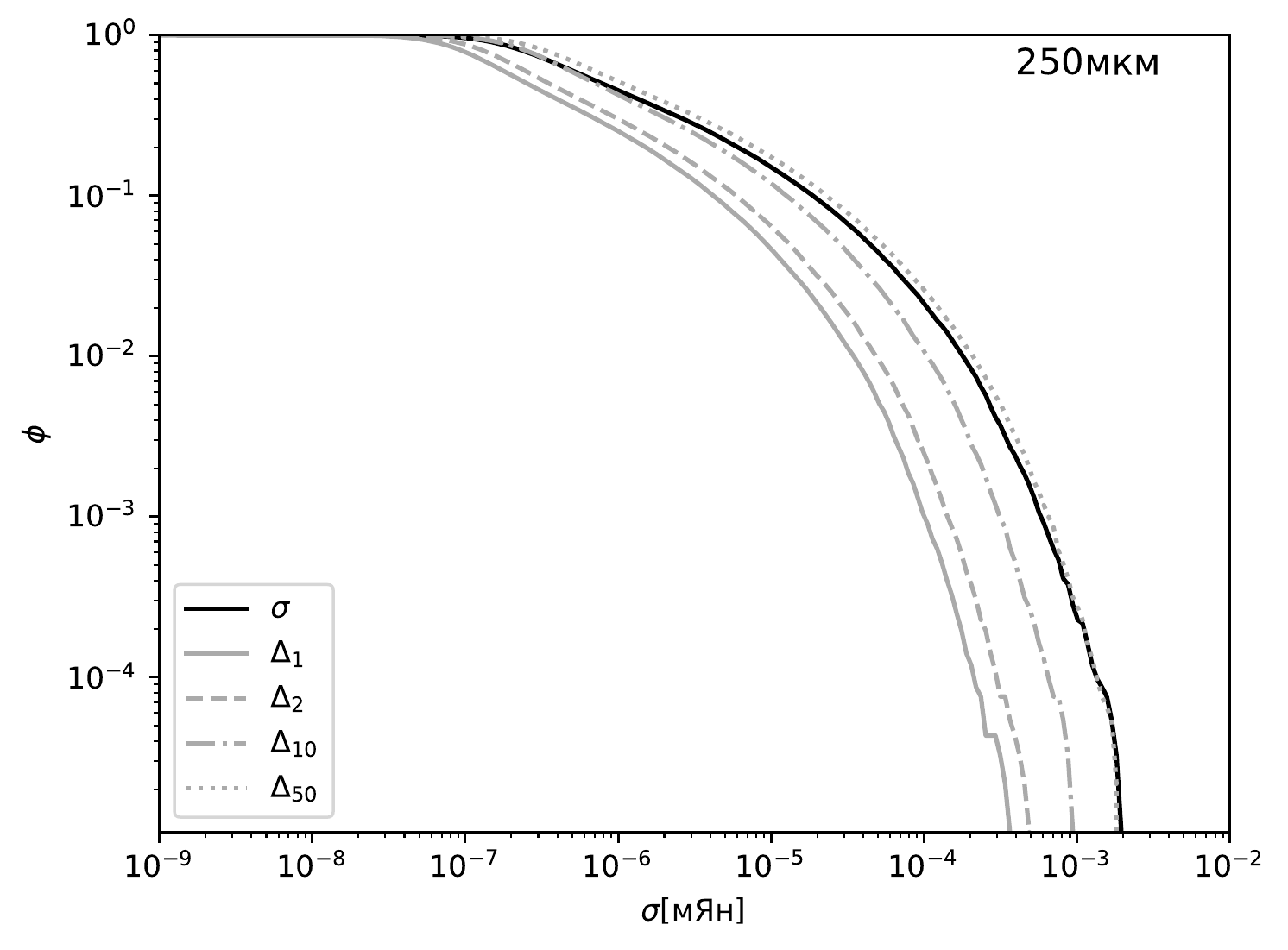}\\
 \includegraphics[width=0.32\textwidth]{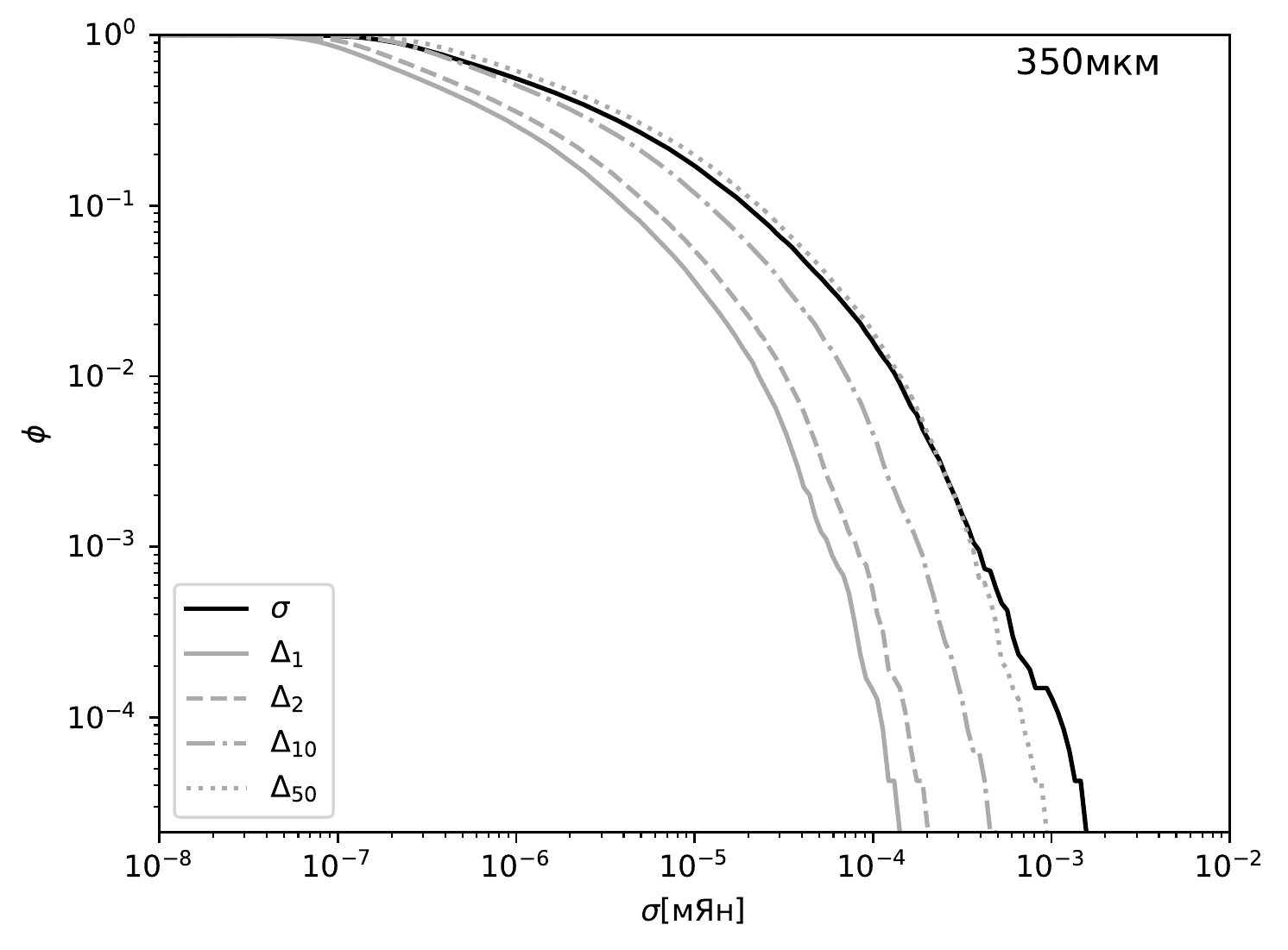}
 \includegraphics[width=0.32\textwidth]{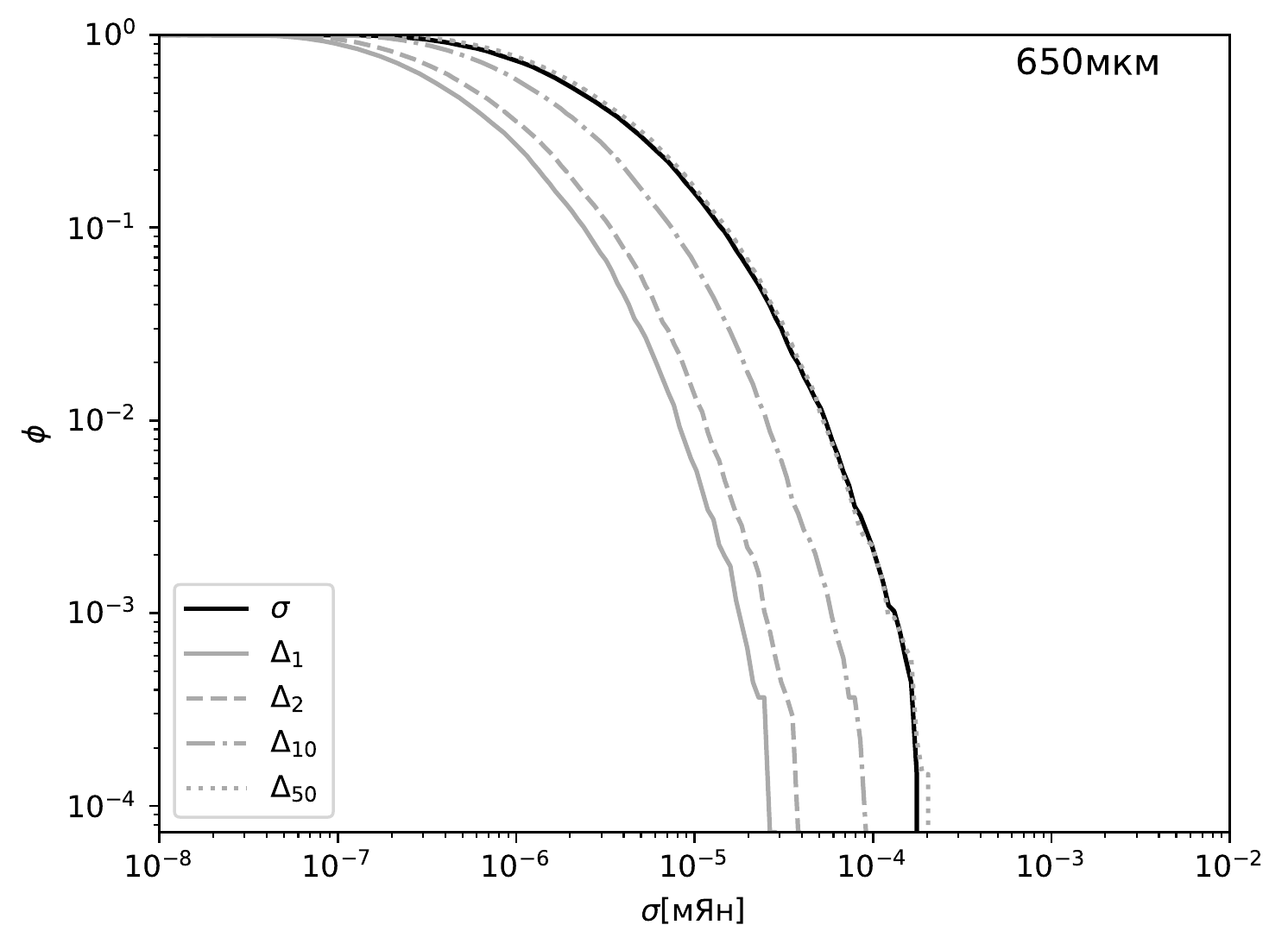}
 \includegraphics[width=0.32\textwidth]{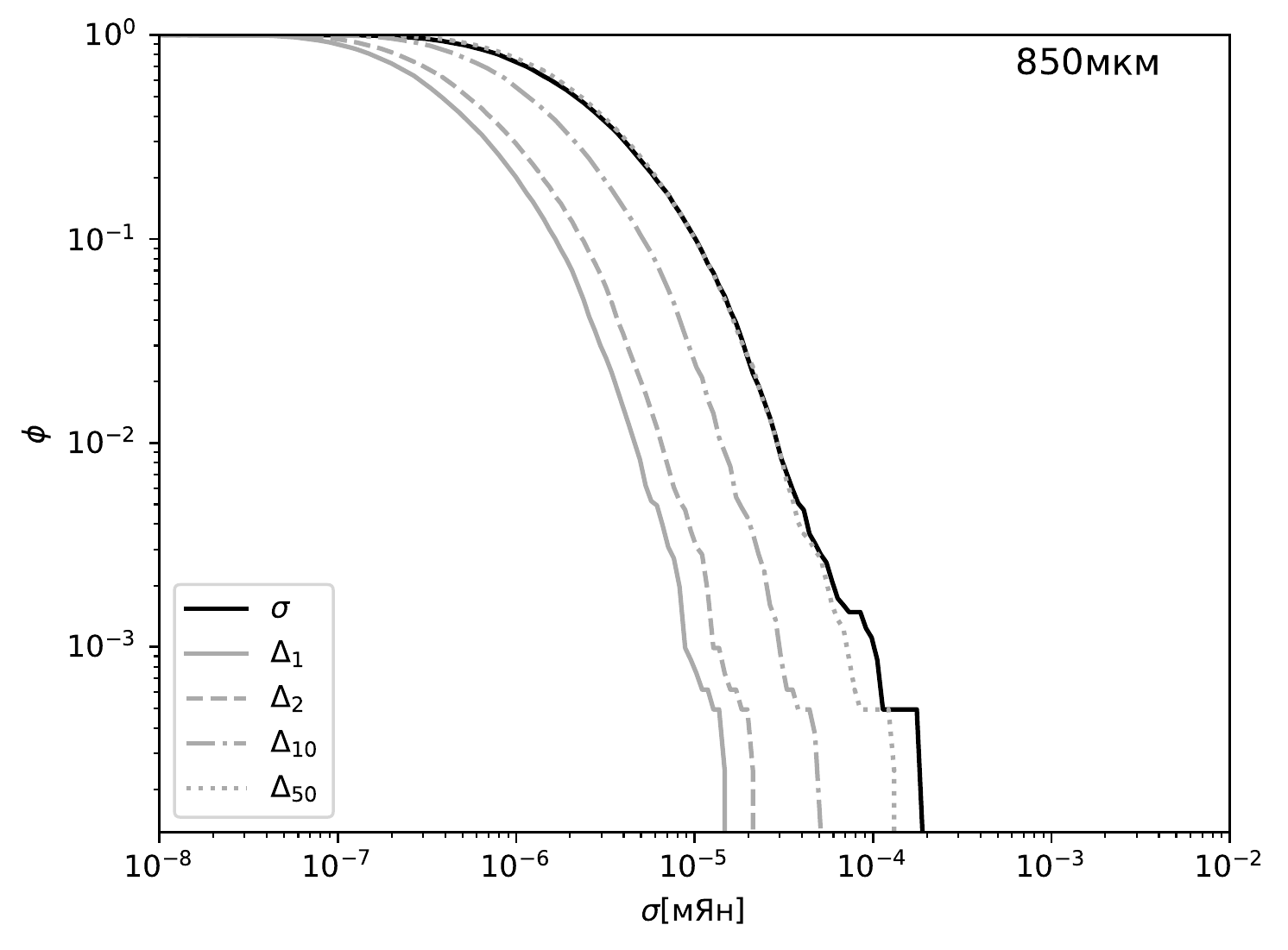}\\
 \includegraphics[width=0.32\textwidth]{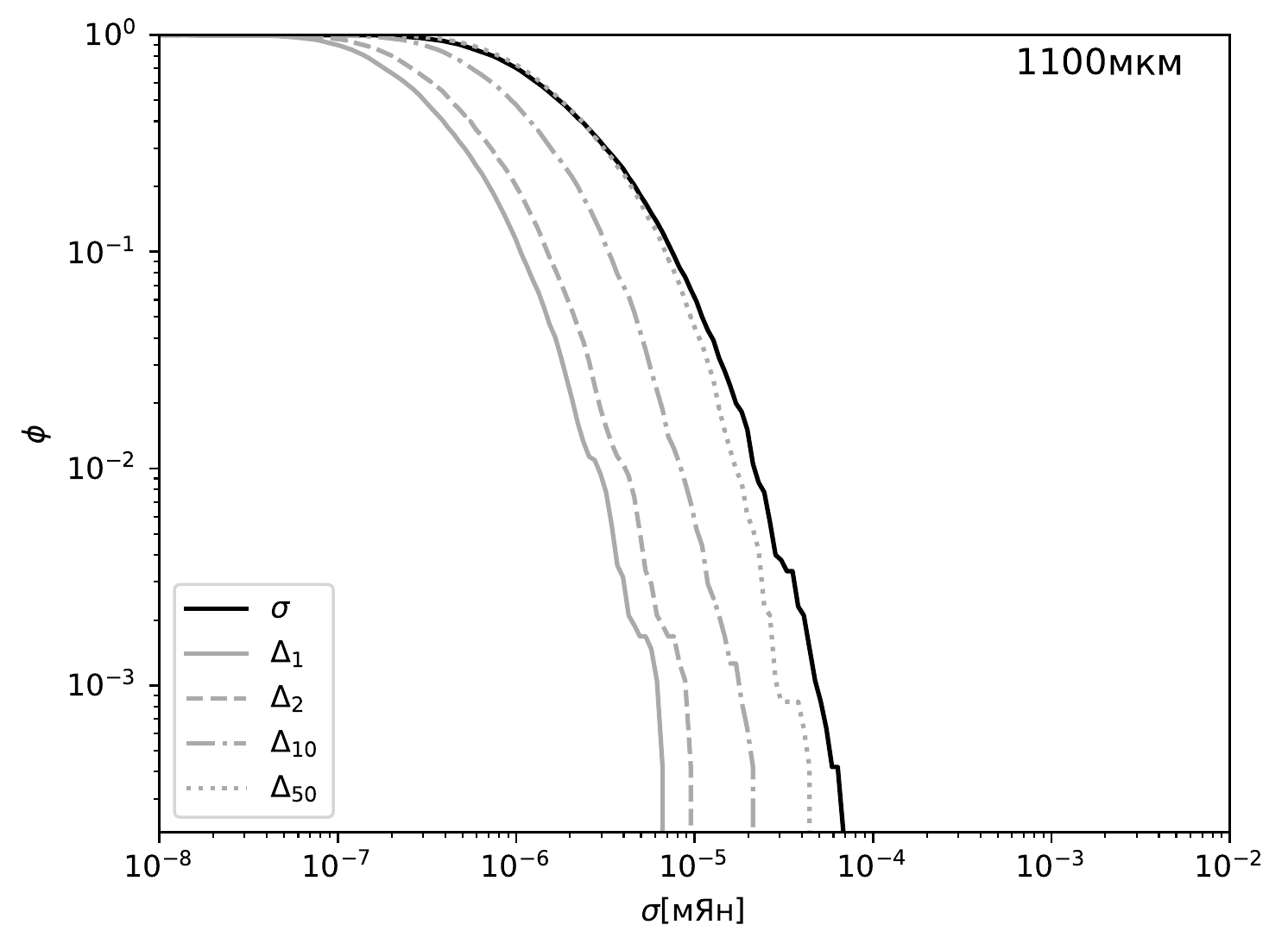}
 \includegraphics[width=0.32\textwidth]{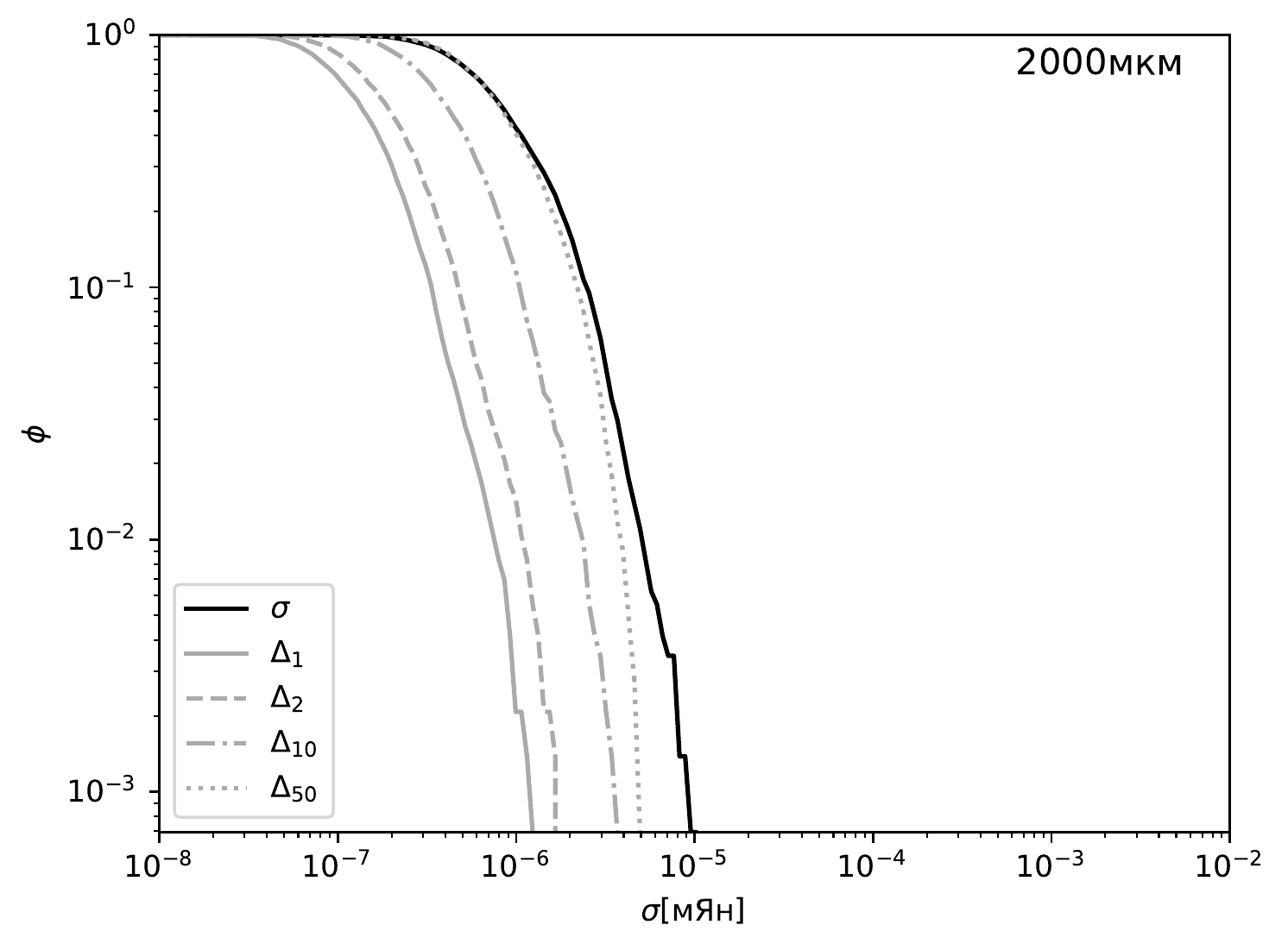}
 \includegraphics[width=0.32\textwidth]{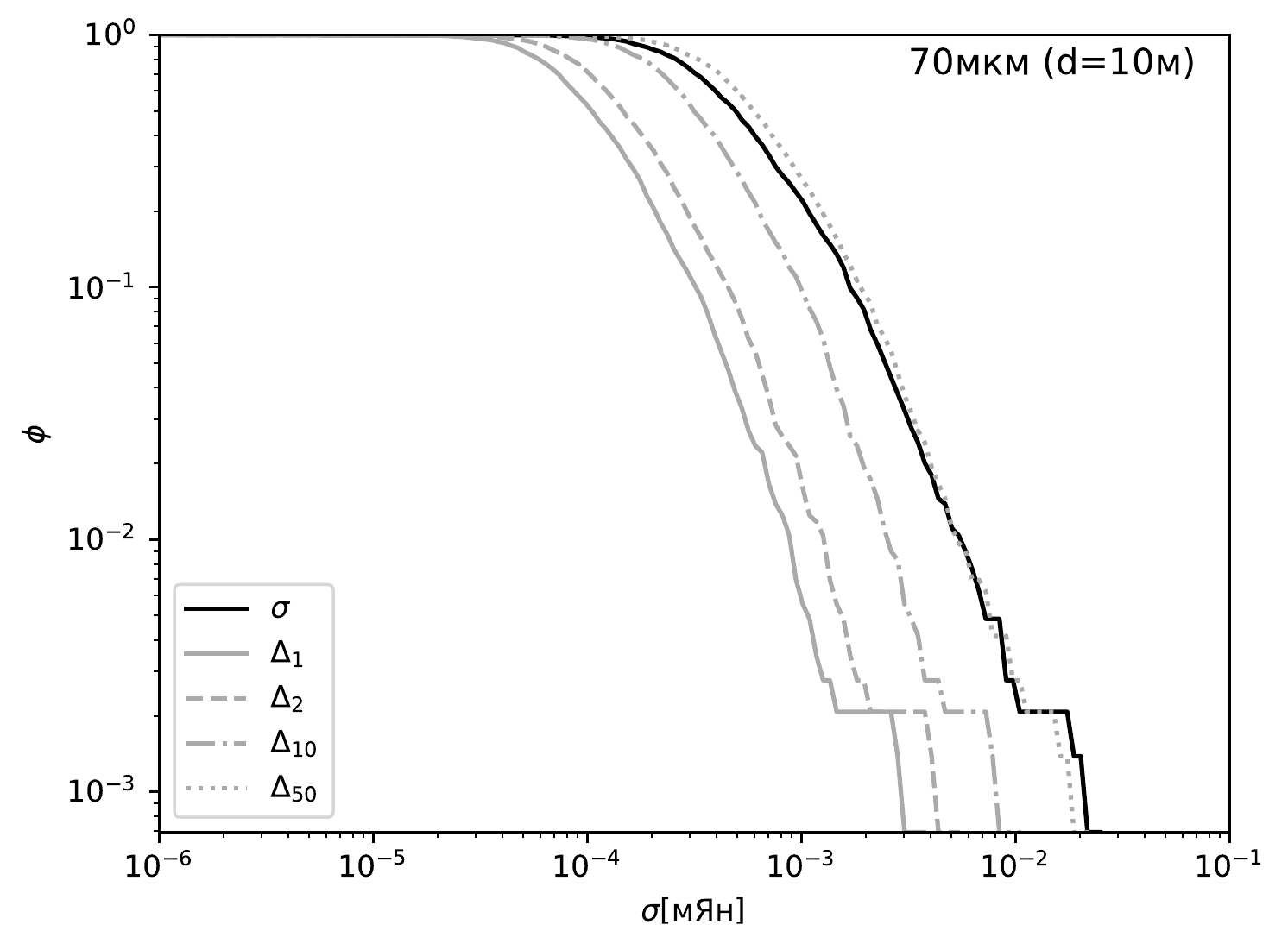}
 \caption{
 Integral distribution of the variability parameters on model map for eight bands of Millimetron detectors.
 From left to right, from top to bottom: 70, 110, 250, 350, 650, 850, 1100 and 2000~$\mu m$.
 The X axis~-- variability value in mJy, the Y axis~-- fraction of area of the model map.
 The Y axis range is chosen from 1 pixel to 100\%.
 The values of dispersion $\sigma$, $\Delta_1$, $\Delta_2$, $\Delta_{10}$ and $\Delta_{50}$ are shown.
 The bottom right panel shows the plot for 70$\mu m$ with angular resolution corresponding to the case of 2000$\mu m$ and 10m main mirror.
 }
 \label{fig:map_int}
\end{figure}

\begin{figure}
 \centering
  \includegraphics[width=0.32\textwidth]{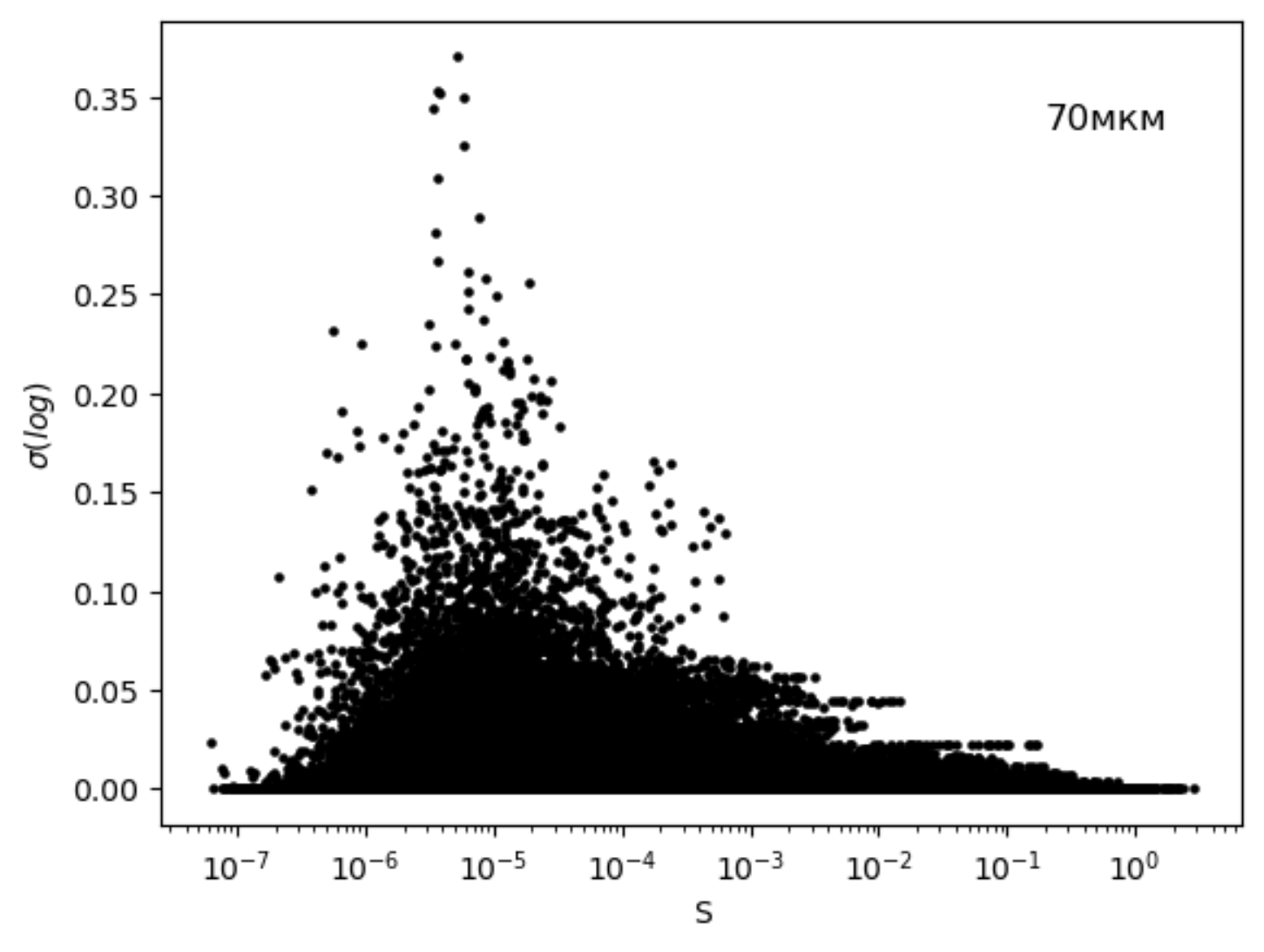}
  \includegraphics[width=0.32\textwidth]{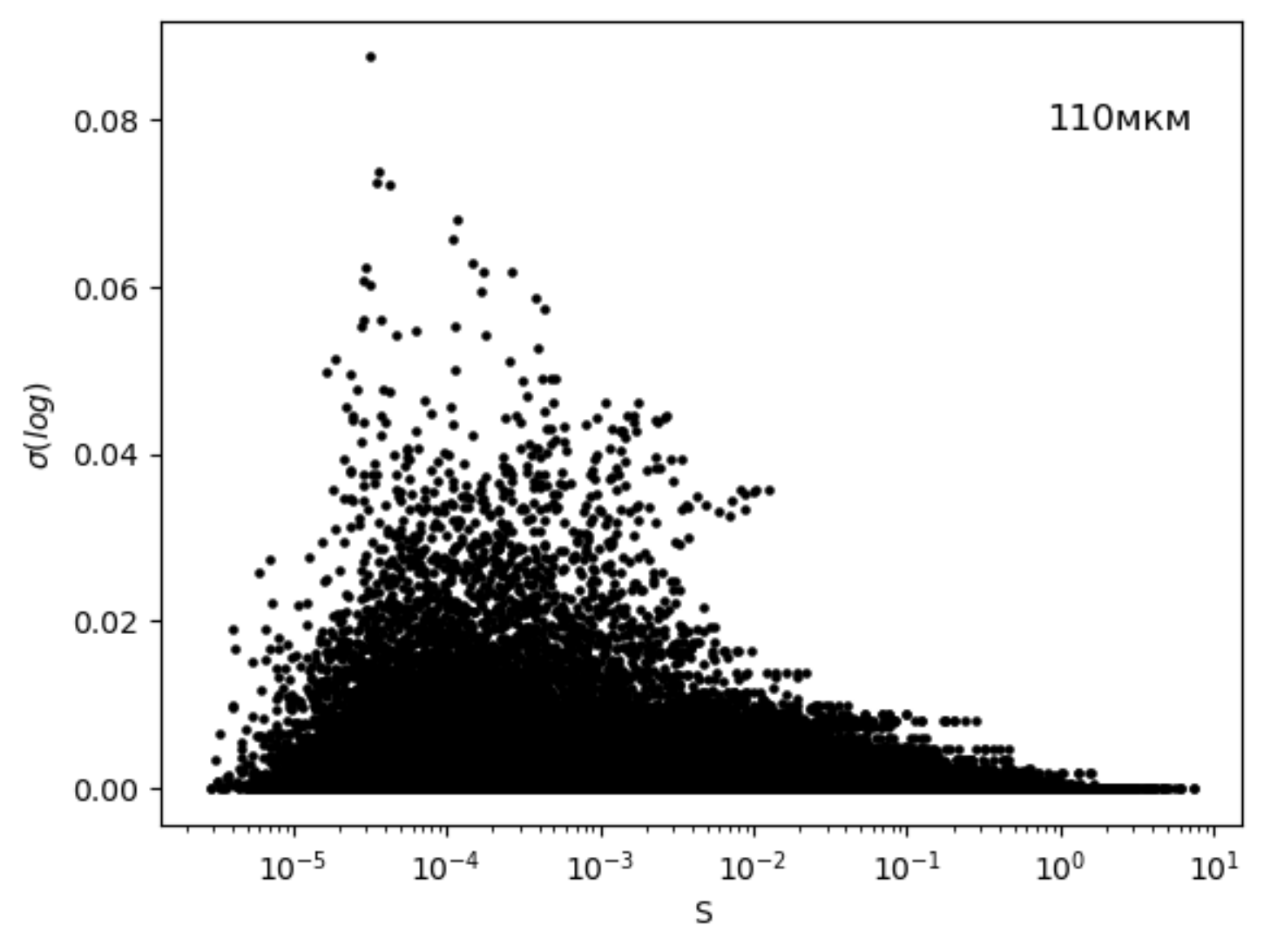}
  \includegraphics[width=0.32\textwidth]{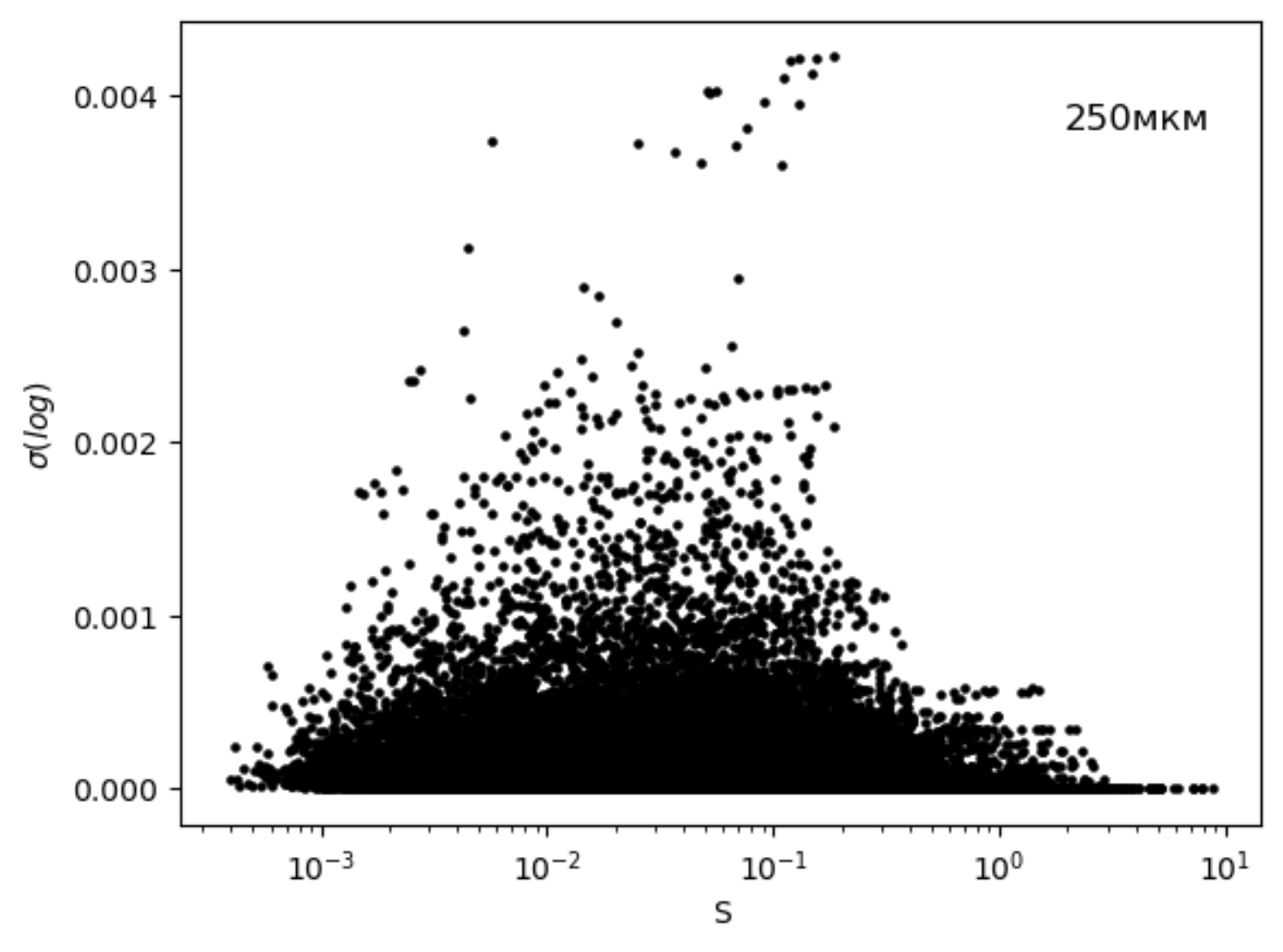}\\
  \includegraphics[width=0.32\textwidth]{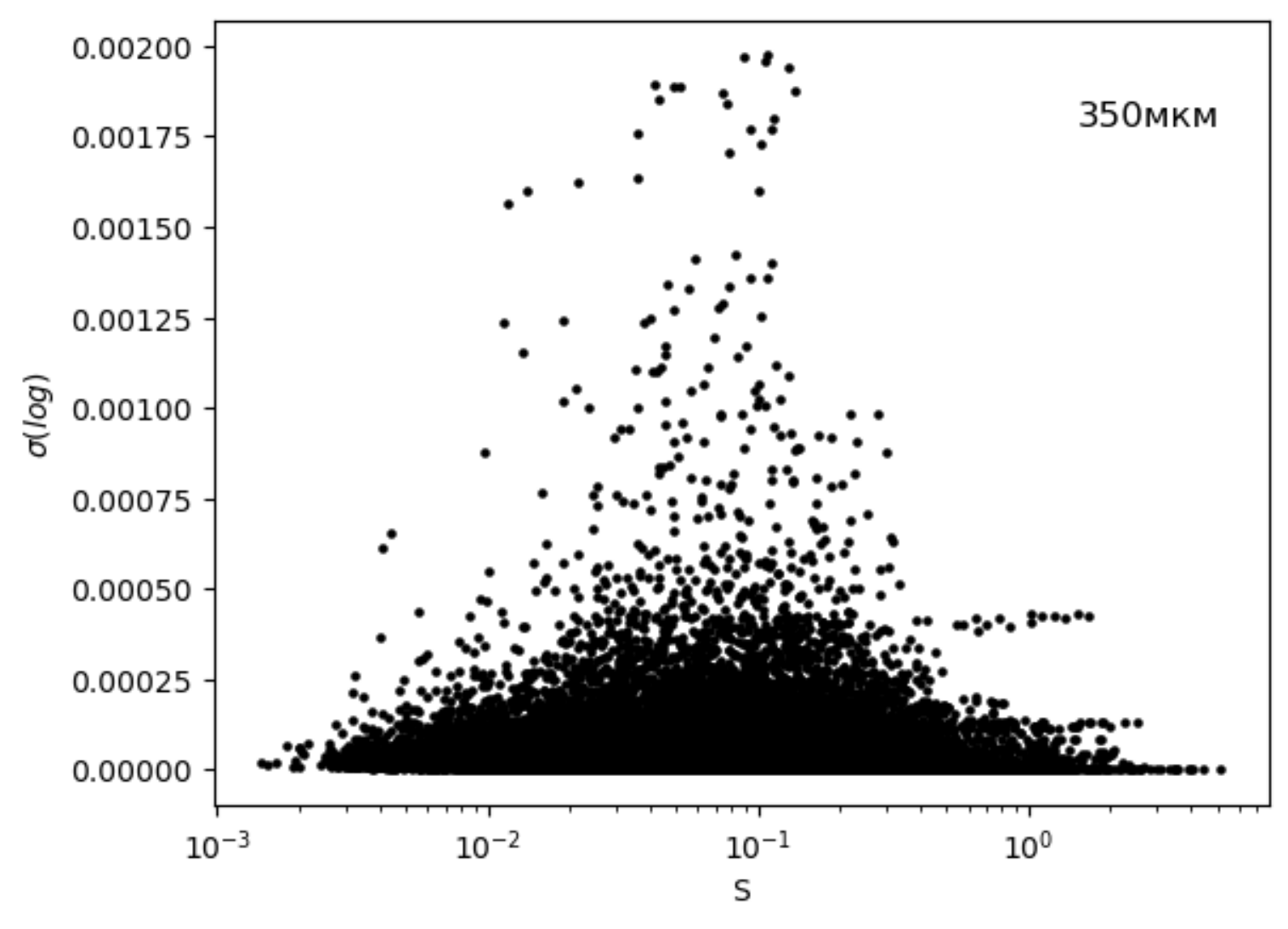}
  \includegraphics[width=0.32\textwidth]{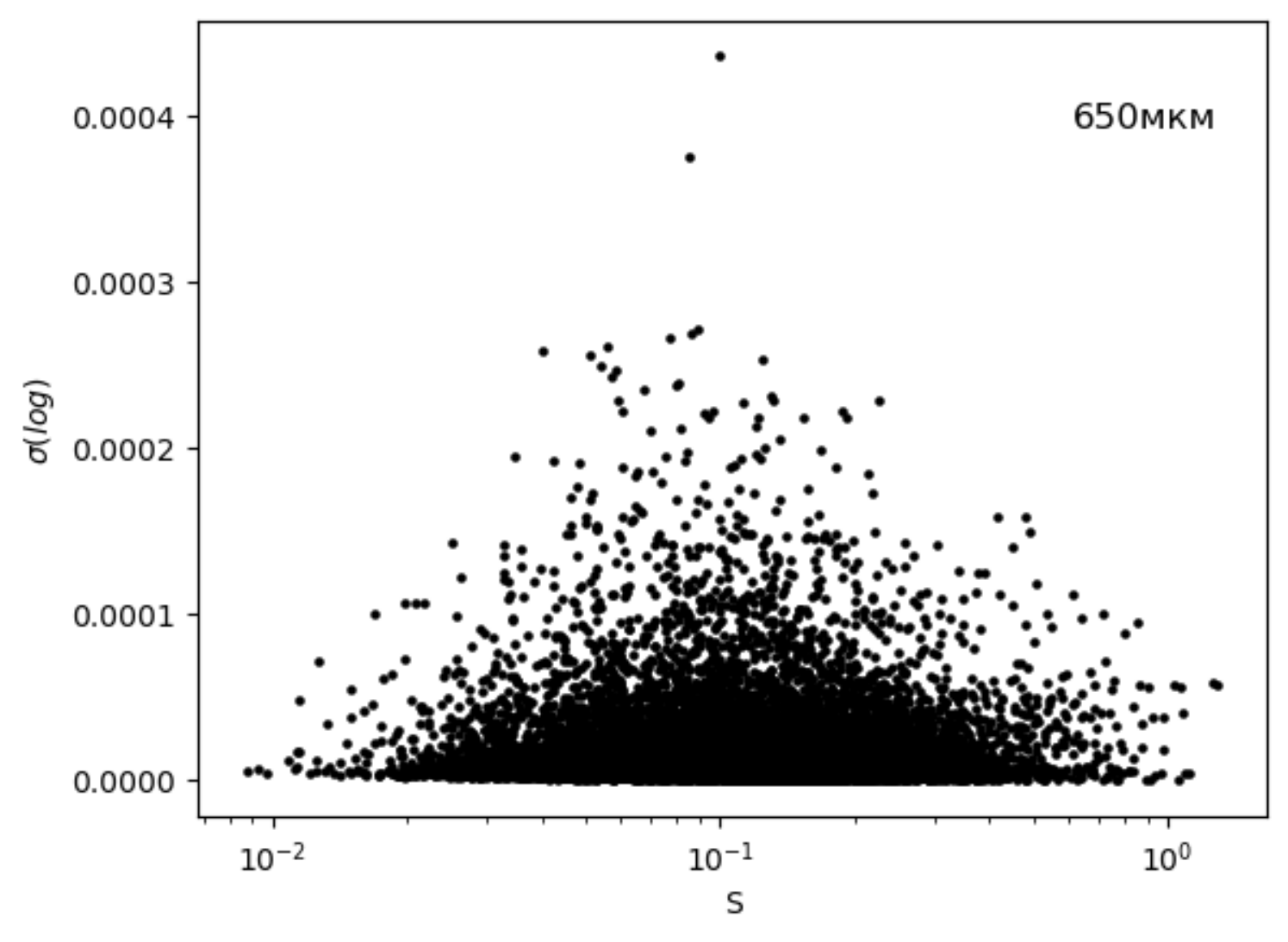}
  \includegraphics[width=0.32\textwidth]{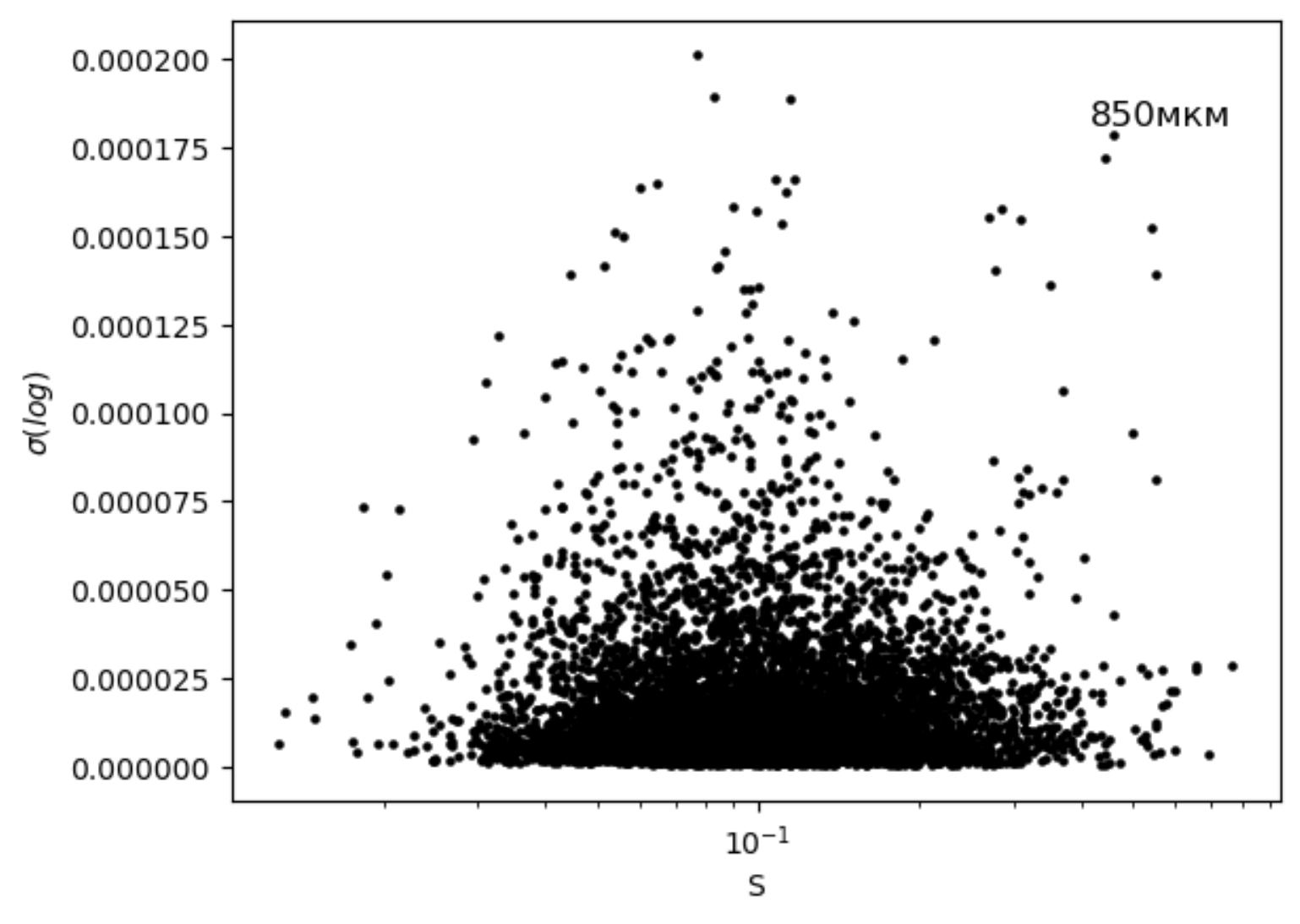}\\
  \includegraphics[width=0.32\textwidth]{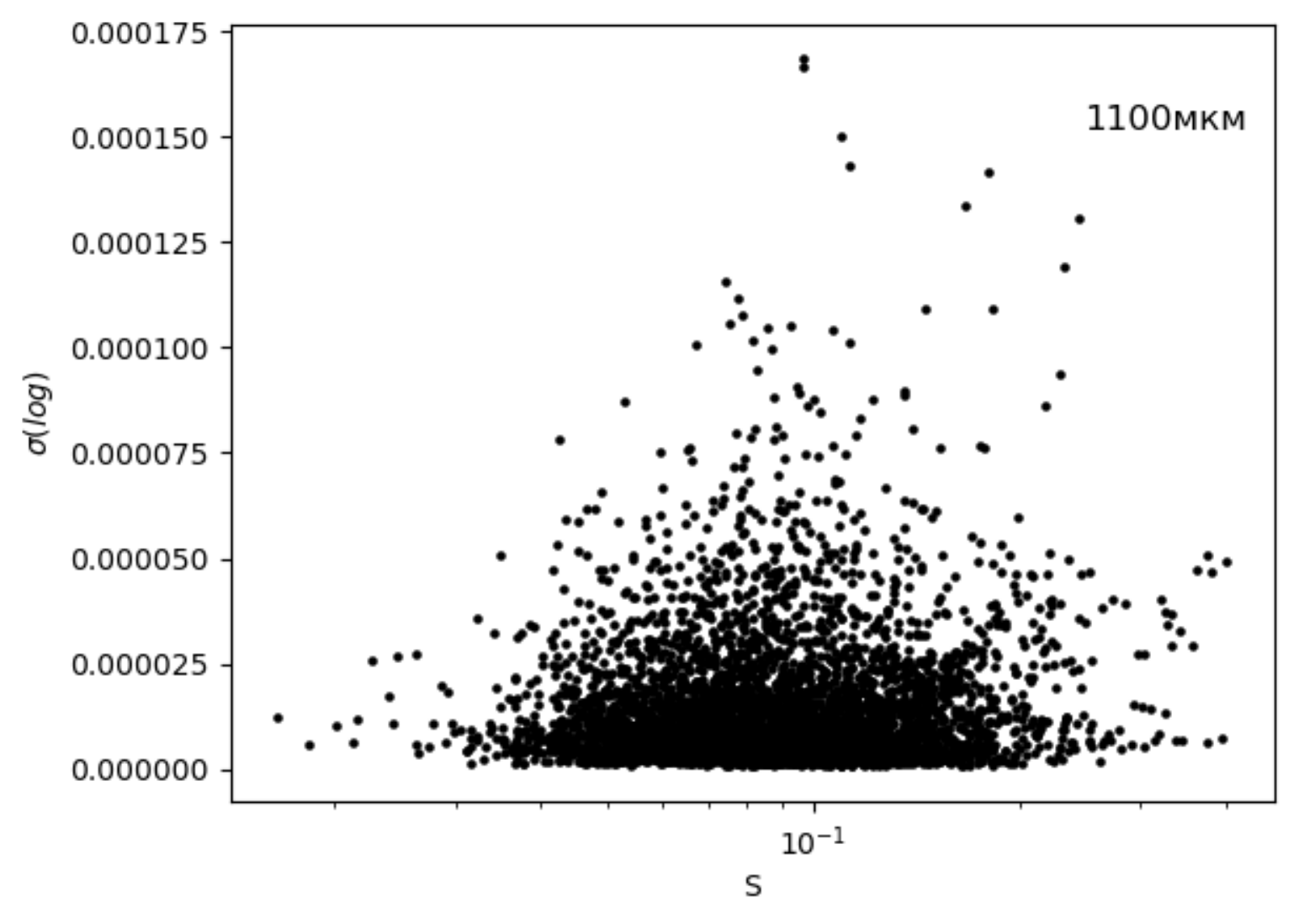}
  \includegraphics[width=0.32\textwidth]{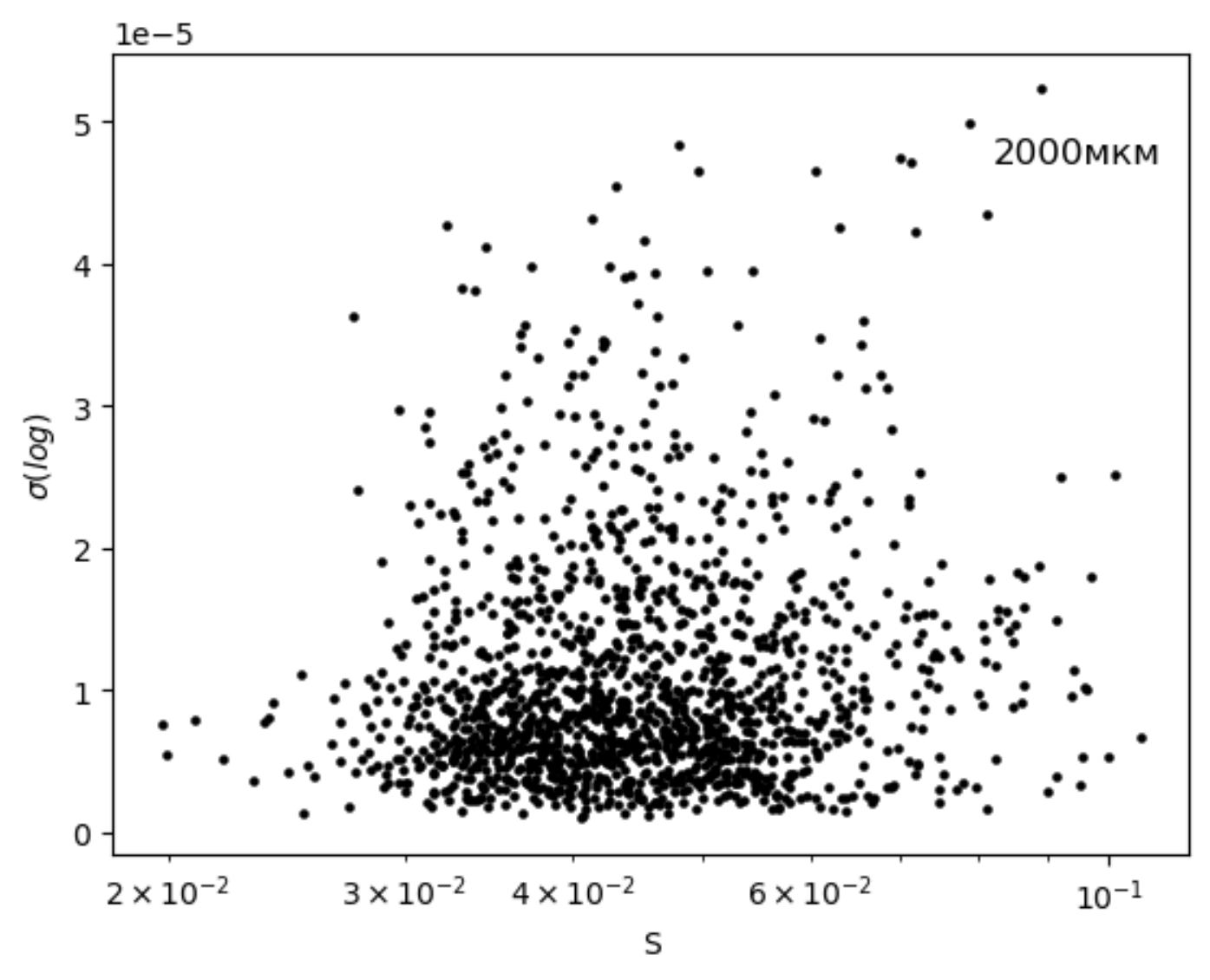}
  \includegraphics[width=0.32\textwidth]{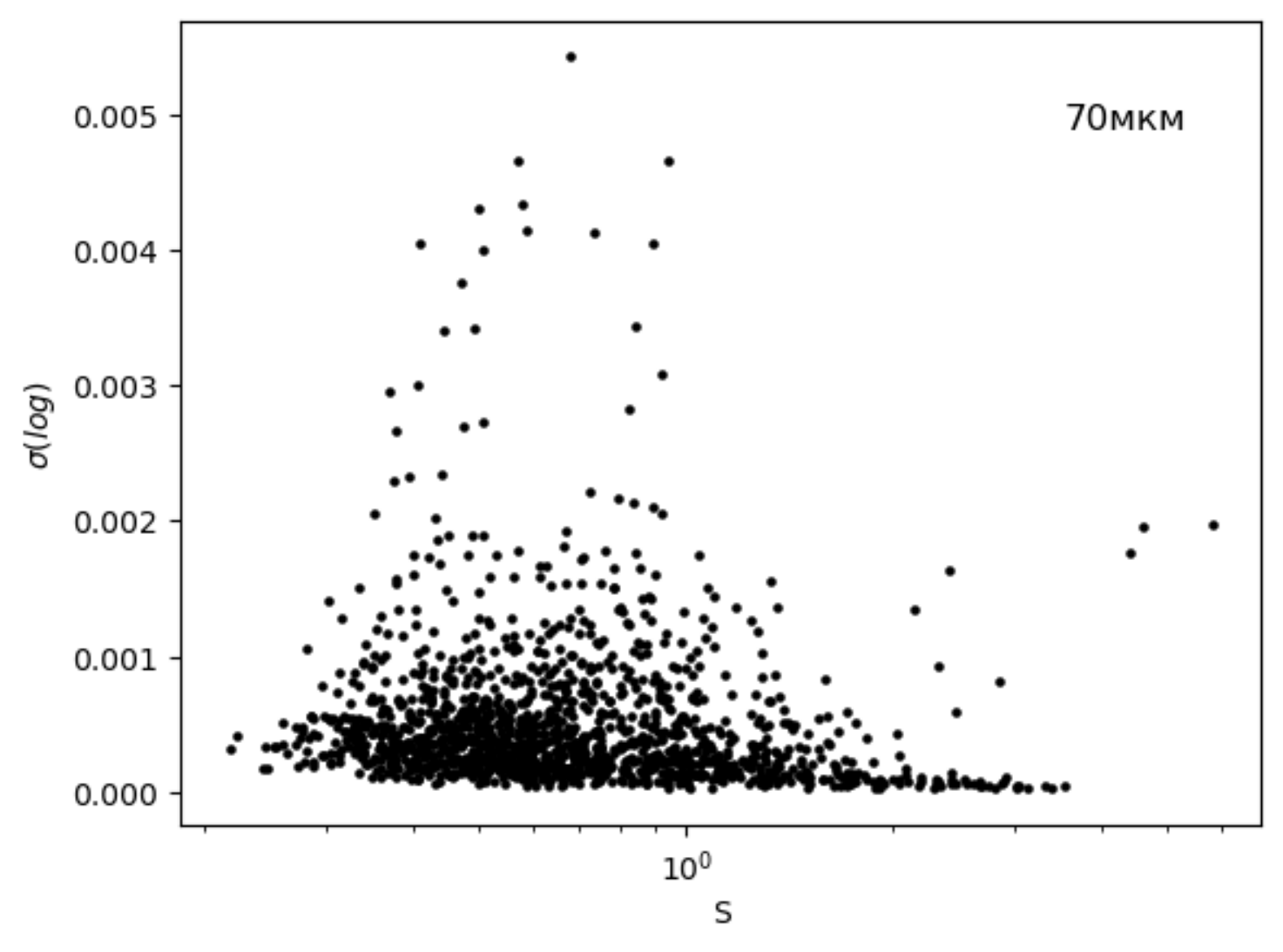}
 \caption{
 The dependence of the dispersion of the logarythmic flux in pixels due to the variability on the pixel flux.
 Eight panels correspond to the eight detectors of the Millimetron.
 From left to right, from top to bottom: 70, 110, 250, 350, 650, 850, 1100 ¨ 2000$\mu m$.
 Lower right panel shows data for the 70$\mu m$ with resolution that corresponds to 2000$\mu m$ with 10m diameter of the mirror (equals 35cm mirror).
 }
 \label{fig:sigma_vs_flux}
\end{figure}

\section*{Conclusions}
\label{sec:results} 
In this paper we have investigated the contribution to the confusion noise from objects on different redshifts, with different luminosities and color characteristics.
Calculations were performed for eight wavebands of the detectors of the Millimetron space mission with the d=10m diameter of the main mirror.
We also obtained the dependence of the confusion noise on the diameter of the main mirror of the telescope for different wavelengths.

The following conclusions can be made.

\begin{itemize}
\item 
 The presence of Large Scale Structure is a crucial factor that must be taken into account.
 Otherwise the confusion noise will be significantly underestimated.
 \item
  Gravitational lensing does not significantly affect the confusion noise.
 \item
  The objects in the following redshift range give the main contribution the the confusion noise.
  The lower redshift boundary does not depend on wavelength and is about $z_{min}\sim 0.5-0.6$, while upper redshift boundary gradually decreases from $\sim4$ to $\sim3$ while the wavelength increases from 70 to 2000$\mu m$.
 \item
  On short wavelengths the confusion noise is created mainly by objects in the luminosity range $10^7L_\odot$ -- $10^9L_\odot$ while at large wavelengths
  it is created by bright galaxies with $L\geq10^{10}L_\odot$.
 \item
  On short wavebands the confusion noise is created by objects with steep spectra that significantly contribute to the adjacent longwave band and give
  low contribution to the adjacent short waveband. 
  Toward longer wavelengths the situation gradually changes to the opposite.
 \item 
  We have evaluated the variability on model maps.
  In order to do so we created a series of maps for a 1 year timespan with step of 1 day.
  It was found that variability is present at four shortwave bands of Millimetron.
  Numerical estimates are given.
\end{itemize}

We would like to stress the possibility of estimation of galaxy evolution parameters from the confusion noise parameters derived from future actual observations.

Authors would like to sincerely thank the referee for useful comments that helped to clarify many aspects of this work.

This study was supported by Lebedev Physical Institute of the Russian Academy of Sciences (project NNG-41-2020).
E.~V.~Miheeva and V.~N.~Lukash were also supported by the Russian Scientific Foundation (project no. 19-02-00199).

\end{document}